%% file: main.tex
\documentclass[fleqn,10pt,twoside,twocolumn]{oldplainarticle-cyr}

\usepackage[T2A]{fontenc}     
\usepackage[utf8]{inputenc}   

\usepackage{paratype}         
\makeatletter
\DeclareFontFamilySubstitution{T2A}{ptm}{PTSerif-TLF}
\makeatother
\makeatletter
\DeclareFontShape{T2A}{PTSerif-TLF}{m}{sc}{<-> ssub * PTSerif-TLF/m/n}{}
\DeclareFontShape{T2A}{PTSerif-TLF}{bx}{sc}{<-> ssub * PTSerif-TLF/bx/n}{}
\makeatother

\usepackage{amsmath,amssymb,amsfonts}
\usepackage{algorithmic}
\usepackage{graphicx}
\usepackage{textcomp}
\usepackage{xcolor}
\usepackage{booktabs}
\usepackage{cite}

\usepackage{adjustbox}
\usepackage{tabularx}
\usepackage{hhline}
\usepackage{longtable}
\usepackage{makecell}

\usepackage{enumitem}
\usepackage{comment}
\usepackage{float}

\usepackage{url}

\usepackage{caption}
\usepackage{subcaption}

\usepackage[hidelinks]{hyperref}
\usepackage{cleveref}

\newcommand{\hx}[1]{\%#1\allowbreak{}}

  \usepackage{cuted}

\begin{document}

\newcommand{\rev}[1]{\textcolor{black}{#1}}
\crefname{figure}{\rev{Figure}}{\rev{Figures}}
\crefname{table}{\rev{Table}}{\rev{Tables}}
\crefname{section}{\rev{Section}}{\rev{Section}}

\title{Comprehensive List of User Deception Techniques in Emails}

\input{main-macros}

\author{Maxime Veit}
\author{Mattia Mossano}
\author{Tobias L\"{a}nge}
\author{Melanie Volkamer}
\affil{Karlsruhe Institute of Technology \\
\texttt{\{maxime.veit, mattia.mossano, tobias.laenge, melanie.volkamer\}@kit.edu}}

\setlength{\tocnumwidth}{2.6em}

\maketitle


\tableofcontents
\input{main-section-inclusion}

\bibliography{biblio/systematic-literature-research_attacks,biblio/main}

\newpage
\appendix
\input{sections/060_appendix}

\end{document}

%% file: main-macros.tex
\newcommand{\trick}{deception technique} 
\newcommand{\Trick}{Deception Technique} 
\newcommand{\TTrick}{Deception technique} 
\newcommand{\trickimpl}{deception technique} 
\newcommand{\Trickimpl}{Deception Technique} 
\newcommand{\TTrickimpl}{Deception technique} 
\newcommand{\correct}{correct}
\newcommand{\correctinfo}{{\correct} information}
\newcommand{\incorrect}{misleading}
\newcommand{\incorrectinfo}{{\incorrect} information}

\newcommand{\casesnum}{176}
\newcommand{\critclasszero}{none}
\newcommand{\critclassone}{low}
\newcommand{\critclasstwo}{medium}
\newcommand{\critclassthree}{concerning}
\newcommand{\critclassfour}{high}
\newcommand{\CritClasszero}{None}
\newcommand{\CritClassone}{Low}
\newcommand{\CritClasstwo}{Medium}
\newcommand{\CritClassthree}{Concerning}
\newcommand{\CritClassfour}{High}

\newcommand{\SenderMailSpoofing}{Sender-Mail Spoofing}
\newcommand{\SenderNameSpoofing}{Sender-Name Spoofing}
\newcommand{\FakeTooltip}{Fake Tooltip}
\newcommand{\HexadecimalEncoding}{Hexadecimal Encoding}
\newcommand{\HTMLBaseTag}{HTML Base-Tag}
\newcommand{\HTMLFormTag}{HTML Form-Tag}
\newcommand{\LinkMismatch}{Link Mismatch}
\newcommand{\MailtoScheme}{Mailto-Scheme}
\newcommand{\OpenRedirect}{Redirect}
\newcommand{\Ropemaker}{Ropemaker}
\newcommand{\URLPosingSuffix}{URL Posing-Suffix}
\newcommand{\URLShortener}{URL Shortener}
\newcommand{\UrlUserinfoField}{URL Userinfo Field}
\newcommand{\DoubleFileExtension}{Double File Extension}
\newcommand{\UnknownFileExtension}{Unknown File Extension}
\newcommand{\DomainPrefix}{Domain Extension}
\newcommand{\Mangle}{Mangle}
\newcommand{\NonASCIICharacters}{Homographic Spoofing} 
\newcommand{\Subdomain}{Subdomain}
\newcommand{\UnrecognizableDomain}{Unrecognizable Domain}
\newcommand{\ExceedinglyLong}{Exceedingly Long}
\newcommand{\RighttoLeftOverride}{Right-to-Left Override}
\newcommand{\UnknownTLD}{Different TLD}
\newcommand{\AnalysisAspects}{aspects}

\newcommand{\EMLCodeAndScreenshots}{Link to online resource}

\newboolean{aggrshortening}  
\setboolean{aggrshortening}{false} 

\newcommand\eatpunct[1]{} 

\newcommand{\newText}[1]{#1}
\definecolor{Green200}{RGB}{1,200,32}
\definecolor{Blue200}{RGB}{1,32,200}
\newcommand{\csrevise}[1]{#1}
\newcommand{\csminrevise}[1]{#1}

%% file: main-section-inclusion.tex
\input{sections/010_intro.tex}
\input{sections/015_fundamentals.tex}
\input{sections/020_findings.tex}

\input{sections/040_conclusion.tex}
\input{sections/050_acks.tex}

%% file: sections/010_intro.tex
\section{Introduction}
Email remains a cornerstone of professional communication, but its enduring ubiquity also exposes longstanding design limitations. Originally built without today's security threats in mind, email has become fertile ground for attackers who exploit its infrastructure and email client features to deceive users.

This research note presents a structured list of email-based deception techniques targeting the sender, link, and attachment security indicators, as well as techniques that exploit properties of the email rendering environment more broadly. For each technique, we describe the underlying mechanism and provide an illustrative example implementation, without assessing effectiveness or real-world severity. Building on our prior systematic literature review~\cite{Veit2024}, which identifies techniques and evaluates email client susceptibility, we extend the documented set of techniques and provide newly developed example implementations.

Our contributions are twofold:

\begin{enumerate}
    \item We present deception techniques from our prior systematic study~\cite{Veit2024} with newly developed example implementations in a consistently structured list.
    \item We introduce novel deception techniques identified through our own examination.
\end{enumerate}

While a systematic evaluation across email clients is outside the scope of this note, the documented deception techniques and example implementations serve as a structured reference for multiple audiences. These include researchers examining email clients or security awareness measures, as well as developers, operators, and designers working in either domain.

%% file: sections/015_fundamentals.tex
\section{Fundamentals}
\label{sec:fundamentals}
This section provides background information and essential terminology. Much of the content is directly sourced from~\cite{Veit2024}.

In this paper, we use the term \textit{security indicator} to refer to interface-defined indicator fields that can convey security-relevant information about an email. We distinguish three security indicators: the \textit{sender security indicator}, the \textit{link security indicator}, and the \textit{attachment security indicator}. A security indicator denotes the designated location in an email client where such information is available (e.g., as a visible label, a truncated string, or content revealed on interaction). Security indicators serve as cues that users may rely on to form trust judgments and to decide whether to perform an email-related action (e.g., replying, clicking a link, or opening an attachment). We refer to these fields as security indicators because they are the interface elements recipients primarily consult to assess whether an email is trustworthy. A security indicator may be fully visible, partially visible (e.g., truncating parts of the sender email address), or not visible in a given client configuration, even though the indicator field exists conceptually.

\subsection{Sender Security Indicator}
\label{sec:senderaddress}
\begin{figure}[H]
    \centering
    \includegraphics[width=0.44\textwidth]{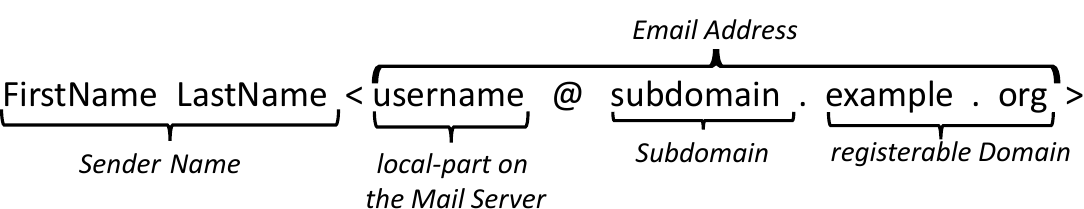}
    \caption{Sender Security Indicator structure and terminology for different components.}
    \label{fig:sender-structure}
\end{figure}

A \textit{sender security indicator} denotes the indicator field in an email client where sender-identifying information is available to be shown, intended to support users in assessing the apparent origin of a message before acting on it.
Technically, the information shown in this indicator is derived from the email header's \texttt{From} field.
As shown in~\cref{fig:sender-structure}, the sender security indicator comprises a freely selectable Sender Name (first/last name) and the email address components, namely the Local-Part (username on the mail server) and the domain hierarchy (Subdomains, Registrable Domain).

\subsection{Link Security Indicator}
\label{sec:targetofurl}

A link is a reference in the form of a URL to another resource (e.g., a website) embedded in the email body, which is typically opened when a user clicks it (or taps it on mobile devices). The \textit{link security indicator} denotes the indicator field in an email client where the link-target information is available to be shown for a link, intended to support users in assessing where a link appears to lead before clicking it. The link security indicator includes the URL as presented to the user.

On desktop and laptop devices, email clients (including web-based email interfaces) commonly reveal the URL when the user hovers over the link via the \textit{status bar} and/or a \textit{tooltip}. On mobile devices (smartphones/tablets), the mechanism for accessing the target URL depends on the operating system and the used email client; most of the time it is displayed in a dialog after a long-press.

The target URL may refer to a legitimate website or to an attacker-controlled malicious website; in the latter case, clicking or tapping on the link leads the user to a malicious website.

\begin{figure}[h]
    \centering
    \includegraphics[width=0.47\textwidth]{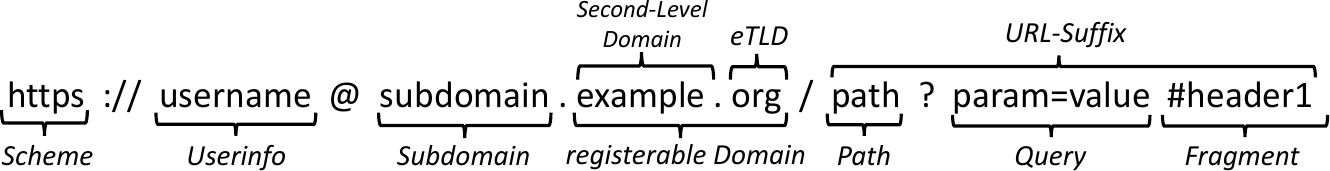}
    \caption{URL structure and nomenclature of its components.~\csrevise{(eTLD: Effective Top-Level-Domain)}}
    \label{fig:URL-structure}
    \vspace{-1em}
\end{figure}

\label{sec:urlstructure}
URLs follow a defined syntactic structure, shown in Figure~\ref{fig:URL-structure}.

The most relevant component of the URL structure for phishing detection is the \textit{Registrable Domain}~\cite{WHATWG-URL}, defined as the Second-Level Domain combined with the Effective Top-Level Domain (eTLD), due to it typically identifying the controlling website that will be contacted when the link is opened.\footnote{RFC~7489~\cite{RFC7489} uses the related term \textit{Organizational Domain} for a similar concept in the context of email authentication policy (DMARC).} Subdomains, Userinfo, and the Path/Query components may provide additional context, but in awareness material \cite{berensBetterTogheter}, the registrable domain is the main feature used to identify the destination organization, because known registrable domains can be associated with specific organizations.

\subsection{Attachment Security Indicator}
\label{sec:fileextension}

An attachment is a file provided as part of an email. The \textit{attachment security indicator} denotes the indicator field in an email client through which attachment-related information is presented to the user, allowing them to assess how an attachment appears to be handled before opening it.

The attachment security indicator, i.e., the \textit{filename}, typically consists of a base name and a file extension separated by a dot (e.g., \texttt{invoice.pdf}), where the extension commonly serves as a user-facing cue for the apparent file type and associated opening behavior. Filenames are not limited to letters and digits and may include other characters (e.g., characters \rev{``}.\rev{''} and \rev{``}-\rev{''}).

File extensions can differ substantially in security risk. In particular, extensions extensions associated with executables or scripts (e.g., \texttt{.exe} on Windows) are commonly considered higher risk than extensions associated with non-executable document or image formats, as opening an executable or script may directly trigger code execution.

%% file: sections/020_findings.tex
\vspace{4pt}

\section{Structure of the report}
\label{sec:structure-of-the-report}

The following sections present a comprehensive list of email-based deception techniques, including those identified through the systematic literature review by Veit et al.~\cite{Veit2024}. To ensure consistency and completeness, we adopt parts of the technique descriptions reported by Veit et al.~\cite{Veit2024}.

We distinguish three provenance cases for the implementations described in the following sections. First, implementations found using the systematic literature review from Veit et al.~\cite{Veit2024} (including their original references). Second, implementations adapted by transferring previously reported deception techniques to a different security indicator. Third, implementations identified during our examination. We indicate the provenance of each implementation explicitly in the beginning of the corresponding implementation description.

Each deception technique is deliberately uniform so that entries can be read independently. Concretely, we first provide a technique-level description that states the attacker’s high-level goal and explains which security indicator is targeted and how it is conceptually manipulated. We then describe one or more concrete implementations that specify the technical mechanism enabling the deception. This structure supports the systematic comparison of the  techniques and allows readers to consult individual entries without having to read the entire list of deception techniques start-to-end. 

The scope of this list is explanatory: we describe deception techniques in isolation to support understanding and to inform countermeasures in both email-clients design and security awareness measures. The examples are not intended to demonstrate or assess real-world attack severity. In practice, severity can be substantially higher when adversaries combine multiple techniques and fully exploit implementation details---for example, by using one technique to hide a security indicator and another technique to display an attacker-controlled fake security indicator instead. While each technique alone may have a more limited effect, their combination can reinforce the deception. Addressing any one of the involved techniques can be sufficient to break such a combined attack chain. Therefore, we document the techniques as individual building blocks so they can be addressed systematically, thereby reducing opportunities for highly optimized combined attacks.

Several deception techniques can look similar, even though they rely on different mechanisms and prerequisites. Since these prerequisites can vary across email clients (e.g., due to truncation or default visibility of security indicators), we do not systematically enumerate client-specific feasibility conditions. Some deception principles also generalize across security indicators (e.g., exceedingly long strings, homoglyphs, or rendering-direction manipulation), which is why similar ideas may reappear across sender-, link-, and attachment-related techniques.

To illustrate the techniques in a consistent manner, we use a narrative in which Eve attempts to phish Bob via email. The narrative begins with a basic phishing email that can be easily identified using standard security awareness practices (see, e.g.,~\cite{berensBetterTogheter}). Each subsequent section then modifies the base scenario according to the structure described above, demonstrating one deception technique (or one implementation) at a time. For clarity and consistency, all screenshots shown in this report are generated illustrations and not taken from any specific email client. To support orientation, each screenshot highlights the relevant security indicator with a red box and highlights the deception-critical part in red. The narrative scenario is set on a desktop client. Where applicable, the screenshots depict the cursor hovering over the link, making link-related security indicators such as the tooltip and status bar visible; on mobile devices, a long press typically serves the equivalent purpose.

\begin{figure}[H]
    \centering
    \includegraphics[width=1\columnwidth]{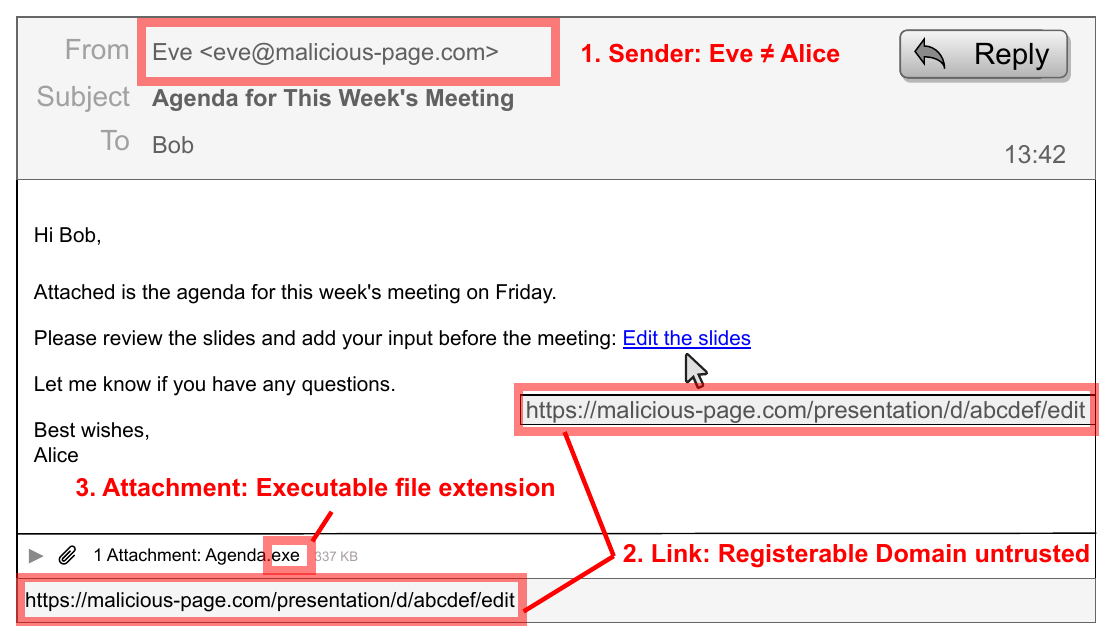}
    \caption{Base scenario - phishing email detectable after security awareness training.}
    \label{fig:nstructure-of-the-report-basic-scenario}
    \vspace{-1em}
\end{figure}

As shown in~\cref{fig:nstructure-of-the-report-basic-scenario}, the base scenario email in which Eve poses as Alice to phish Bob is easily recognized as phishing due to security indicators such as the sender email address, the link, and the attached executable file.

\section{Sender-related Deception Techniques}
\label{sec:sender-related-deception-techniques}

Security awareness training commonly instructs recipients to inspect the sender security indicator when assessing the origin of an email~\cite{berensBetterTogheter}. In principle, the sender can define the email header's \texttt{From} field freely, giving an attacker direct influence over how this indicator is displayed. In practice, however, infrastructure-level mechanisms may constrain the feasibility of certain approaches. Although infrastructure-level email authentication technologies such as DMARC can prevent sender email address spoofing by rejecting emails during transmission~\cite{dmarctobias}, we cannot rely on other protective measures being in place, so this remains an issue for email clients. The following deception techniques demonstrate distinct ways to manipulate this indicator, with feasibility depending on the deployment context. To keep each demonstration focused, the examples modify only the element under discussion and do not combine multiple deception techniques; note that when the attacker controls the Registrable Domain, the Local-Part can be set to any value.

\subsection{Sender Email Address Spoofing}
\label{sec:sender-email-address-spoofing}
In this deception technique, the attacker aims to impersonate a trusted sender by spoofing the sender's email address.

To achieve this, the attacker targets the sender security indicator by replacing the displayed sender email address with a trustworthy-looking one. This conceals the attacker’s identity and the message’s true origin.

A concrete implementation of this deception technique (mentioned in literature such as~\cite{pirocca_toolkit_2020,maroofi_adoption_2021,soussi_feasibility_2020,du_research_2013,weaver_training_2021,jakobsson_user_2016,shen_weak_2021}) exploits that the sender security indicator is derived from the email header's \texttt{From} field, which is defined by the sender and can therefore be freely modified. 

\begin{figure}[H]
    \centering
    \includegraphics[width=1\columnwidth]{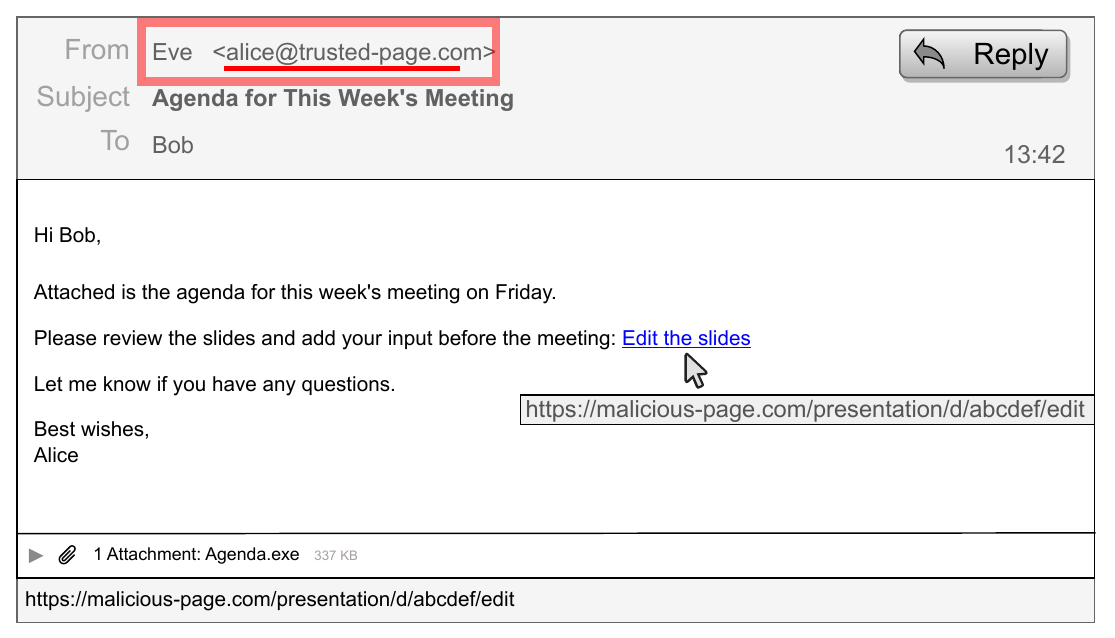}
    \caption{Example of Sender Email Address Spoofing.}
    \label{fig:nsender-related-deception-techniques}
    \vspace{-1em}
\end{figure}

 As shown in~\cref{fig:nsender-related-deception-techniques}, the attacker could replace a revealing address such as \texttt{eve@malicious-page.com} with an apparently legitimate address like \texttt{alice@trusted-page.com} in the \texttt{From} field.

\subsection{Sender Email Address Local-Part Replacement}
\label{sec:sender-mail-address-local-part-replacement}
In this deception technique, the attacker aims to impersonate a trusted sender by modifying the Local-Part of the sender email address.

To achieve this, the attacker targets the sender security indicator by crafting the Local-Part (the part before the \texttt{@} symbol) to resemble a trustworthy identity, while the Registrable Domain remains attacker-controlled. This deception technique exploits that some recipients may focus on the beginning of the email address as the primary identifying cue. As a result, they may not fully inspect the Registrable Domain after the \texttt{@} symbol.

A concrete implementation of this deception technique (identified during our examination) exploits that the sender security indicator is derived from the email header's \texttt{From} field, which is defined by the sender and can therefore be freely chosen. In particular, the attacker can select an arbitrary Local-Part at a Registrable Domain under their control or at a mail provider where the desired username (i.e., Local-Part) is available. 

\begin{figure}[H]
    \centering
    \includegraphics[width=1\columnwidth]{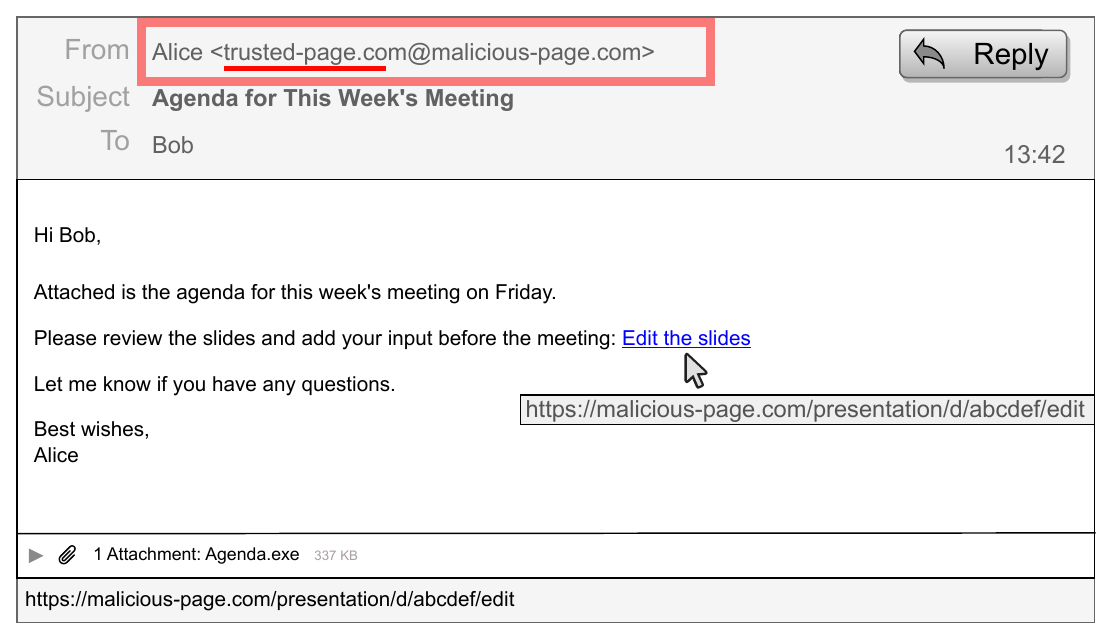}
    \caption{Example of Sender Email Address Local-Part Replacement.}
    \label{fig:nsender-mail-address-local-part-replacement}
    \vspace{-1em}
\end{figure}

 As shown in~\cref{fig:nsender-mail-address-local-part-replacement}, the attacker uses \texttt{alice@malicious-page.com} or crafts a domain-like Local-Part such as \texttt{trusted-page.com@malicious-page.com} to appear familiar at the beginning of the address, although the mailbox and the Registrable Domain \texttt{malicious-page.com} belong to the attacker.

\subsection{Sender Name Replacement}
\label{sec:Sender-Name-Replacement}
In this deception technique, the attacker aims to impersonate a trusted sender by replacing the Sender Name of the sender security indicator. 

To achieve this, the attacker targets the sender security indicator by choosing a trustworthy-looking Sender Name (typically first and last name) that is shown to the recipient as the apparent sender identity, thereby concealing the attacker’s identity and the true origin of the message.

We describe the following implementations of this deception technique.

\subsubsection{Implementation: Different Sender Name}
\label{sec:sender-name-replacement-different-name}

This implementation (mentioned in literature such as~\cite{jakobsson_user_2016}) exploits that the Sender Name displayed by the email client is derived from the email header. It can therefore be freely chosen by the sender. 

\begin{figure}[H]
    \centering
    \includegraphics[width=1\columnwidth]{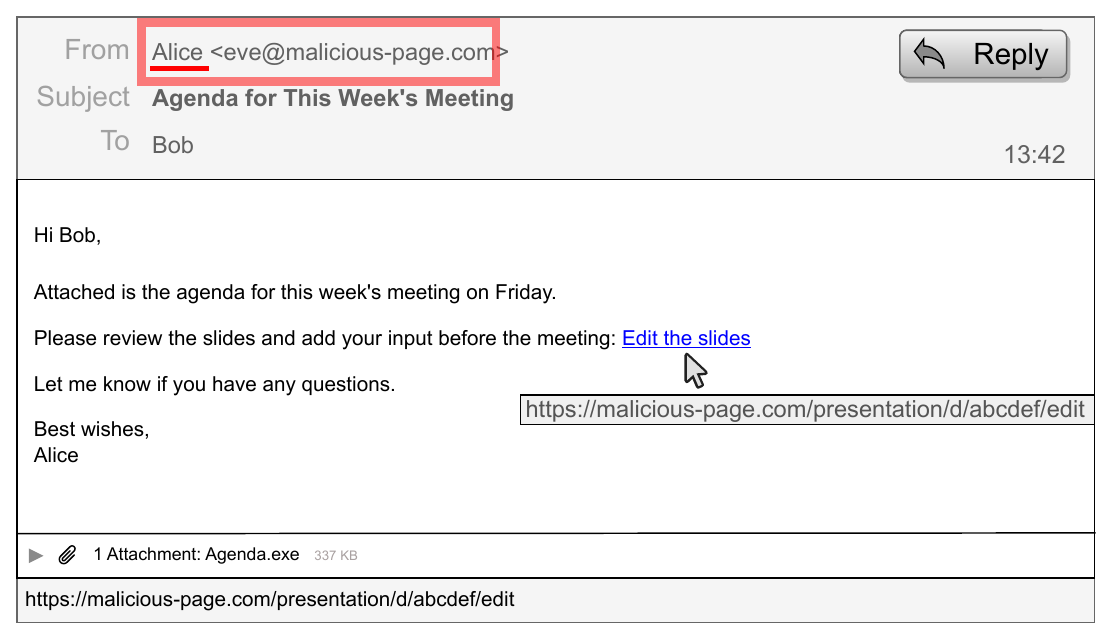}
    \caption{Example of Sender Name Replacement (Implementation: Different Sender Name).}
    \label{fig:nsender-name-replacement-different-name}
    \vspace{-1em}
\end{figure}

As shown in~\cref{fig:nsender-name-replacement-different-name}, the attacker can replace a revealing Sender Name such as \texttt{Eve} with a legitimate-looking name such as \texttt{Alice}.

\subsubsection{Implementation: Email Address in the Sender Name}
\label{sec:sender-name-replacement-email-address-in-the-sender-name}

This implementation (identified during our examination) exploits that the Sender Name of the \texttt{From} field can contain arbitrary text, including an email address instead of a personal name. Some email clients display only the Sender Name by default when it is present and hide the email address in the default view.

\begin{figure}[H]
    \centering
    \begin{subfigure}[b]{\columnwidth}
        \centering
        \includegraphics[width=1\columnwidth]{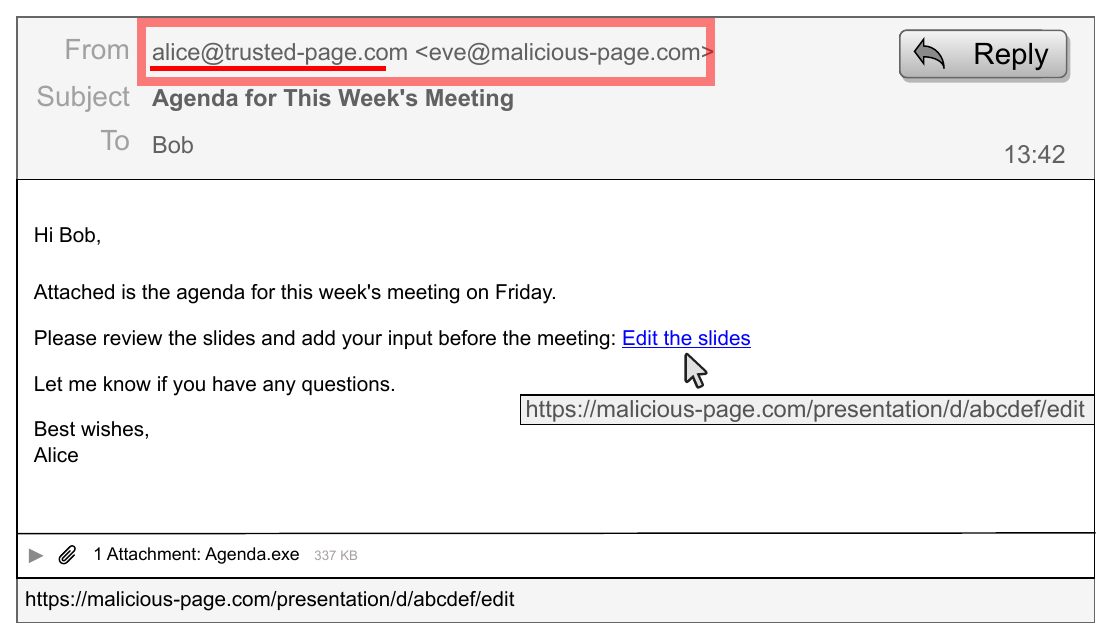}
        \caption{Email client shows both Sender Name and Sender Email Address.}
        \label{fig:nsender-name-replacement-email-address-in-the-sender-name-shown}
    \end{subfigure}
    
    \vspace{1ex} 
    
    \begin{subfigure}[b]{\columnwidth}
        \centering
        \includegraphics[width=1\columnwidth]{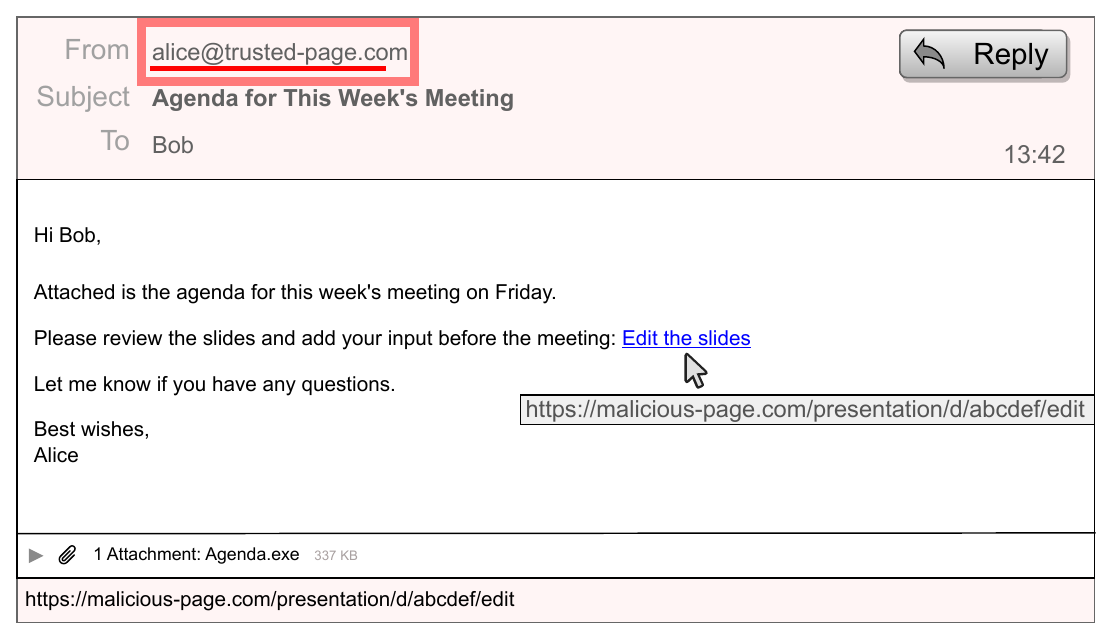}
        \caption{Depicted with an email client that only shows the Sender Name.}
        \label{fig:nsender-name-replacement-email-address-in-the-sender-name-hidden}
    \end{subfigure}

    \caption{Example of Sender Name Replacement (Implementation: Email Address in the Sender Name).}
    \label{fig:nsender-name-replacement-email-address-in-the-sender-name}
    \vspace{-1em}
\end{figure}

 As shown in~\cref{fig:nsender-name-replacement-email-address-in-the-sender-name-shown}, the attacker can set the Sender Name to \texttt{alice@trusted-page.com}. If the email client shows only the Sender Name when available, only the deceptive address is displayed to the recipient, as shown in~\cref{fig:nsender-name-replacement-email-address-in-the-sender-name-hidden}.

\subsubsection{Implementation: Look-alike Email Address in the Sender Name}
\label{sec:sender-name-replacement-look-alike-email-address-in-the-sender-name}
This implementation (identified during our examination) exploits that the Sender Name can contain arbitrary text and that some email clients apply client-side heuristics to detect email-address-like strings in Sender Name. By using Unicode homoglyphs of ASCII characters, an attacker can craft an email-address-like string that bypasses such detection. Otherwise, the email client may recognize that the Sender Name contains an email address and consequently display the email address (i.e., the revealing underlying sender address).

\begin{figure}[H]
    \centering
    \includegraphics[width=1\columnwidth]{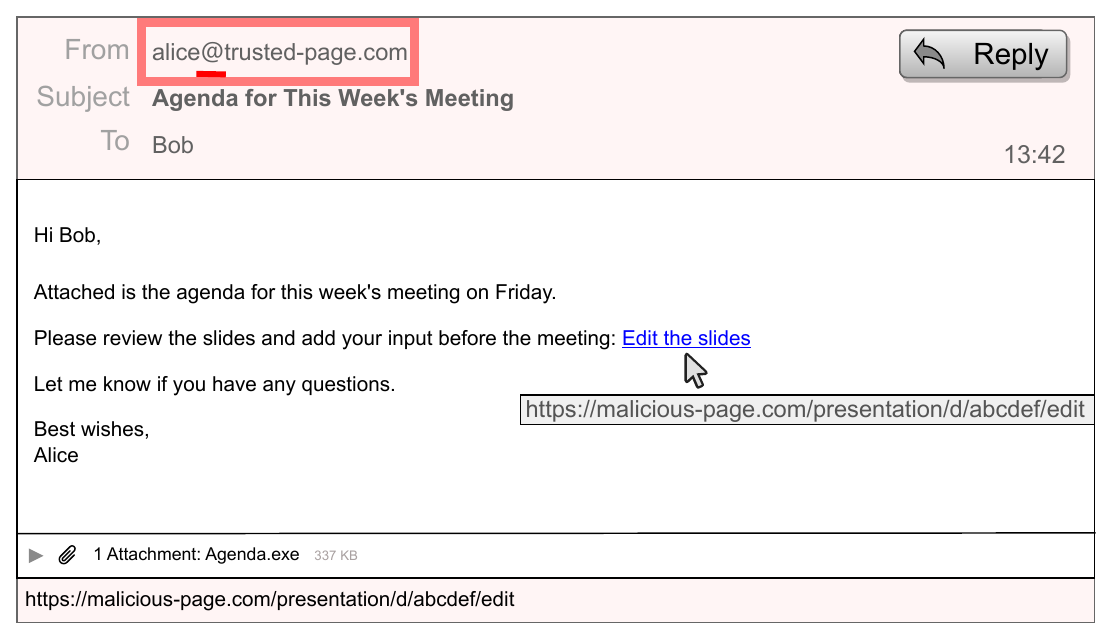}
   \caption{Example of Sender Name Replacement (Implementation: Look-alike Email Address in the Sender Name). Depicted with an email client that only shows the Sender Name. }
    \label{fig:nsender-name-replacement-look-alike-email-address-in-the-sender-name}
    \vspace{-1em}
\end{figure}

 As shown in~\cref{fig:nsender-name-replacement-look-alike-email-address-in-the-sender-name}, an attacker can replace the ASCII ``@'' (U+0040) in the spoofed email address with the Unicode \textit{Fullwidth Commercial At} (U+FF20), yielding a string that appears visually similar to an email address but may be misinterpreted by the email client. As a result, the email client may display only the deceptive string in the Sender Name, thereby concealing the attacker-controlled sender email address in the default view.

\subsection{Sender Spoofing using Notification Services}
\label{sec:sender-spoofing-using-notification-services}
In this deception technique, the attacker aims to impersonate a trusted service by abusing legitimate notification services to deliver attacker-controlled content via service-generated notification emails.

To achieve this, the attacker exploits the free-text input fields offered by some notification services to inject malicious content (e.g., phishing links) into notifications that are subsequently sent to recipients under the trusted service’s identity.

A concrete implementation of this deception technique (mentioned in literature such as~\cite{Hossein2017}) exploits that some notification services allow senders to include free-text content in service-generated notification emails. This functionality enables embedding attacker-chosen content (e.g., links to malicious websites) in such text, while recipients may interpret the message as part of a legitimate notification from a trusted service.

\begin{figure}[H]
    \centering
    \includegraphics[width=1\columnwidth]{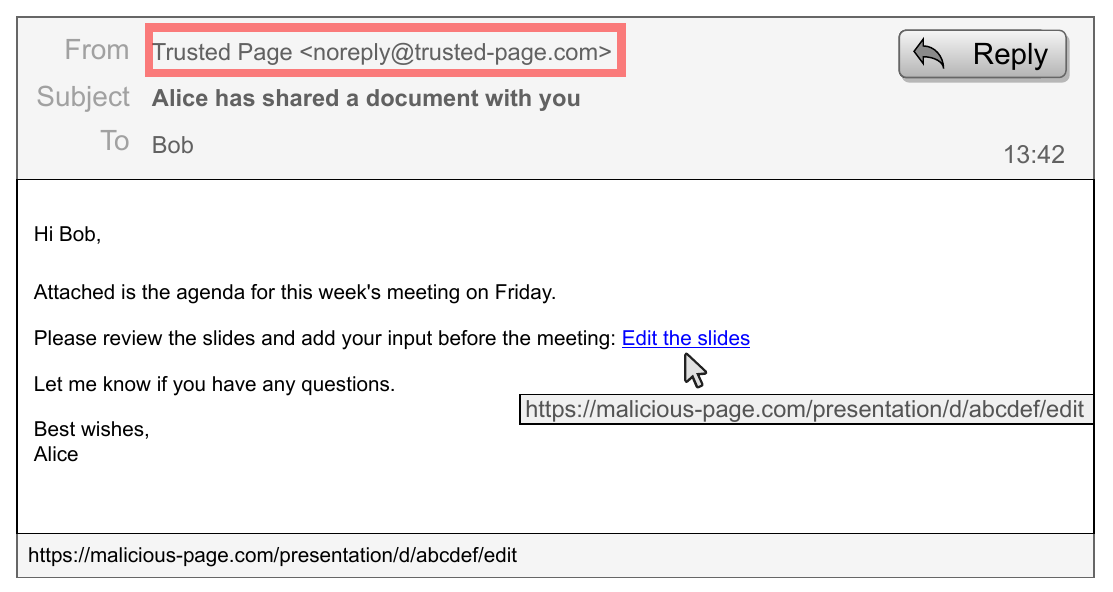}
    \caption{Example of Sender Spoofing using Notification Services.} 
    \label{fig:nsender-spoofing-using-notification-services}
    \vspace{-1em}
\end{figure}

As shown in~\cref{fig:nsender-spoofing-using-notification-services}, the attacker shares a document via the file-sharing service \textit{Trusted Page} and injects a phishing link into the optional invitation message field. The notification email is sent by \textit{Trusted Page}'s infrastructure and appears as a legitimate service notification, while the embedded link points to an attacker-controlled destination (\texttt{https://malicious-page.com/[...]}).

\subsection{Sender Exceedingly Long}
\label{sec:sender-exceedingly-long}

In this deception technique, the attacker aims to impersonate a trusted sender by hiding the Registrable Domain of the sender security indicator outside the visible area of the email client.

To achieve this, the attacker inflates the Subdomain. This can push the Registrable Domain of the sender email address outside the visible area in the email client’s default view. At the same time, the visible prefix is crafted to resemble a trusted identity (e.g., by starting with a trusted-looking name or domain string). Since the non-visible Registrable Domain determines the actual mail domain and is attacker-controlled, the attacker can freely choose the Local-Part.

A concrete implementation of this deception technique (adapted from link-related deception techniques~\cref{sec:link-url-exceedingly-long}) exploits that the Subdomain can be padded with a long filler string and a trusted-looking prefix, so that the actual Registrable Domain is pushed outside the visible area.

\begin{figure}[H]
    \centering
    \includegraphics[width=1\columnwidth]{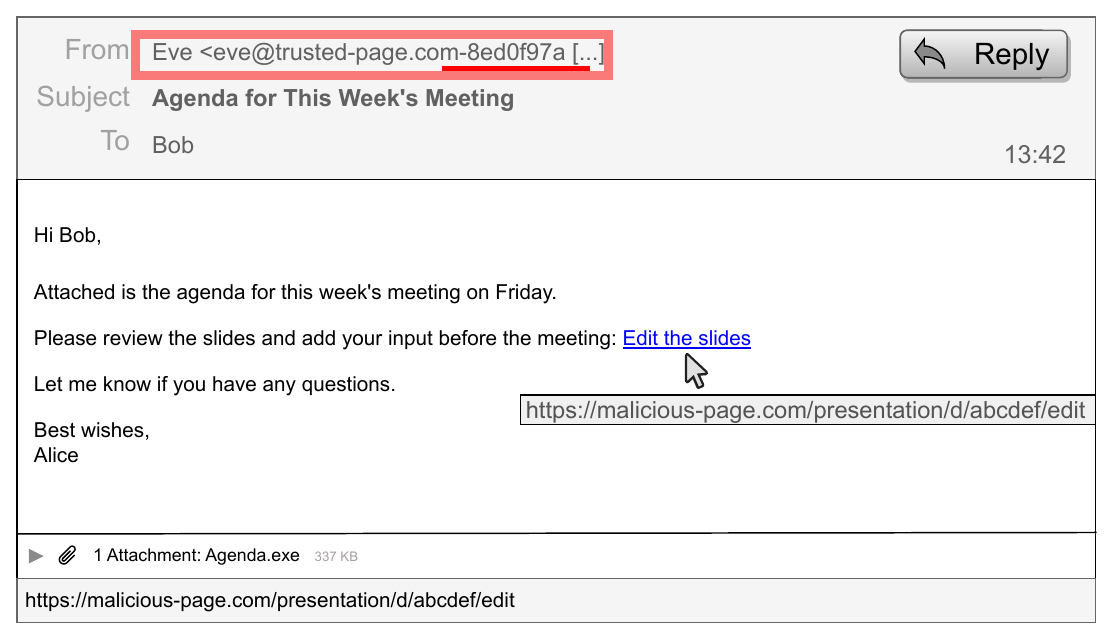}
    \caption{Example of Sender Exceedingly Long.} 
    \label{fig:nsender-exceedingly-long}
    \vspace{-1em}
\end{figure}

As shown in~\cref{fig:nsender-exceedingly-long}, the attacker inserts a long filler string such as \texttt{-8ed0f97a45dfd4gf5} and a trusted-looking prefix such as \texttt{trusted-page.com} into the Subdomain to push the actual Registrable Domain \texttt{malicious-page.com} outside the visible area of the email client. 


\subsection{Sender Homographic Spoofing}
\label{sec:Sender-Homographic-Spoofing}
In this deception technique, the attacker aims to impersonate a trusted sender by using a visually confusable Registrable Domain in the sender email address.

To achieve this, the attacker registers a Registrable Domain that appears visually identical to a legitimate domain by substituting ASCII characters with Unicode homoglyphs (e.g., from Cyrillic or Greek alphabets). Since the Registrable Domain (i.e., the deceptive homoglyph domain) is attacker-controlled, the attacker can freely choose the Local-Part.

A concrete implementation of this deception technique (mentioned in literature such as~\cite{shen_weak_2021}) exploits that internationalized domain names can contain Unicode characters. For domain registration, such domains are encoded in Punycode, but many email clients display the decoded Unicode form~\cite{Veit2024}. By substituting characters with homoglyphs, the attacker can craft a domain that is visually indistinguishable from a legitimate domain.

\begin{figure}[H]
    \centering
    \includegraphics[width=1\columnwidth]{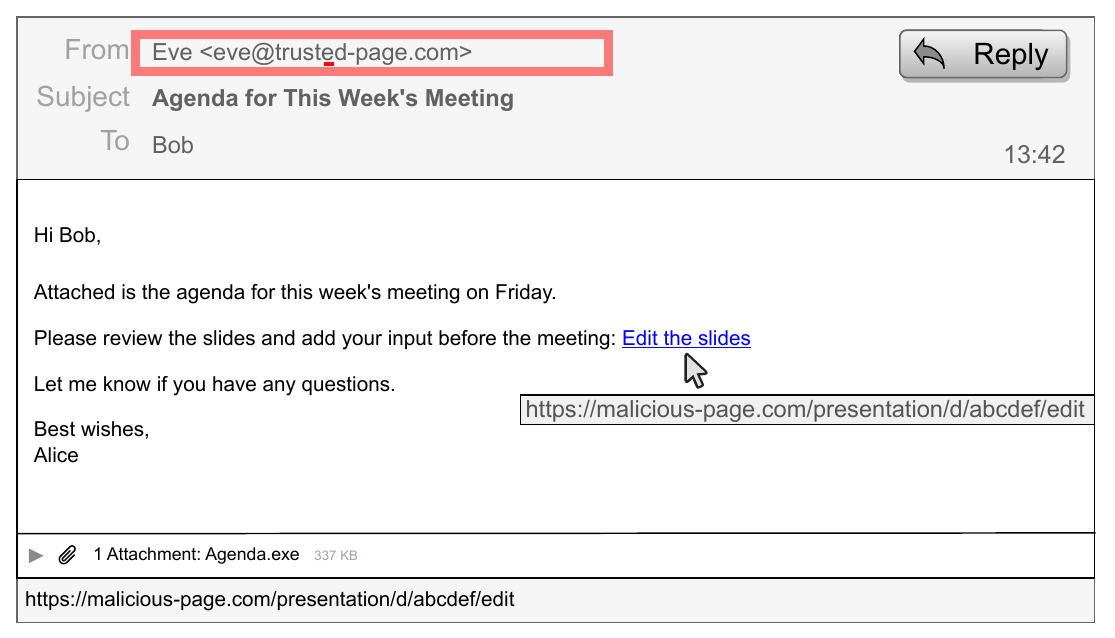}
    \caption{Example of Sender Homographic Spoofing.}
    \label{fig:nSender-Homographic-Spoofing}
    \vspace{-1em}
\end{figure}

As shown in~\cref{fig:nSender-Homographic-Spoofing}, the attacker registers a domain that is rendered in Unicode as \foreignlanguage{russian}{trustеd-page.com}. This domain is distinct from \texttt{trusted-page.com} because the first \texttt{e} is replaced with the Cyrillic character \foreignlanguage{russian}{е}. 


\subsection{Sender Right-to-Left Override}
\label{sec:Sender-Right-to-Left-Override}

In this deception technique, the attacker aims to impersonate a trusted sender by manipulating the text direction within the sender email address so that the displayed sender email address appears visually similar to a trusted one, even though the underlying string is technically different.

To achieve this, the attacker manipulates how the sender email address is rendered by the email client so that the sender email address is displayed in reverse order. As a result, a technically different Registrable Domain can appear visually similar to a trusted Registrable Domain.

A concrete implementation of this deception technique (adapted from attachment-related deception techniques~\cref{sec:attachment-right-to-left-override}) exploits that some email clients apply Unicode text-direction rules when displaying the sender email address without highlighting bidirectional control characters~\cite{Veit2024}. The attacker inserts the Unicode \textit{Right-to-Left Override} (RLO, U+202E) early in the sender email address. As a result, subsequent characters may be displayed in reverse order. This allows crafting an address that appears visually similar to a trusted address, even though the underlying string is technically different.

\begin{figure}[H]
    \centering
    \includegraphics[width=1\columnwidth]{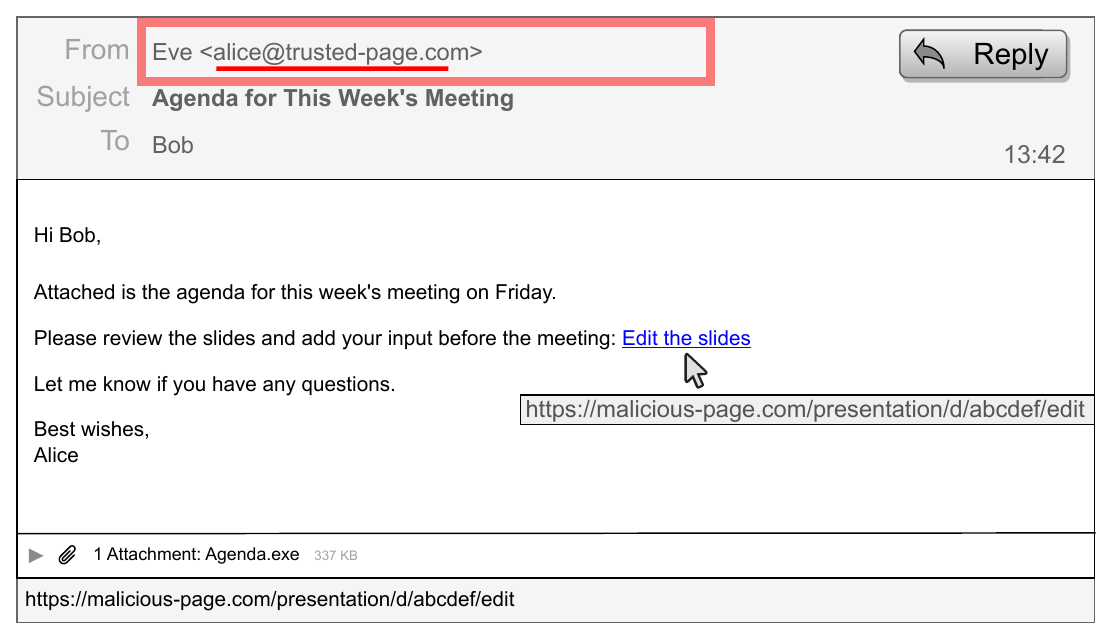}
    \caption{Example of Sender Right-to-Left Override.}
    \label{fig:nSender-Right-to-Left-Override}
    \vspace{-1em}
\end{figure}

 As shown in~\cref{fig:nSender-Right-to-Left-Override}, the attacker can use \texttt{[RLO]ed.egap-detsurt@ecila}, which is displayed as \texttt{alice@trusted-page.com}.


\subsection{Sender Unrecognizable Domain}
\label{sec:sender-unrecognizable-domain}

In this deception technique, the attacker aims to impersonate a trusted sender by using an unrecognizable or unfamiliar-looking domain or IP in the sender email address.

To achieve this, the attacker selects a Registrable Domain that does not provide meaningful cues to recipients (e.g., because it is a numeric address or resembles a random character sequence). Since such domains are difficult to interpret at a glance, recipients may rely on other email cues when deciding whether the message is trustworthy.

We describe the following implementations of this deception technique. 

\subsubsection{Implementation: IP Address}
\label{sec:sender-unrecognizable-domain-ip-address}

This implementation (adapted from link-related deception techniques~\cref{sec:link-url-unrecognizable-domain-ip-address}) exploits that a numeric address (e.g., IPv4 or IPv6) can be used in place of the Registrable Domain, which can appear unfamiliar and hard to assess.

\begin{figure}[H]
    \centering
    \includegraphics[width=1\columnwidth]{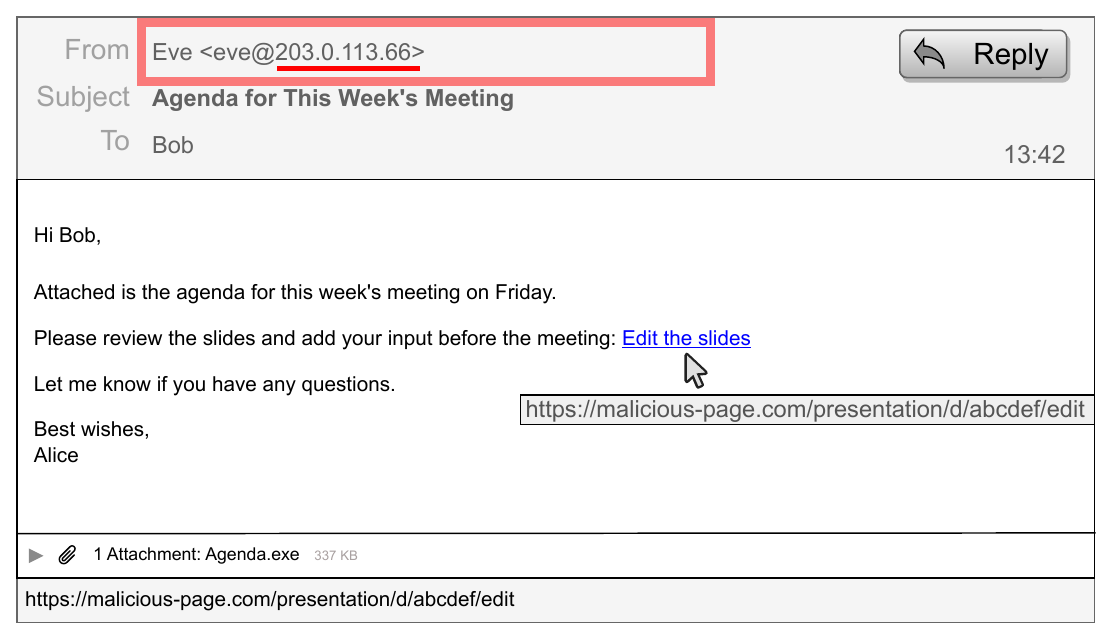}
    \caption{Example of Sender Unrecognizable Domain (Implementation: IP Address).}
    \label{fig:nSender-Unrecognizable-Domain-IP-Address}
    \vspace{-1em}
\end{figure}

 As shown in~\cref{fig:nSender-Unrecognizable-Domain-IP-Address}, the attacker can use an IP \texttt{203.0.113.66} instead of the Registrable Domain.

\subsubsection{Implementation: Random Characters}
\label{sec:sender-unrecognizable-domain-random-characters}

This implementation (adapted from link-related deception techniques~\cref{sec:link-url-unrecognizable-domain-random-characters}) exploits that a Registrable Domain can be chosen to resemble a random character sequence. Such domains provide little semantic context and may therefore be overlooked or under-scrutinized by recipients.

\begin{figure}[H]
    \centering
    \includegraphics[width=1\columnwidth]{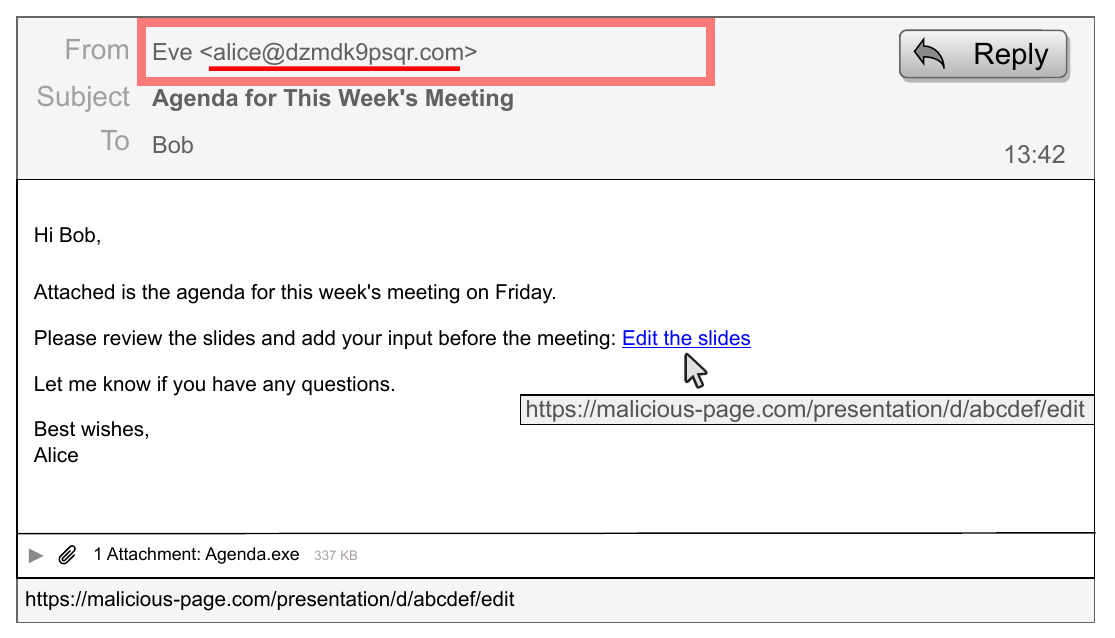}
    \caption{Example of Sender Unrecognizable Domain (Implementation: Random Characters).}
    \label{fig:nSender-Unrecognizable-Domain-Random-Characters}
    \vspace{-1em}
\end{figure}

 As shown in~\cref{fig:nSender-Unrecognizable-Domain-Random-Characters}, the attacker can use \texttt{dzmdk9psqr.com} as Registrable Domain.

\subsubsection{Implementation: Use of Fitting Keywords}
\label{sec:sender-unrecognizable-domain-use-of-fitting-keywords}

This implementation (adapted from link-related deception techniques~\cref{sec:link-url-unrecognizable-domain-use-of-fitting-keywords}) exploits that an attacker-controlled domain can incorporate fitting keywords that make it appear plausible, despite being unrelated to the targeted service.

\begin{figure}[H]
    \centering
    \includegraphics[width=1\columnwidth]{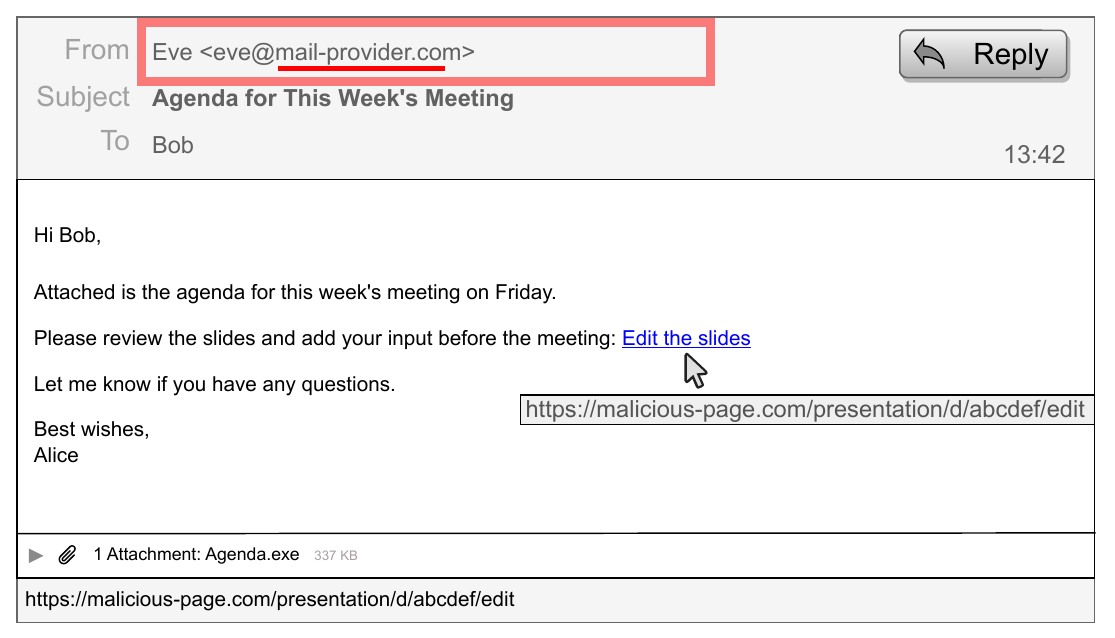}
    \caption{Example of Sender Unrecognizable Domain (Implementation: Use of Fitting Keywords).}
    \label{fig:nSender-Unrecognizable-Domain-Use-of-Fitting-Keywords}
    \vspace{-1em}
\end{figure}

 As shown in~\cref{fig:nSender-Unrecognizable-Domain-Use-of-Fitting-Keywords}, the attacker can use a domain such as \texttt{mail-provider.com} for \texttt{trusted-page.com}, which looks descriptive despite being unrelated to the real service.

\subsection{Sender Domain Extension}
\label{sec:sender-domain-extension}

In this deception technique, the attacker aims to impersonate a trusted sender by using a Registrable Domain that embeds the targeted brand as part of an extended Registrable Domain.

To achieve this, the attacker registers a look-alike domain that contains the legitimate brand name plus additional terms (e.g., security-related words) so that the resulting Registrable Domain appears plausible at a glance. This exploits that recipients may have difficulty judging whether such an extended domain is actually owned by, or affiliated with, the legitimate brand.

A concrete implementation of this deception technique (adapted from link-related deception techniques~\cref{sec:link-url-domain-extension}) exploits that a domain can be registered that appends or prepends (i.e., add before) additional strings to a trusted brand name, and used as the Registrable Domain of the sender email address.

\begin{figure}[H]
    \centering
    \includegraphics[width=1\columnwidth]{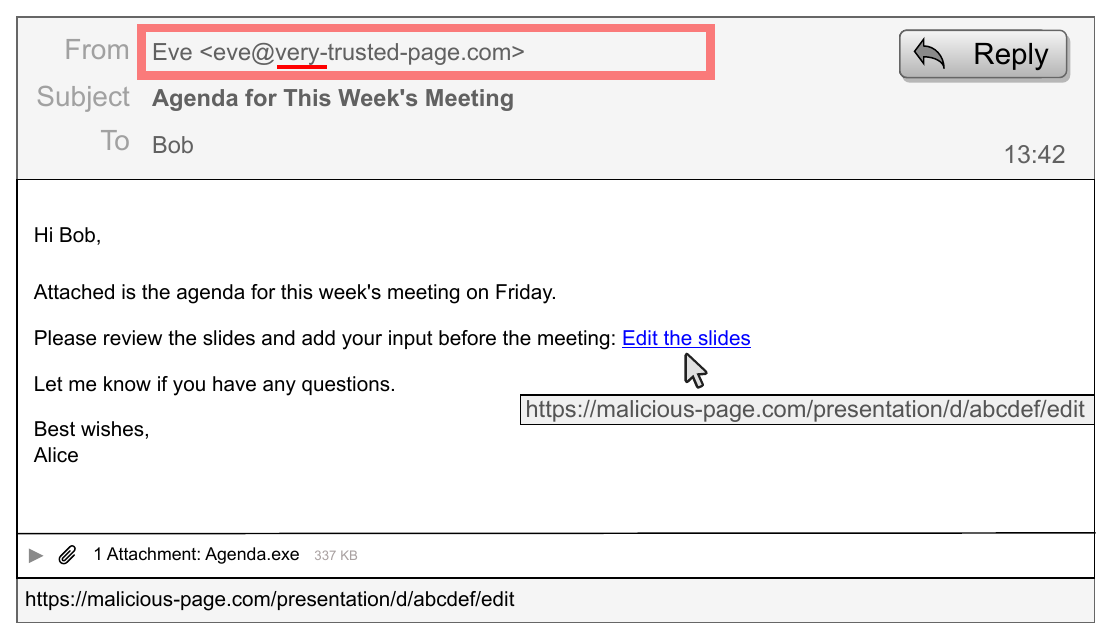}
    \caption{Example of Sender Domain Extension.}
    \label{fig:nSender-Domain-Extension}
    \vspace{-1em}
\end{figure}

As shown in~\cref{fig:nSender-Domain-Extension}, the attacker prepends the trusted brand domain \texttt{trusted-page.com} with the prefix \texttt{very-}, resulting in \texttt{very-trusted-page.com}. This can make the Registrable Domain appear affiliated with the legitimate brand at a glance, even though it is a different Registrable Domain.


\subsection{Sender Subdomain}
\label{sec:sender-subdomain}

In this deception technique, the attacker aims to impersonate a trusted sender by placing a trusted-looking Registrable Domain string into the Subdomain of the sender email address.

To achieve this, the attacker crafts the sender email address such that a familiar brand or trusted domain appears early in the Subdomain, while the actual Registrable Domain belongs to the attacker. This exploits that recipients may focus on the beginning of a domain-like string and may not fully inspect the Registrable Domain.

A concrete implementation of this deception technique (adapted from link-related deception techniques~\cref{sec:link-url-subdomain}) exploits that the attacker can place an arbitrary Subdomain under an attacker-controlled domain, allowing a targeted brand or trusted domain string to appear at the beginning of the sender email address.

\begin{figure}[H]
    \centering
    \includegraphics[width=1\columnwidth]{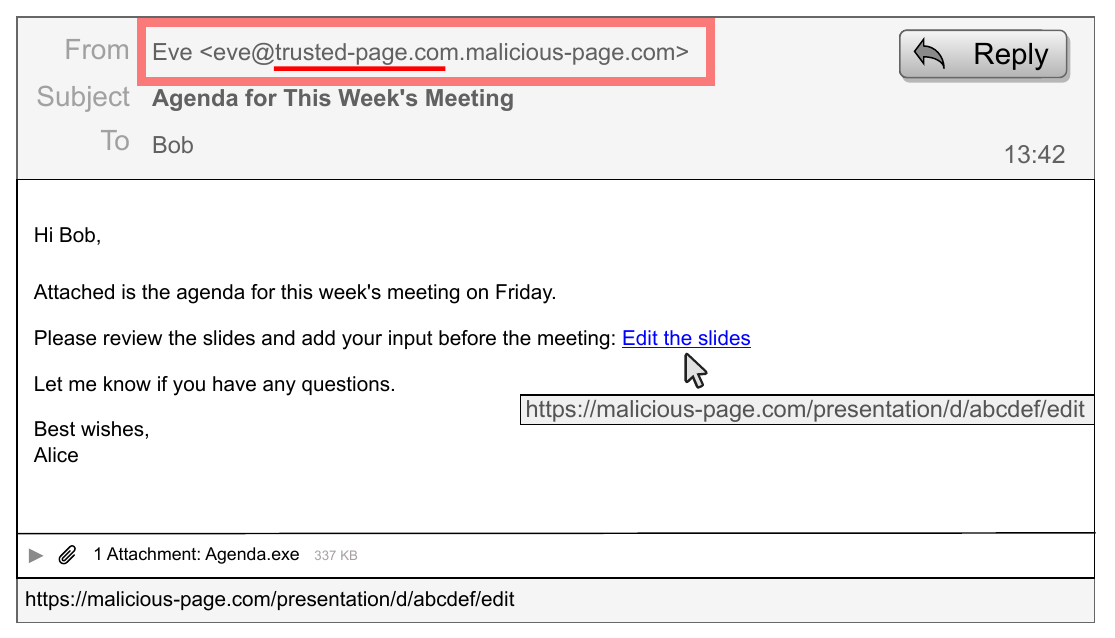}
    \caption{Example of Sender Subdomain.}
    \label{fig:nSender-Subdomain}
    \vspace{-1em}
\end{figure}

 As shown in~\cref{fig:nSender-Subdomain}, the attacker can use the sender email address \texttt{eve@trusted-page.com.malicious-page.com}, where the trusted domain string \texttt{trusted-page.com} appears as part of the Subdomain while the Registrable Domain \texttt{malicious-page.com} belongs to the attacker. This can make the Registrable Domain appear legitimate at a glance, even though the actual email domain is different.

\subsection{Sender Different eTLD}
\label{sec:sender-different-tld}

In this deception technique, the attacker aims to impersonate a trusted sender by using a Registrable Domain that contains the targeted brand name but uses a different eTLD.

To achieve this, the attacker registers a domain that matches the brand name while substituting the eTLD. This exploits that brands cannot practically register all available eTLD variants and that recipients may have difficulty judging which eTLDs are legitimately used by a given brand.

A concrete implementation of this deception technique (adapted from link-related deception techniques ~\cref{sec:link-url-different-etld}) exploits that the targeted brand name can be registered  under an alternative eTLD and used as the Registrable Domain of the sender email address.

\begin{figure}[H]
    \centering
    \includegraphics[width=1\columnwidth]{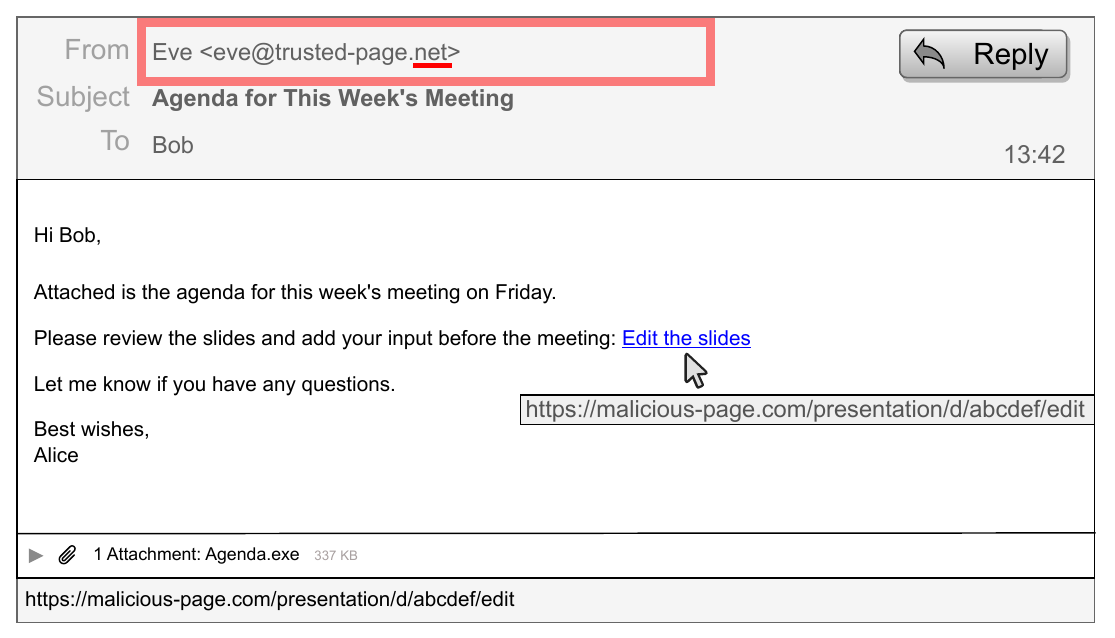}
    \caption{Example of Sender Different eTLD.}
    \label{fig:nSender-Different-eTLD}
    \vspace{-1em}
\end{figure}

 As shown in~\cref{fig:nSender-Different-eTLD}, the attacker can register \texttt{trusted-page.net} instead of the legitimate domain \texttt{trusted-page.com} and use it as the Registrable Domain for the sender email address. 

\subsection{Sender Mangle} 
\label{sec:sender-mangle}

In this deception technique, the attacker aims to impersonate a trusted sender by using a Registrable Domain that is a slightly modified (mangled) variant of a legitimate domain.

To achieve this, the attacker uses a domain whose spelling differs only subtly from the legitimate domain. This exploits that recipients may overlook small character differences when scanning the Registrable Domain of the sender email address.

A concrete implementation of this deception technique (mentioned in literature such as~\cite{jakobsson_user_2016}) exploits that attackers can register domains with minor variations of a trusted domain, for example through intentional typos (e.g., \textit{mircosoft} instead of \textit{microsoft}), character substitutions with visually similar patterns (e.g., \textit{m} vs. \textit{rn} or \textit{p} vs. \textit{q}), or small edits such as doubling characters.
 
\begin{figure}[H]
    \centering
    \includegraphics[width=1\columnwidth]{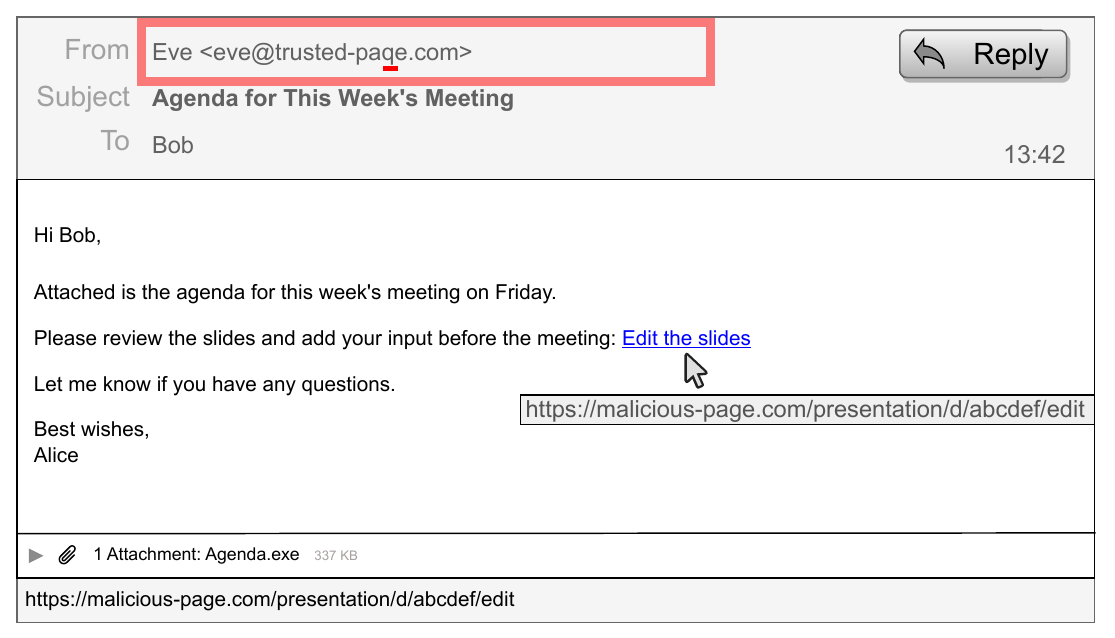}
    \caption{Example of Sender Mangle.}
    \label{fig:nSender-Mangle}
    \vspace{-1em}
\end{figure}

 As shown in~\cref{fig:nSender-Mangle}, the attacker replaces the \texttt{g} with the visually similar \texttt{q} in \texttt{trusted-page.com}, resulting in \texttt{trusted-paqe.com}.

\subsection{Sender External Content Manipulation}
\label{sec:sender-external-content-manipulation}

In this deception technique, the attacker aims to impersonate a trusted sender by inducing the trusted sender to forward an attacker-prepared email and then changing the displayed message content afterwards.

To achieve this, the attacker induces a trusted sender to forward an attacker-prepared email and controls which email content is visible to the final recipient after forwarding. This exploits that the forwarded email is displayed under the trusted sender’s identity, while externally controlled rendering can hide the originally visible content and reveal attacker-chosen content, thereby concealing the original content of the email.

A concrete implementation of this deception technique (adapted from link-related deception
techniques) exploits that externally loaded content (e.g., remotely hosted CSS) can change which parts of an email are visible after delivery. The attacker composes an email that contains multiple content blocks and an external CSS reference such that the trusted sender initially sees only safe content and is induced to forward the email. After the email is forwarded, the attacker changes the external CSS so that the final recipient is shown attacker-chosen content, which can make the email appear like it was authored by the trusted sender. As a side note, this can also apply when the forwarded email is signed (e.g., using S/MIME), because the visible content can still be controlled through externally loaded rendering.

\begin{figure}[H]
     \centering
     \begin{subfigure}[b]{\columnwidth}
         \centering
         \includegraphics[width=1\columnwidth]{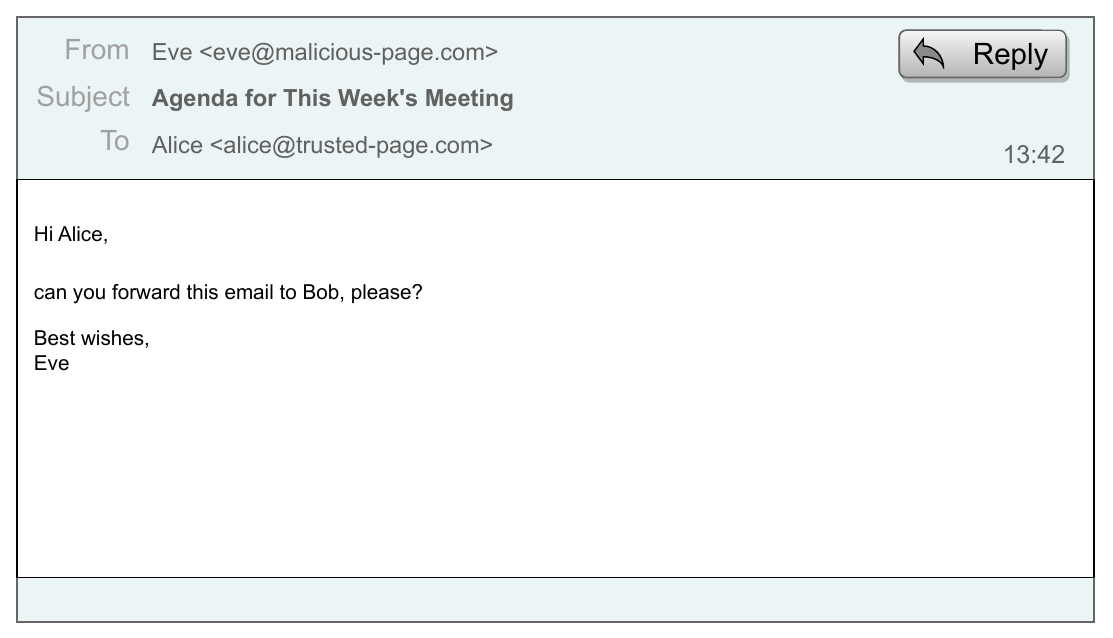}
         \caption{Email from Attacker to Alice. (Email client of Alice)}
         \label{fig:nSender-External-Content-Manipulation-original}
     \end{subfigure}

     \vspace{1ex}

     \begin{subfigure}[b]{\columnwidth}
         \centering
         \includegraphics[width=1\columnwidth]{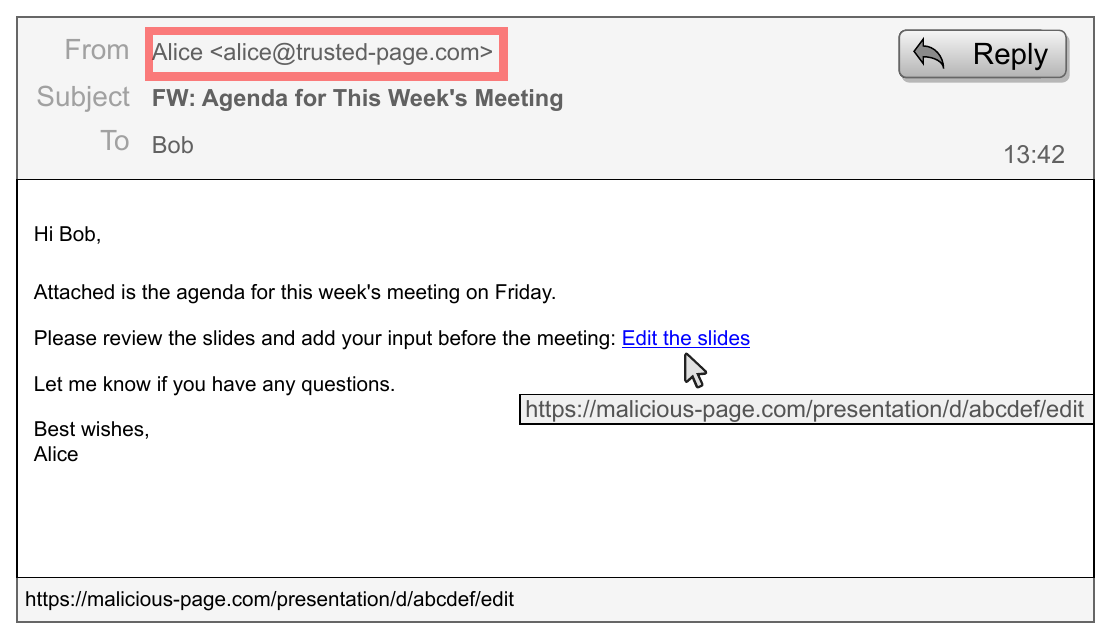}
         \caption{Forwarded email with content replaced.}
         \label{fig:nSender-External-Content-Manipulation-replaced}
     \end{subfigure}

     \caption{Example of Sender External Content Manipulation.}
     \label{fig:nSender-External-Content-Manipulation}
     \vspace{-1em}
 \end{figure}

 As shown in~\cref{fig:nSender-External-Content-Manipulation}, the attacker sends an email in which an attacker-chosen message is initially hidden and only safe content is shown to the trusted sender. After the trusted sender forwards the email, the previously hidden attacker-chosen content (including the malicious link) is shown again to the final recipient. For this demonstration, the attachment was removed.

\section{Link-related Deception Techniques}
\label{sec:link-related-deception-techniques}

Security awareness training commonly instructs recipients to verify the link security indicator before following a link -- by hovering over the link to inspect the Registrable Domain of the target URL~\cite{berensBetterTogheter}. An attacker can influence how this indicator is displayed through HTML structure, URL composition, and externally loaded content. Regardless of the approach chosen, however, the attacker must ensure that following the link leads to an attacker-controlled destination. The following deception techniques demonstrate how attackers manipulate this indicator while satisfying this constraint.

\subsection{Link HTML Fake Tooltip}
\label{sec:link-html-fake-tooltip}

In this deception technique, the attacker aims to mislead recipients about the target of a link by displaying a fake tooltip with a trustworthy-looking URL when the recipient hovers over the link.

To achieve this, the attacker presents a tooltip that suggests a safe destination, although the actual target URL of the link is different. As a result, recipients may rely on the displayed tooltip content instead of the actual link destination.

A concrete implementation of this deception technique (mentioned in literature such as~\cite{reinheimer_investigation_2020}) exploits that HTML/CSS mechanisms (e.g., the \texttt{title} attribute or hover-triggered elements) can be used to display an attacker-chosen URL on hover within an email body.

\begin{figure}[H]
    \centering
    \includegraphics[width=1\columnwidth]{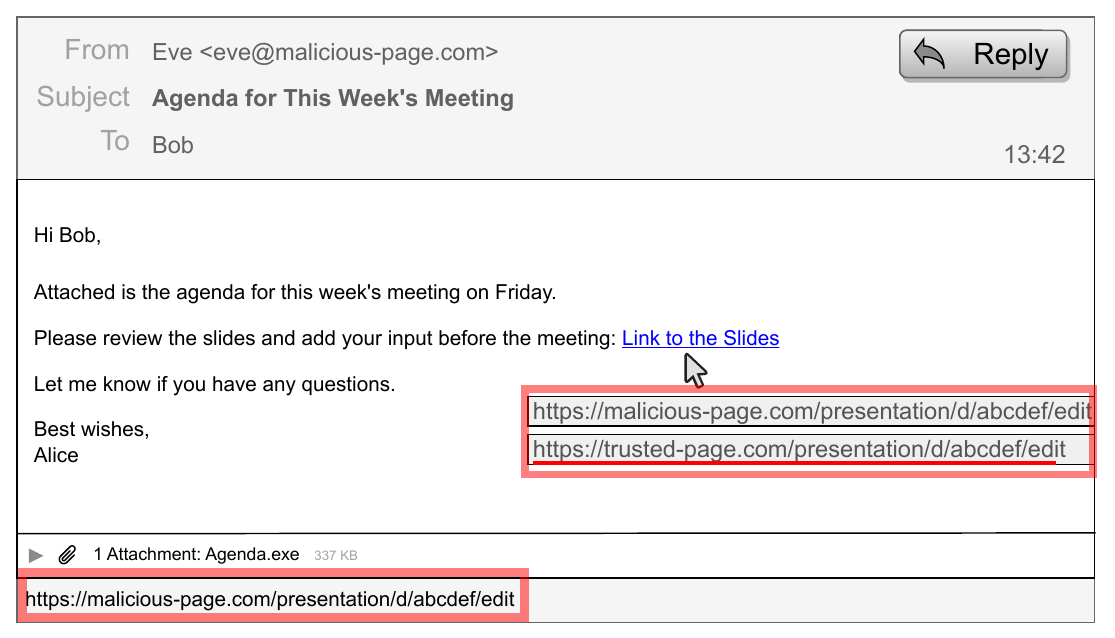}
    \caption{Example of Link HTML Fake Tooltip.}
    \label{fig:nLink-Fake-Tooltip}
    \vspace{-1em}
\end{figure}

As shown in~\cref{fig:nLink-Fake-Tooltip}, the attacker displays a trustworthy-looking URL in a fake tooltip with the Registrable Domain \texttt{trusted-page.com} upon hover, while the link actually points to the different destination \texttt{malicious-page.com}.


\subsection{Link URL Percent Encoding}
\label{sec:Link-URL-Percent-Encoding}

In this deception technique, the attacker aims to mislead recipients about the destination of a link by using URL percent encoding in the Registrable Domain part of the URL.

To achieve this, the attacker encodes characters in the Registrable Domain so that it is not easily readable by humans, but is decoded and resolved normally by the browser when the link is opened. This exploits that recipients may not recognize the Registrable Domain when it is percent-encoded and may therefore misjudge the link destination.

A concrete implementation of this deception technique (mentioned in literature such as~\cite{sankhwar_novel_2018,du_research_2013}) exploits that percent-encoded characters in a URL are automatically decoded when the link is opened, while the Registrable Domain remains
unreadable at a glance in the link security indicator. By encoding the Registrable Domain and prepending a trusted-looking Subdomain, the attacker constructs a URL that appears to begin with a recognizable domain while the actual Registrable Domain is concealed in encoded form.

\begin{figure}[H]
    \centering
    \includegraphics[width=1\columnwidth]{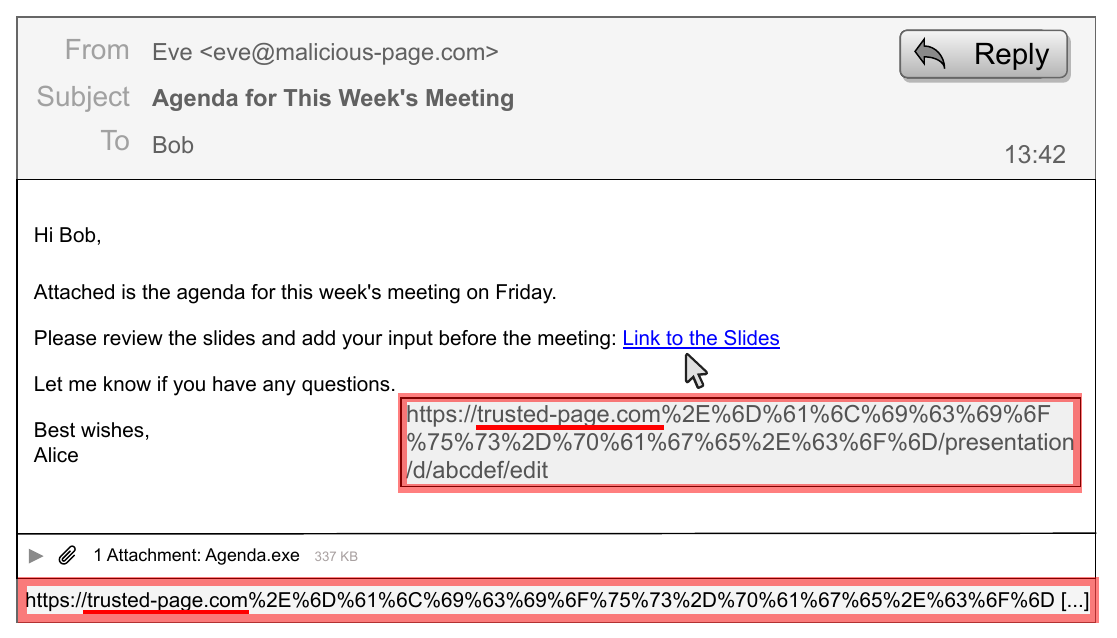}
    \caption{Example of Link URL Percent Encoding.}
    \label{fig:nLink-URL-Percent-Encoding}
    \vspace{-1em}
\end{figure}

As shown in~\cref{fig:nLink-URL-Percent-Encoding}, the attacker uses \texttt{.malicious-page.com}, encodes it as:
\begin{center}
\texttt{\hx{2E}\hx{6D}\hx{61}\hx{6C}\hx{69}\hx{63}\hx{69}\hx{6F}\hx{75}\hx{73}\hx{2D}\hx{70}\hx{61}\hx{67}\hx{65}\hx{2E}\hx{63}\hx{6F}\hx{6D}}
\end{center}
and prepends the Subdomain \texttt{trusted-page.com}, resulting in:
\begin{center}
\texttt{trusted-page.com\hx{2E}\hx{6D}\hx{61}\hx{6C}\hx{69}\hx{63}\hx{69}\hx{6F}\hx{75}\hx{73}\hx{2D}\hx{70}\hx{61}\hx{67}\hx{65}\hx{2E}\hx{63}\hx{6F}\hx{6D}}
\end{center}

\subsection{Link HTML Base Tag}
\label{sec:Link-HTML-Base-Tag}

In this deception technique, the attacker aims to mislead recipients about the destination of a link by using the HTML \texttt{base} tag so that a relative link appears to point to a trustworthy domain.

To achieve this, the attacker sets a base URL and then uses relative URLs for links in the email content. This exploits that some email clients show only the relative part of a link on hover (e.g., in the status bar or tooltip), while the base URL is silently applied when the link is opened.

A concrete implementation of this deception technique (mentioned in literature such as~\cite{orman_towards_2012}) exploits that the HTML \texttt{base} tag specifies the base URL used to resolve all relative URLs in the email. By setting the base URL to an attacker-controlled domain and using a trusted-looking relative URL, the displayed link can appear benign although the resulting destination is attacker-controlled.

\begin{figure}[H]
    \centering
    \includegraphics[width=1\columnwidth]{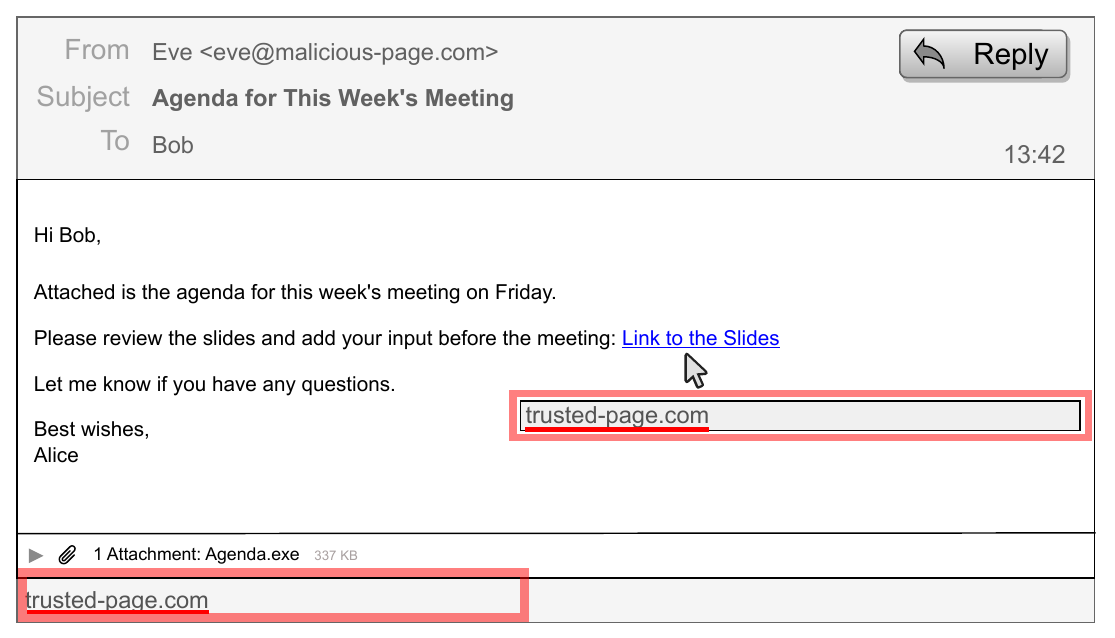}
    \caption{Example of Link HTML Base Tag.}
    \label{fig:nLink-HTML-Base-Tag}
    \vspace{-1em}
\end{figure}

 As shown in~\cref{fig:nLink-HTML-Base-Tag}, the attacker can set the base URL to \texttt{https://malicious-page.com/} and then uses a relative URL such as \texttt{trusted-page.com}. When the recipient hovers over the link, the email client displays only \texttt{trusted-page.com}. When the link is opened, the base URL is applied, resulting in \texttt{https://malicious-page.com/trusted-page.com}.

\subsection{Link HTML Form Tag}
\label{sec:link-html-form-tag}

In this deception technique, the attacker aims to hide the destination of a link by using an HTML form submit button instead of a standard link.

To achieve this, the attacker uses the destination URL as the form submission URL and presents a submit button that looks like a normal link. This exploits that many email clients do not show the destination URL in the tooltip or status bar when hovering over a form submit button.

A concrete implementation of this deception technique (mentioned in literature such as~\cite{qabajeh_experimental_2014,du_research_2013}) exploits that submitting an HTML form in an email opens the form submission URL, similar to following a link. Unlike a normal link, however, the destination URL is typically not shown on hover for form submit buttons. By styling the form submit button using CSS, the form can be made to resemble a typical link.

\begin{figure}[H]
    \centering
    \includegraphics[width=1\columnwidth]{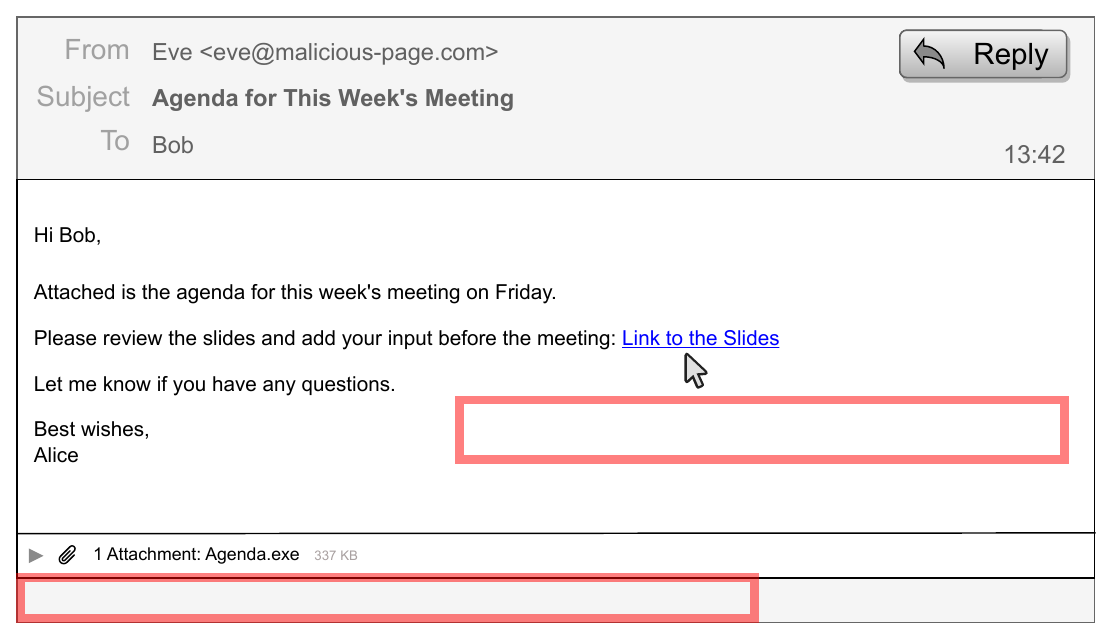}
    \caption{Example of Link HTML Form Tag.}
    \label{fig:nlink-html-form-tag}
    \vspace{-1em}
\end{figure}

 As shown in~\cref{fig:nlink-html-form-tag}, the attacker uses a form submit element so that the destination URL is not shown on hover. 


\subsection{Link Mismatch}
\label{sec:link-mismatch}

In this deception technique, the attacker aims to mislead recipients about the destination of a link by displaying link text that suggests a trustworthy destination, while the actual target URL is different.

To achieve this, the attacker uses a trustworthy-looking link text, but sets the actual target URL to an attacker-controlled website. This exploits that recipients may rely on the displayed link text when assessing a link and may not verify the actual target URL (e.g., by hovering).

We describe the following implementations of this deception technique.

\subsubsection{Implementation: Entire URL as Link Text}
\label{sec:link-mismatch-entire-url-text}

This implementation (mentioned in literature such as~\cite{pearson_click_2017,sankhwar_novel_2018,suriya_integrated_2009,li_detection_2020,khursheed_microtargeting_2020}) exploits that link text does not need to match the link's actual target URL and that recipients may not perform further checks such as hovering.

\begin{figure}[H]
    \centering
    \includegraphics[width=1\columnwidth]{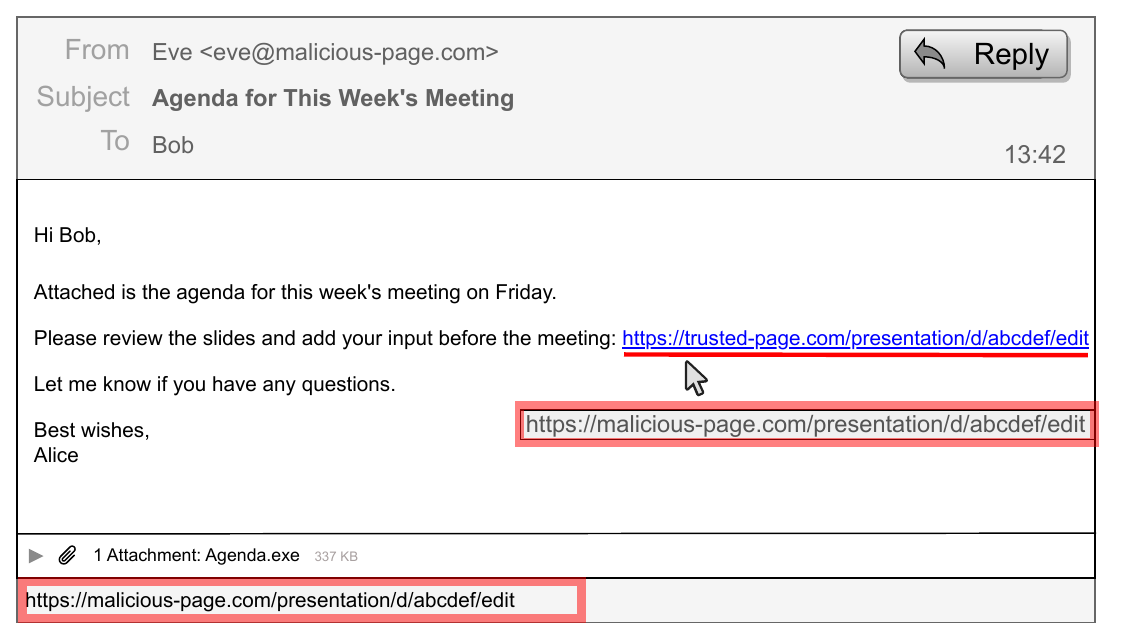}
    \caption{Example of Link Mismatch (Implementation: Entire URL as Link Text).}
    \label{fig:nLink-Mismatch-Entire-URL-As-Link-Text}
    \vspace{-1em}
\end{figure}

 As shown in~\cref{fig:nLink-Mismatch-Entire-URL-As-Link-Text}, the attacker displays the trustworthy-looking URL \texttt{https://trusted-page.com/presentation/d/}[...] as link text. When the link is hovered, the tooltip reveals that the actual target URL is different.

\subsubsection{Implementation: Domain or Brand as Link Text}
\label{sec:link-mismatch-partial-url-text}

This implementation (mentioned in literature such as~\cite{pearson_click_2017,sankhwar_novel_2018,suriya_integrated_2009,li_detection_2020,khursheed_microtargeting_2020}) exploits that link text does not need to match the link's actual target URL. It uses only the domain or a brand-like string as link text (e.g., \texttt{trusted-page.com} or \texttt{Trusted Page}) while the actual target URL points elsewhere.

\begin{figure}[H]
    \centering
    \includegraphics[width=1\columnwidth]{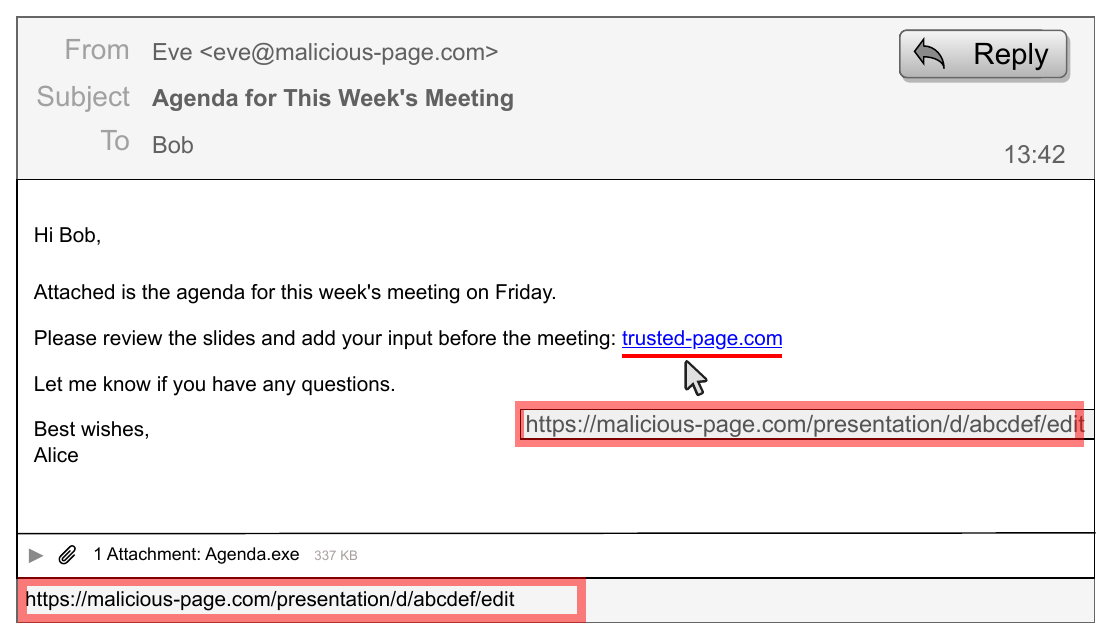}
    \caption{Example of Link Mismatch (Implementation: Domain or Brand as Link Text).}
    \label{fig:nLink-Mismatch-Domain-or-Brand-As-Linktext}
    \vspace{-1em}
\end{figure}

 As shown in~\cref{fig:nLink-Mismatch-Domain-or-Brand-As-Linktext}, the attacker uses \texttt{trusted-page.com} as link text while the actual target URL points to \texttt{malicious-page.com}.

\subsubsection{Implementation: Look-alike URL Scheme to Bypass Email Client Warnings}
\label{sec:link-mismatch-lookalike-url-scheme}

This implementation (identified during our examination) exploits that some email clients use the URL Scheme string (e.g., \texttt{http://} or \texttt{https://}) as a detection signal to warn users when link text resembles a URL. By replacing characters in the URL Scheme string with visually similar Unicode characters, the attacker can craft link text that still appears as a URL to the recipient but no longer matches the pattern used for client-side detection.

\begin{figure}[H]
    \centering
    \includegraphics[width=1\columnwidth]{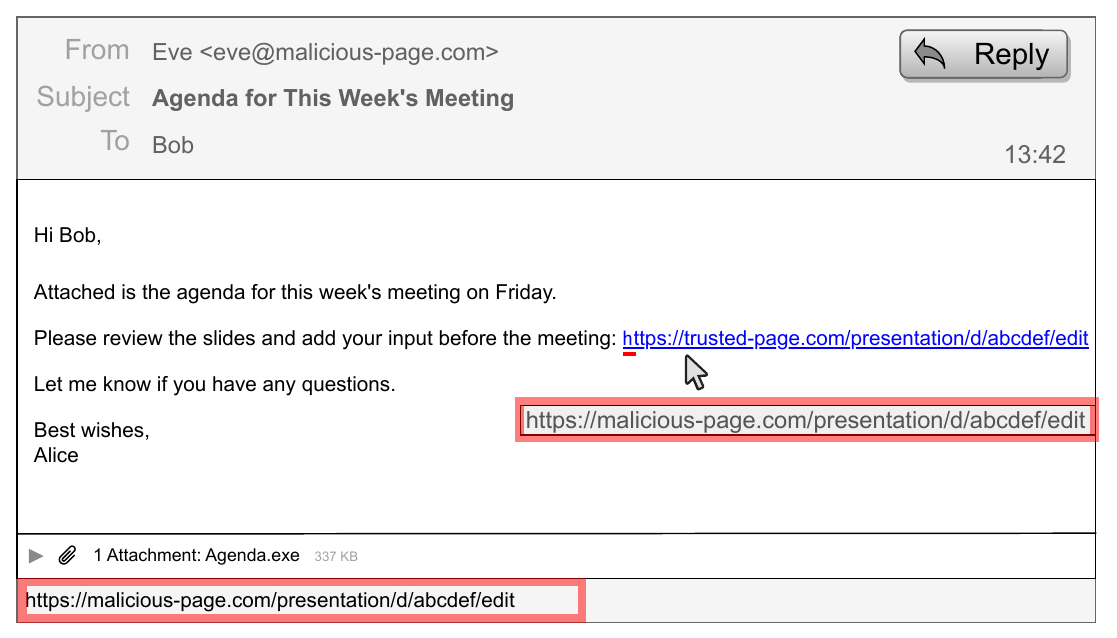}
    \caption{Example of Link Mismatch (Implementation: Look-alike URL Prefix to Bypass Client Warnings).}
    \label{fig:nlink-mismatch-lookalike-url-scheme}
    \vspace{-1em}
\end{figure}

 As shown in~\cref{fig:nlink-mismatch-lookalike-url-scheme}, the email client shows a warning for a URL-like link text in the default case. After substituting the \texttt{h} in the \texttt{https://} URL Scheme string with the visually similar Unicode character \texttt{\foreignlanguage{russian}{һ}}, the link text still appears to start with \texttt{https://} to the recipient, but the warning is no longer displayed.

\subsection{Link External Content Manipulation}
\label{sec:link-external-content-manipulation}

In this deception technique, the attacker aims to mislead recipients about the destination of a link by changing externally loaded content after the email has been delivered.

To achieve this, the attacker embeds references to external content (e.g., remotely hosted CSS) that affects how the email is rendered when it is opened. This exploits that the displayed link destination can change between different openings of the same email, while recipients may assume that the email content remains stable.

A concrete implementation of this deception technique (mentioned in literature such as~\cite{muller_re_2019}) exploits that some email clients load and apply externally hosted CSS when rendering an email. Since the attacker can change the external CSS at will, the attacker can decide which of two links is shown or hidden. For the recipient, there may be no visual difference at first glance; only when hovering over the link is the different target URL revealed. However, if the link is checked once, recipients may not check it a second time.

\begin{figure}[H]
    \centering
    \begin{subfigure}[b]{\columnwidth}
        \centering
        \includegraphics[width=1\columnwidth]{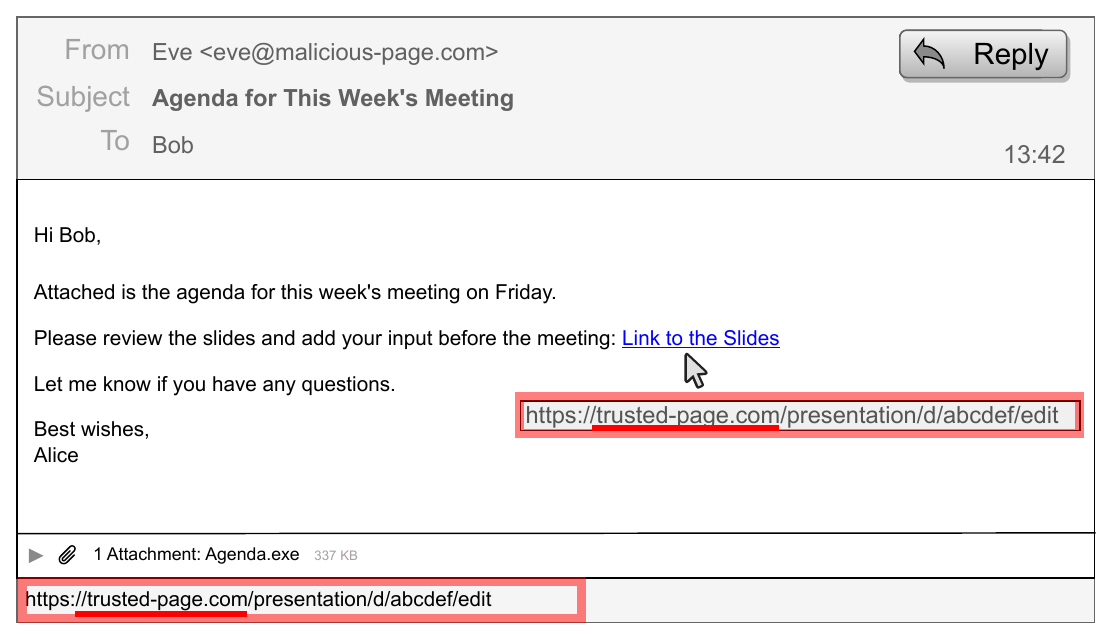}
        \caption{Before external CSS change.}
        \label{fig:nLink-External-Content-Manipulation-Change-of-the-Link-Destination-before}
    \end{subfigure}

    \vspace{1ex}

    \begin{subfigure}[b]{\columnwidth}
        \centering
        \includegraphics[width=1\columnwidth]{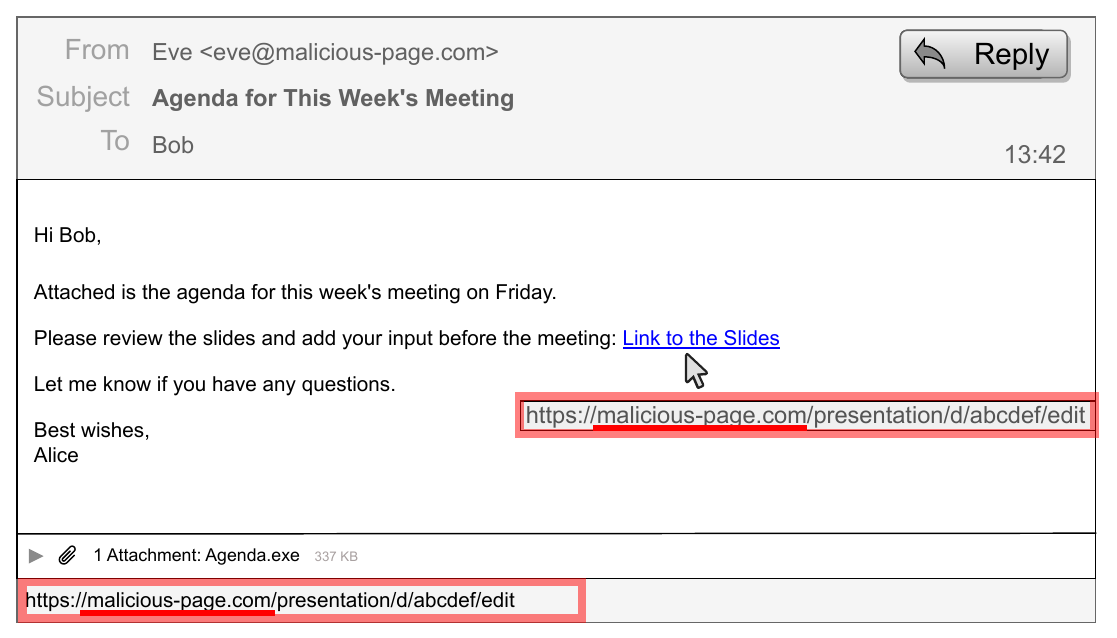}
        \caption{After external CSS change.}
        \label{fig:nLink-External-Content-Manipulation-Change-of-the-Link-Destination-after}
    \end{subfigure}

    \caption{Example of Link External Content Manipulation.}
    \label{fig:nLink-External-Content-Manipulation}
    \vspace{-1em}
\end{figure}

 As shown in~\cref{fig:nLink-External-Content-Manipulation-Change-of-the-Link-Destination-before,fig:nLink-External-Content-Manipulation-Change-of-the-Link-Destination-after}, the attacker initially serves external content such that the link appears to lead to a trustworthy website when the email is first opened. When the email is opened again, the attacker changes the external content so that the same link directs to an attacker-controlled website. This enables scenarios where a first recipient verifies the link and forwards the email, while later recipients follow a different (malicious) destination.

\subsection{Link URL Suffix}
\label{sec:link-url-suffix}

In this deception technique, the attacker aims to mislead recipients about the destination of a link by appending trustworthy-looking strings to a URL.

To achieve this, the attacker places these strings in the URL Path, Query, or Fragment, while keeping the Registrable Domain attacker-controlled. This exploits that recipients may focus on familiar strings that appear later in the URL and may not fully inspect which Registrable Domain determines the actual website that is opened.

We describe the following implementations of this deception technique.

\subsubsection{Implementation: Domain in URL Suffix}
\label{sec:link-url-suffix-domain-in-Suffix}

This implementation (mentioned in literature such as~\cite{soni_phishing_2011,jampen_dont_2020}) exploits that strings placed in the URL Path, Query, or Fragment do not change which Registrable Domain is contacted (see URL structure in~\cref{sec:urlstructure}). The attacker therefore appends a trusted-looking Registrable Domain string as the suffix.

\begin{figure}[H]
     \centering
     \includegraphics[width=1\columnwidth]{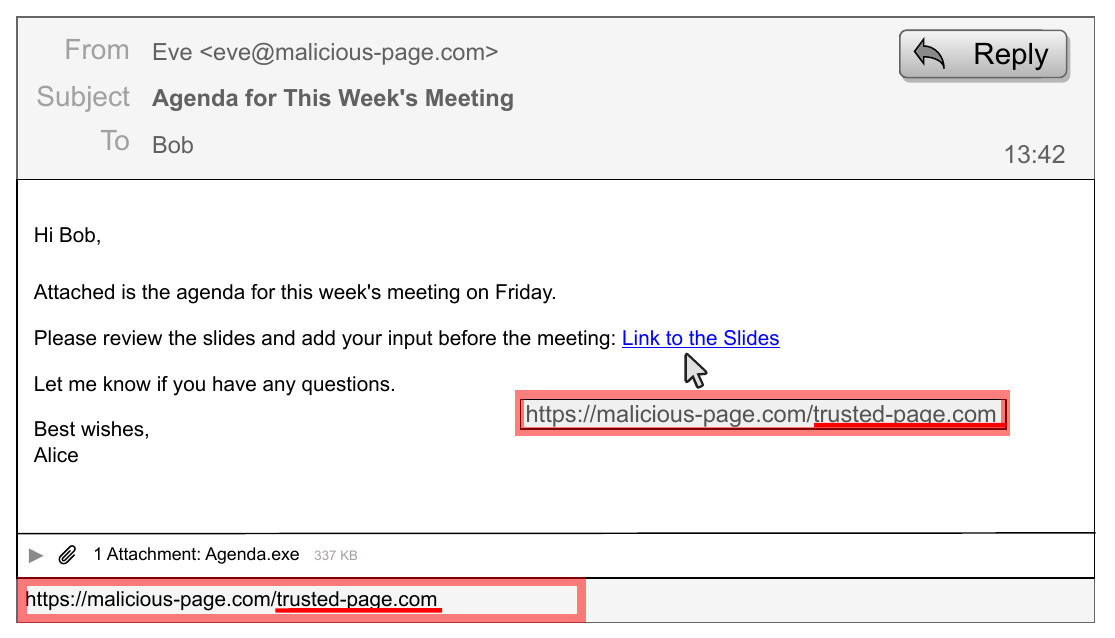}
    \caption{Example of Link URL Suffix (Implementation: Domain in URL Suffix).}
     \label{fig:nlink-url-suffix-domain-in-Suffix}
     \vspace{-1em}
 \end{figure}

 As shown in~\cref{fig:nlink-url-suffix-domain-in-Suffix}, the attacker uses \texttt{https://malicious-page.com/trusted-page.com}. Equivalent variants place the same suffix in the Query or Fragment, e.g., \texttt{https://malicious-page.com/?trusted-page.com} or \texttt{https://malicious-page.com/\#trusted-page.com}.

\subsubsection{Implementation: Entire URL in URL Suffix}
\label{sec:link-url-suffix-entire-url-in-URL-Suffix}

This implementation (mentioned in literature such as~\cite{soni_phishing_2011,jampen_dont_2020}) exploits that an attacker can append an entire trustworthy-looking URL as the suffix.

\begin{figure}[H]
     \centering
     \includegraphics[width=1\columnwidth]{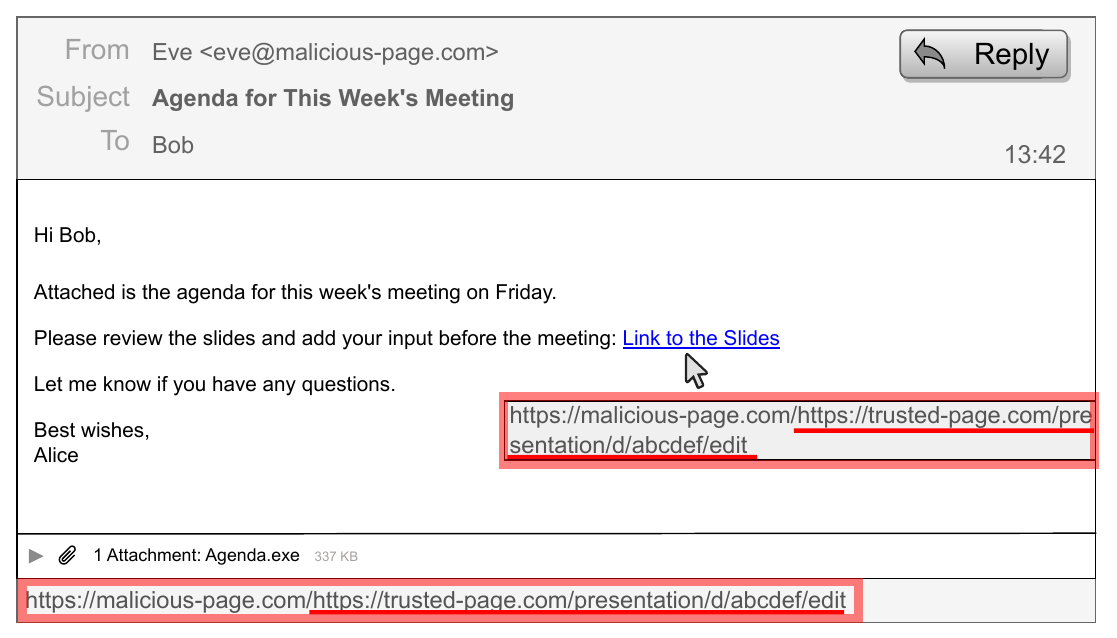}
    \caption{Example of Link URL Suffix (Implementation: Entire URL in URL Suffix).}
     \label{fig:nlink-url-suffix-entire-url-in-URL-Suffix}
     \vspace{-1em}
 \end{figure}

As shown in~\cref{fig:nlink-url-suffix-entire-url-in-URL-Suffix}, the attacker uses \texttt{https://malicious-\allowbreak{}page.com/\allowbreak{}https://trusted-\allowbreak{}page.com\allowbreak{}/presentation/[...]}. Equivalent variants place the suffix in the Query or Fragment:
\begin{itemize}[noitemsep]
    \item \texttt{https://malicious-\allowbreak{}page.com\allowbreak{}/?https://trusted-\allowbreak{}page.com\allowbreak{}/presentation/[...]}
    \item \texttt{https://\allowbreak{}malicious-\allowbreak{}page.com/\#https://\allowbreak{}trusted-\allowbreak{}page.com\allowbreak{}/presentation/[...]}
\end{itemize}

\subsubsection{Implementation: Luring Label with Domain in URL Suffix}
\label{sec:link-url-suffix-luring-domain-with-Domain-in-URL-Suffix}

This implementation (identified during our examination) exploits that an attacker can prepend a luring label (e.g., \texttt{redirection-to}) before a trusted-looking Registrable Domain string in the suffix to suggest a redirection service.

\begin{figure}[H]
     \centering
     \includegraphics[width=1\columnwidth]{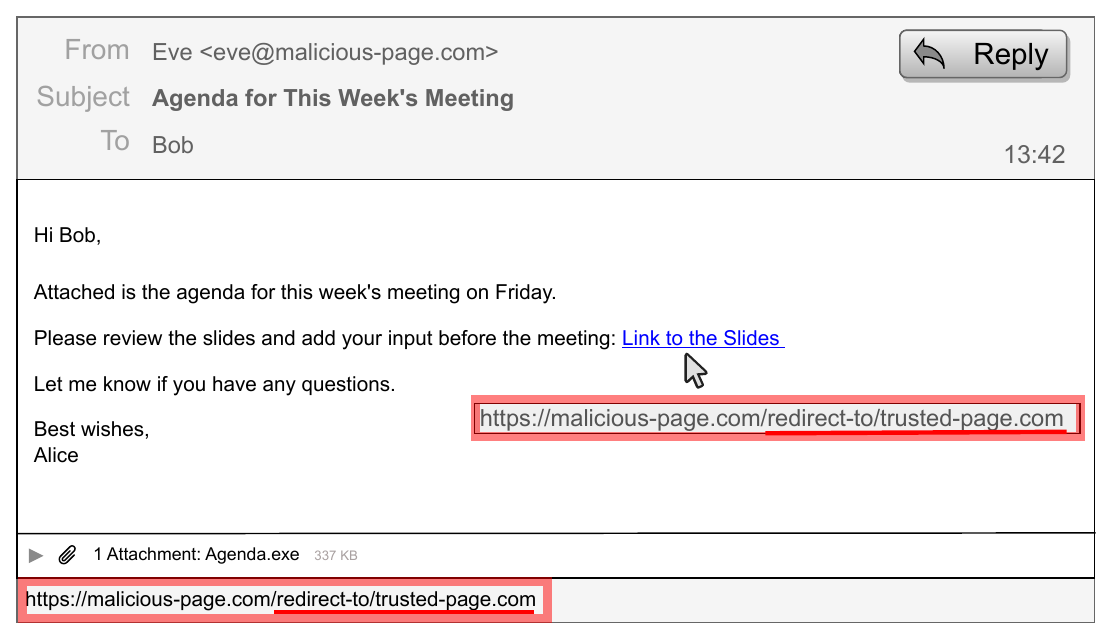}
    \caption{Example of Link URL Suffix (Implementation: Luring Label with Domain in URL Suffix).}
     \label{fig:nlink-url-suffix-luring-domain-with-Domain-in-URL-Suffix}
     \vspace{-1em}
 \end{figure}

As shown in~\cref{fig:nlink-url-suffix-luring-domain-with-Domain-in-URL-Suffix}, the attacker uses \texttt{https://malicious-\allowbreak{}page.com\allowbreak{}/redirect-to\allowbreak{}/trusted-\allowbreak{}page.com}. Equivalent variants place the suffix in the Query or Fragment:
\begin{itemize}[noitemsep]
    \item \texttt{https://\allowbreak{}malicious-\allowbreak{}page.com\allowbreak{}/?redirect-\allowbreak{}to=\allowbreak{}trusted-page.com}
    \item \texttt{https://\allowbreak{}malicious-\allowbreak{}page.com\allowbreak{}/\#redirect-\allowbreak{}to\#trusted-\allowbreak{}page.com}
\end{itemize}

\subsubsection{Implementation: Luring Label with URL in URL Suffix}
\label{sec:link-url-suffix-luring-url-in-URL-Suffix}

This implementation (identified during our examination) exploits that an attacker can prepend a luring label (e.g., \texttt{redirection-to}) before a trusted-looking URL in the suffix, while the Registrable Domain remains attacker-controlled.

\begin{figure}[H]
     \centering
     \includegraphics[width=1\columnwidth]{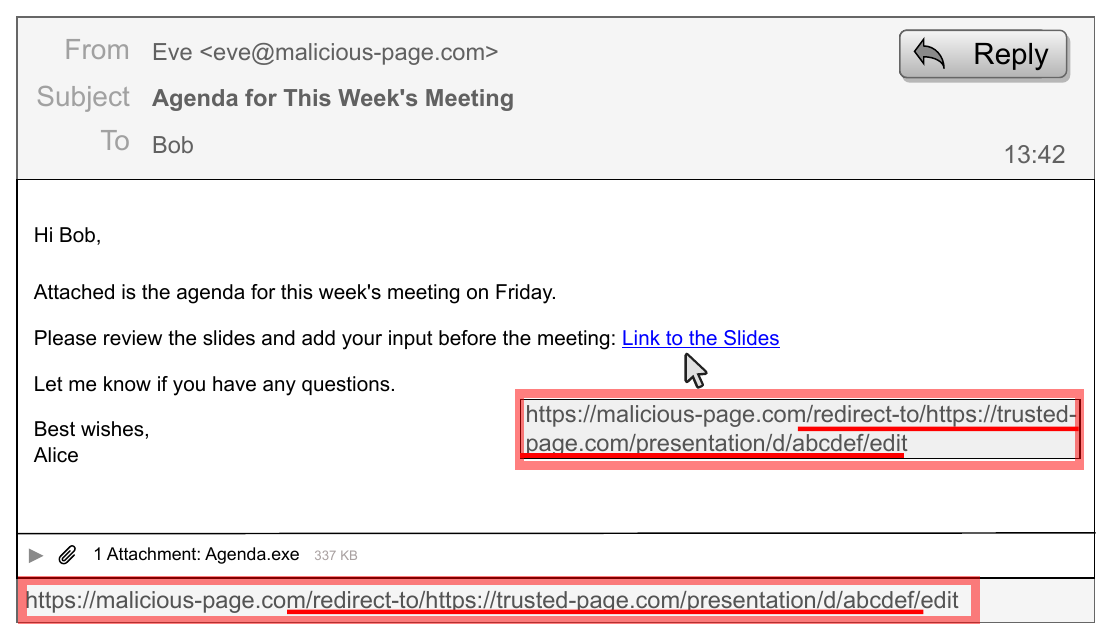}
    \caption{Example of Link URL Suffix (Implementation: Luring Label with URL in URL Suffix).}
     \label{fig:nlink-url-suffix-luring-url-in-URL-Suffix}
     \vspace{-1em}
 \end{figure}

As shown in~\cref{fig:nlink-url-suffix-luring-url-in-URL-Suffix}, the attacker uses \texttt{https://malicious-\allowbreak{}page.com\allowbreak{}/redirection-\allowbreak{}to/https://trusted-\allowbreak{}page.com/presentation/[...]}. Equivalent variants place the suffix in the Query or Fragment:
\begin{itemize}[noitemsep]
    \item \texttt{https://\allowbreak{}malicious-\allowbreak{}page.com/?redirect-to=\allowbreak{}https://trusted-\allowbreak{}page.com\allowbreak{}/presentation/[...]}
    \item \texttt{https://\allowbreak{}malicious-\allowbreak{}page.com/\allowbreak{}\#redirect\allowbreak{}/https://\allowbreak{}trusted-\allowbreak{}page.com/\allowbreak{}presentation/[...]}
\end{itemize}

\subsubsection{Implementation: Look-alike Domain String in URL Suffix}
\label{sec:link-url-suffix-look-alike-Domain-String-in-URL-Suffix}

This implementation (identified during our examination) exploits that the suffix can contain look-alike domain strings that resemble the targeted domain but are slightly modified (e.g., using separators such as \texttt{-} instead of \texttt{.}). Such suffixes can appear like a redirection hint while the Registrable Domain remains attacker-controlled.

\begin{figure}[H]
     \centering
     \includegraphics[width=1\columnwidth]{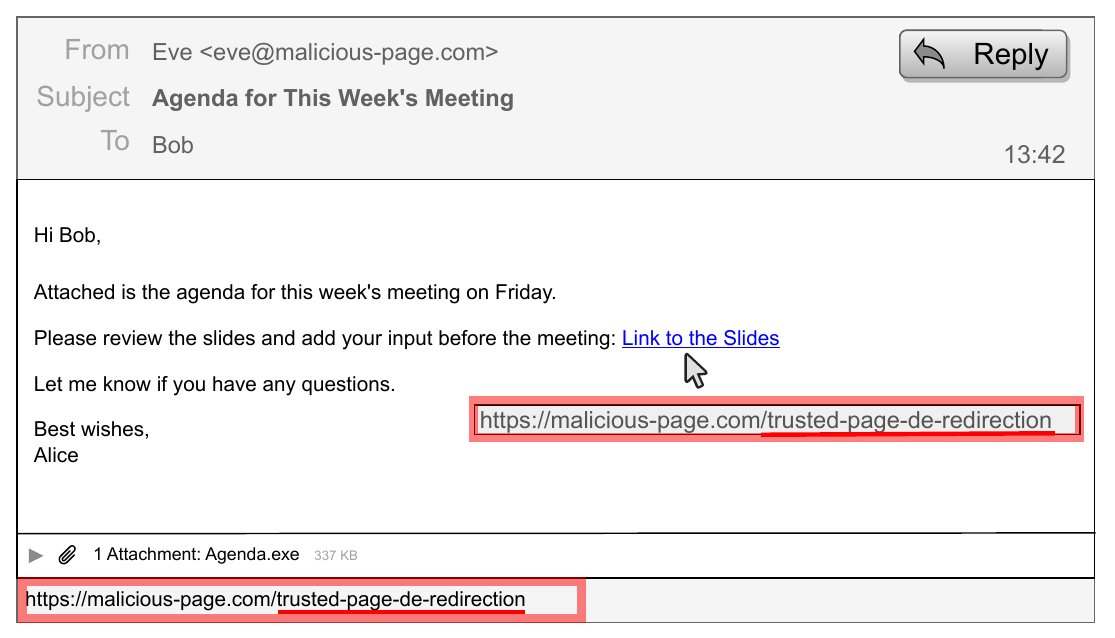}
    \caption{Example of Link URL Suffix (Implementation: Look-alike Domain String in URL Suffix).}
     \label{fig:nlink-url-suffix-look-alike-Domain-String-in-URL-Suffix}
     \vspace{-1em}
 \end{figure}

 As shown in~\cref{fig:nlink-url-suffix-look-alike-Domain-String-in-URL-Suffix}, the attacker uses \texttt{https://malicious-page.com/trusted-page-de-redirection}. Equivalent variants place the same suffix in the Query or Fragment, e.g., \texttt{https://malicious-page.com/?trusted-page-de-redirection} or \texttt{https://malicious-page.com/\#trusted-page-de-redirection}.

\subsection{Link URL Omitting Slash}
\label{sec:link-url-omitting-slash}

In this deception technique, the attacker aims to mislead recipients about the destination of a link by crafting a URL whose structure makes the Registrable Domain boundary difficult to identify.

To achieve this, the attacker omits a slash in the URL, creating ambiguity about where the Registrable Domain ends. This exploits that recipients may use simple positional heuristics to locate the Registrable Domain and may therefore misidentify it.

A concrete implementation of this deception technique (identified during our examination) exploits that recipients may identify the Registrable Domain using the third forward slash as an orientation cue (as taught in some security awareness trainings~\cite{berensBetterTogheter}). By omitting the slash before the Fragment delimiter (i.e.,~\#), the attacker ensures the first slash after the Registrable Domain appears only after the Fragment. Since the Fragment does not affect which website is opened, a trusted-looking string placed there does not change the actual destination, while a slash following the Fragment can be mistaken as the domain boundary.

\begin{figure}[H]
    \centering
    \includegraphics[width=1\columnwidth]{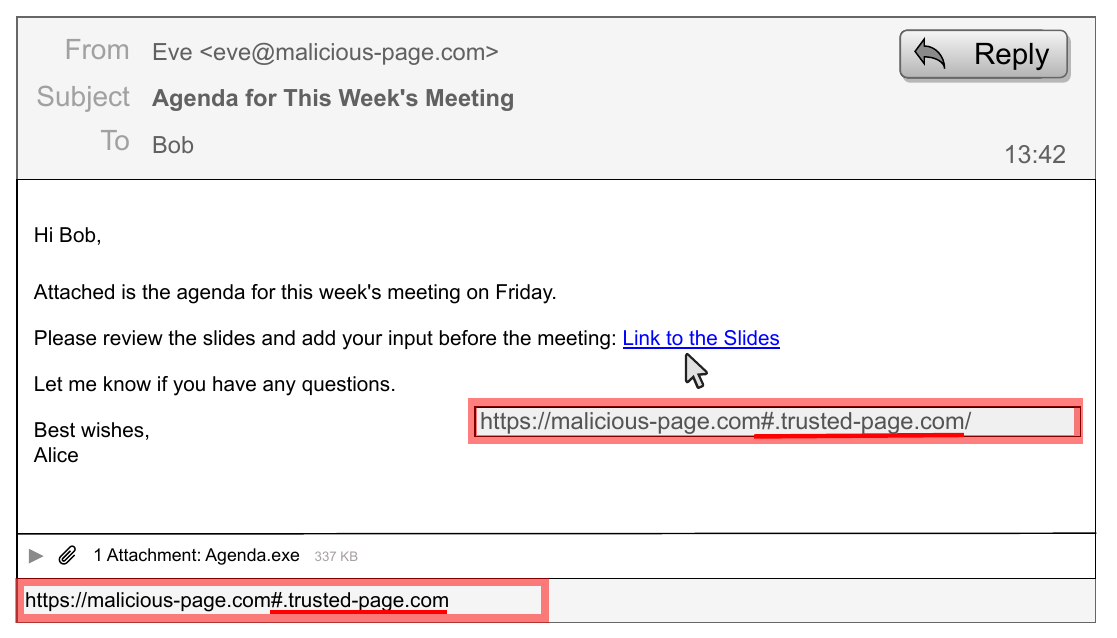}
    \caption{Example of Link URL Omitting Slash.}
    \label{fig:nlink-url-omitting-slash}
    \vspace{-1em}
\end{figure}

 As shown in~\cref{fig:nlink-url-omitting-slash}, the attacker uses \texttt{https://malicious-page.com\#.trusted-page.com/}. The Fragment string \texttt{.trusted-page.com/} is followed by a forward slash, which recipients applying the third-slash heuristic may misread as the Registrable Domain boundary.

\subsection{Link URL Userinfo}
\label{sec:link-url-userinfo}

In this deception technique, the attacker aims to mislead the recipient about the destination of a link by placing a trustworthy-looking URL string into the Userinfo of the target URL.

To achieve this, the attacker places a trusted-looking Registrable Domain string in the Userinfo of the URL. This exploits that recipients may scan URLs from left to right and may not recognize that the Registrable Domain follows the Userinfo (after the \texttt{@} symbol) and determines the actual link destination.

A concrete implementation of this deception technique (mentioned in literature such as~\cite{sankhwar_novel_2018,qabajeh_experimental_2014,suriya_integrated_2009,vrbancic_datasets_2020,zhu_dtof-ann_2020,awasthi_generating_2021,priya_gravitational_2020,azeez_identifying_2020-1,abedin_phishing_2020,salloum_phishing_2021-2,singh_phishing_2020,swarnalatha_real-time_2021,du_research_2013,bhardwaj_why_2020}) exploits that URLs can contain an optional Userinfo (see URL structure in~\cref{sec:urlstructure}) whose content can be freely chosen by the attacker. By placing a trusted-looking Registrable Domain string in the Userinfo before the \texttt{@} symbol, the attacker makes the URL begin with a trusted-looking string while the actual Registrable Domain after \texttt{@} remains attacker-controlled.

\begin{figure}[H]
    \centering
    \includegraphics[width=1\columnwidth]{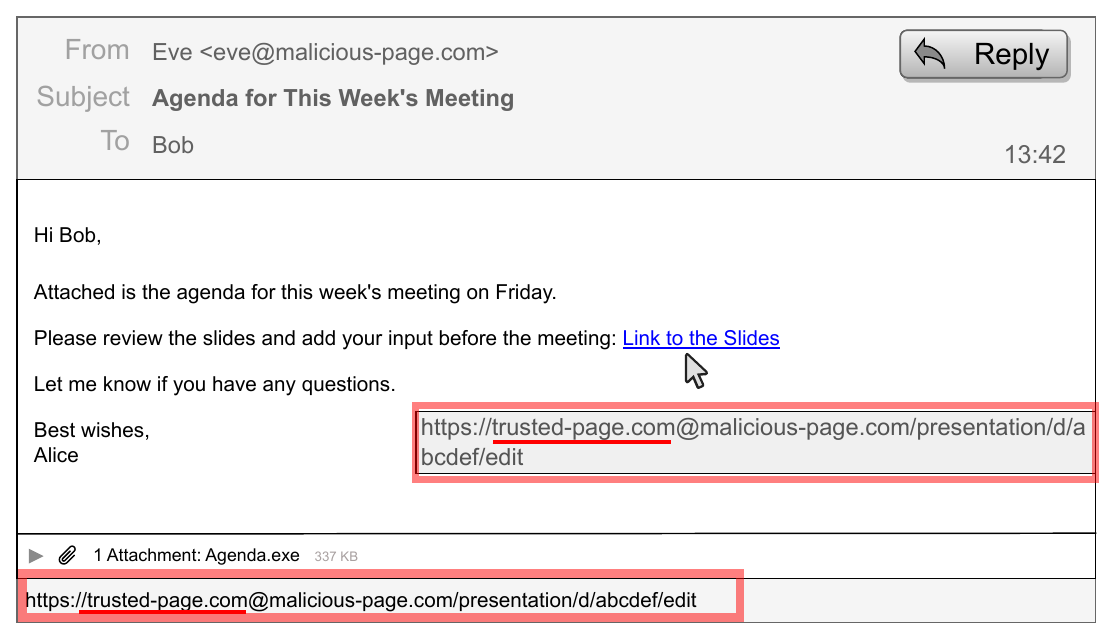}
    \caption{Example of Link URL Userinfo.}
    \label{fig:nlink-url-userinfo}
    \vspace{-1em}
\end{figure}

As shown in~\cref{fig:nlink-url-userinfo}, the attacker uses \texttt{https://trusted-\allowbreak{}page.com@malicious-\allowbreak{}page.com/[...]}. The URL appears to begin with \texttt{trusted-page.com}, while the actual Registrable Domain \texttt{malicious-page.com} follows after \texttt{@}.

\subsection{Link URL Shorteners}
\label{sec:link-url-shorteners}

In this deception technique, the attacker aims to mislead recipients about the destination of a link by using a URL shortening service.

To achieve this, the attacker replaces the original URL with a shortened URL whose Registrable Domain belongs to a known shortening service. This exploits that recipients may be more neutral towards a shortener domain than towards an unknown domain.

We describe the following implementations of this deception technique.

\subsubsection{Implementation: Basic Shortened URL}
\label{sec:link-url-shorteners-basic-shortened-url}

This implementation (mentioned in literature such as~\cite{volkamer_user_2017}) exploits that a URL shortener redirects to an attacker-controlled destination while the displayed URL only reveals the shortener domain and an identifier.

\begin{figure}[H]
     \centering
     \includegraphics[width=1\columnwidth]{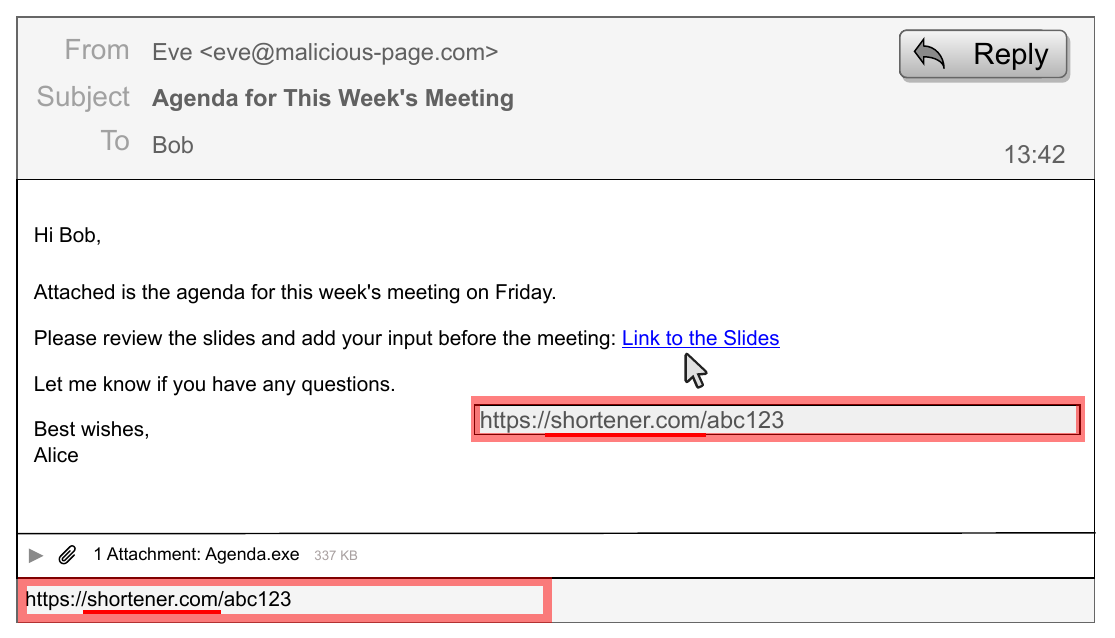}
     \caption{Example of Link URL Shorteners (Implementation: Basic Shortened URL).}
     \label{fig:nlink-url-shorteners-basic-shortened-url}
     \vspace{-1em}
 \end{figure}

 As shown in~\cref{fig:nlink-url-shorteners-basic-shortened-url}, the attacker uses a shortened URL such as \texttt{https://shortener.com/abc123}, which conceals that the final destination is \texttt{malicious-page.com}.

\subsubsection{Implementation: Suffix}
\label{sec:link-url-shorteners-shortened-url-suffix}

This implementation (identified during our examination) exploits that, depending on the URL shortener, an attacker can append additional strings in the Path, Query, or Fragment of a shortened URL without changing the final redirection destination, analogous to the Link URL Suffix implementations (cf.~\cref{sec:link-url-suffix}).

\begin{figure}[H]
     \centering
     \includegraphics[width=1\columnwidth]{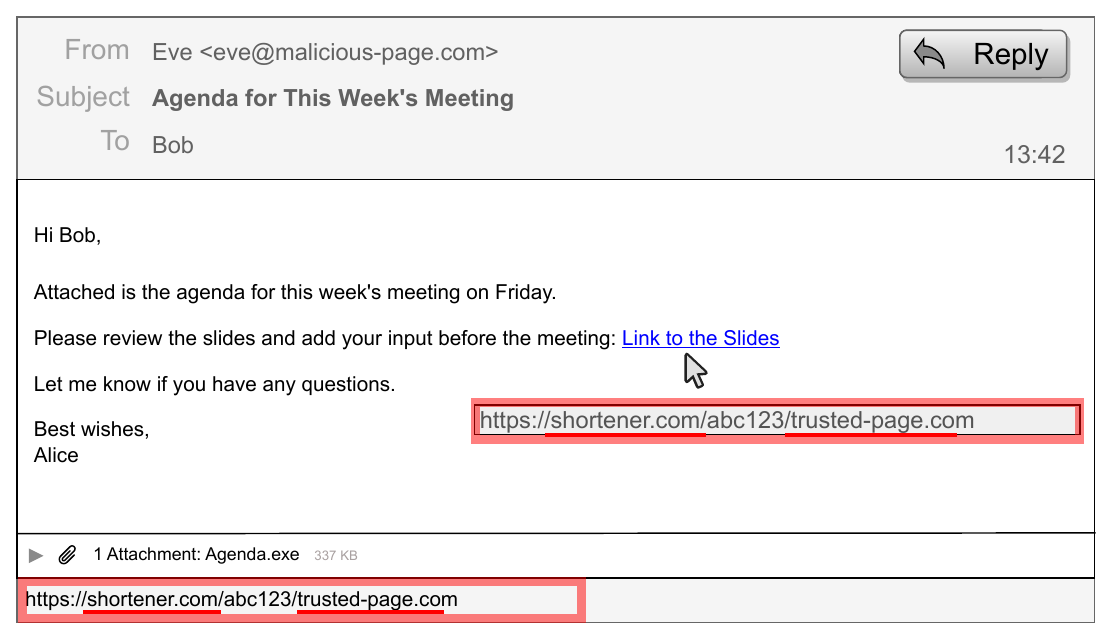}
     \caption{Example of Link URL Shorteners (Implementation: Shortened URL Suffix).}
     \label{fig:nlink-url-shorteners-shortened-url-suffix}
     \vspace{-1em}
 \end{figure}

 As shown in~\cref{fig:nlink-url-shorteners-shortened-url-suffix}, the attacker uses a shortened URL such as \texttt{https://shortener.com/abc123/trusted-page.com}.

\subsubsection{Implementation: Identifier}
\label{sec:link-url-shorteners-custom-shortener-identifier}

This implementation (identified during our examination) exploits that some URL shorteners allow choosing a custom identifier string. The attacker can select an identifier that contains a trusted domain string, optionally combined with a luring label, even though the actual destination remains attacker-controlled.

\begin{figure}[H]
     \centering
     \includegraphics[width=1\columnwidth]{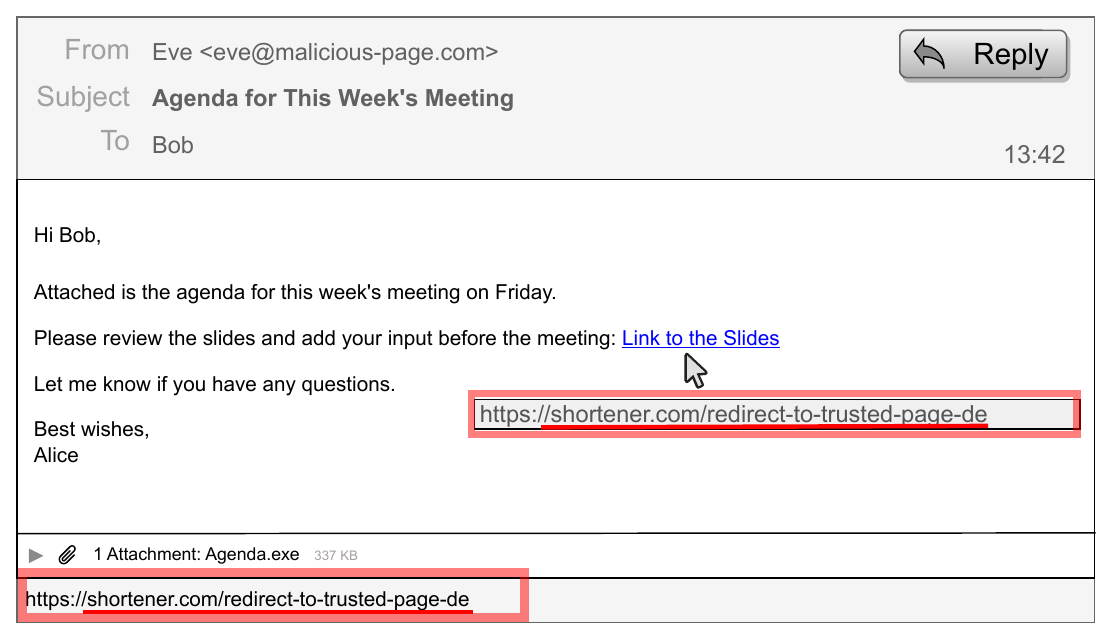}
     \caption{Example of Link URL Shorteners (Implementation: Custom Shortened URL Identifier).}
     \label{fig:nlink-url-shorteners-custom-shortener-identifier}
     \vspace{-1em}
 \end{figure}

 As shown in~\cref{fig:nlink-url-shorteners-custom-shortener-identifier}, the attacker uses a shortened URL such as \texttt{https://shortener.com/redirect-to-trusted-page-de}. The identifier suggests a redirection to \texttt{trusted-page.com}, although the shortener redirects to an attacker-controlled destination.

\subsection{Link URL Redirects}
\label{sec:link-url-redirects}

In this deception technique, the attacker aims to mislead recipients about the destination of a link by using a redirect URL whose Registrable Domain belongs to a trusted website, although the URL forwards to an attacker-controlled website.

To achieve this, the attacker uses a redirect endpoint on a trusted website and embeds the actual destination as part of the URL (e.g., as a Query parameter or Path component). This exploits that recipients may assess the link mainly based on the trusted Registrable Domain and may not recognize that the URL redirects to a different destination.

We describe the following implementations of this deception technique.

\subsubsection{Implementation: Misuse of Legitimate Redirect Services}
\label{sec:link-url-redirects-misuse-of-legitimate-redirect-service}

This implementation (mentioned in literature such as~\cite{volkamer_user_2017}) exploits that some websites provide redirect URLs as an intended feature (e.g., for tracking or link protection), where the destination URL is encoded as part of the redirect URL (e.g., \texttt{https://www.google\allowbreak{}.com/url?\allowbreak{}url=<destination>}). Unlike unintended open-redirect vulnerabilities, these redirect endpoints are deliberate, but they may still only partially mitigate abuse, so the trusted Registrable Domain can still be used to disguise the actual destination.

\begin{figure}[H]
    \centering
    \includegraphics[width=1\columnwidth]{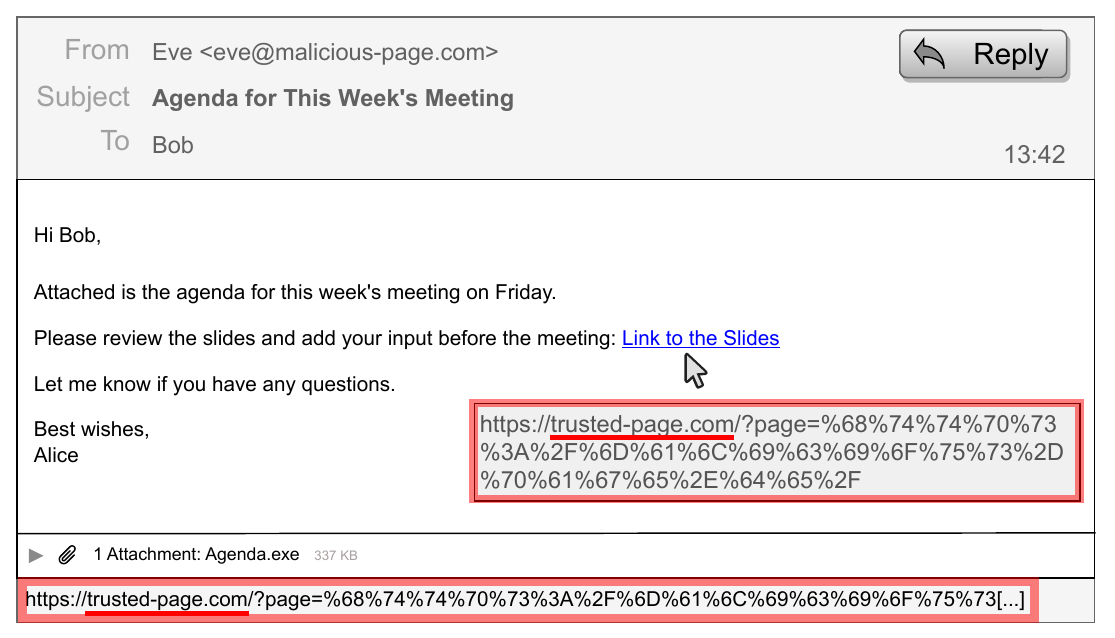}
    \caption{Example of Link URL Redirects (Implementation: Misuse of Legitimate Redirect Services).}
    \label{fig:nlink-url-redirects-misuse-of-legitimate-redirect-service}
    \vspace{-1em}
\end{figure}

 As shown in~\cref{fig:nlink-url-redirects-misuse-of-legitimate-redirect-service}, the attacker uses a redirect URL under \texttt{trusted-page.com} where the destination is provided as a URL parameter and appears in encoded form. In the example, the encoded parameter value contains the attacker-controlled destination \texttt{malicious-page.com}:
\texttt{https://trusted-\allowbreak{}page.com/?u\allowbreak{}rl=\%68\%74[...]}.
When the link is hovered, the link security indicator shows the trusted Registrable Domain \texttt{trusted-page.com}, although opening the link forwards to \texttt{malicious-page.com}.

\subsubsection{Implementation: Open Redirect on a Trusted Domain}
\label{sec:link-url-redirects-open-redirect-on-a-trusted-domain}

This implementation (mentioned in literature such as~\cite{shue2008exploitable}) exploits unintended open redirects caused by misconfigurations or vulnerabilities on otherwise trusted domains. The attacker crafts a redirect URL on the trusted domain that forwards to an attacker-controlled destination.

\begin{figure}[H]
     \centering
     \includegraphics[width=1\columnwidth]{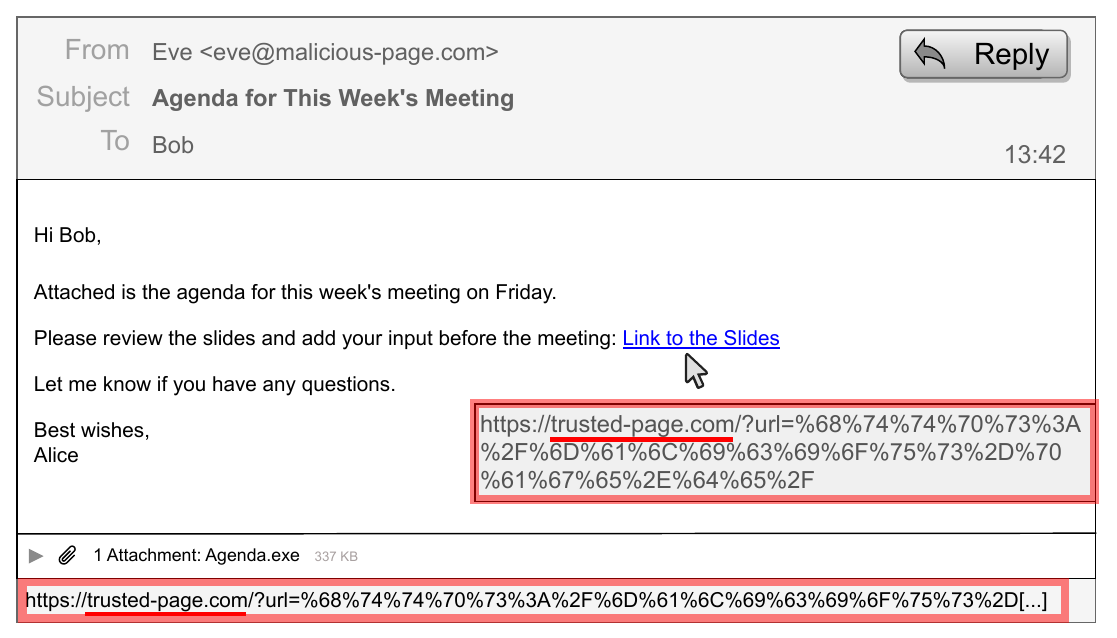}
     \caption{Example of Link URL Redirects (Implementation: Open Redirect on a Trusted Domain).}
     \label{fig:nlink-url-redirects-open-redirect-on-a-trusted-domain}
     \vspace{-1em}
 \end{figure}

 As shown in~\cref{fig:nlink-url-redirects-open-redirect-on-a-trusted-domain}, the attacker uses a URL under a trusted domain (e.g., \texttt{https://trusted-\allowbreak{}page.com/?pa\allowbreak{}ge=\%68\%74[...]}). When the link is hovered, the URL with the trusted Registrable Domain \texttt{trusted-page.com} is shown, although opening the link redirects to \texttt{malicious-page.com}.

\subsection{Link URL Homographic Spoofing}
\label{sec:link-url-homographic-spoofing}

In this deception technique, the attacker aims to mislead recipients about the destination of a link by using a visually confusable Registrable Domain in the link URL.

To achieve this, the attacker registers a Registrable Domain that appears visually identical to a legitimate domain by substituting ASCII characters with Unicode homoglyphs (e.g., from Cyrillic or Greek alphabets). 

A concrete implementation of this deception technique (mentioned in literature such as~\cite{suriya_integrated_2009,hu_assessing_2021,hannay_assessment_2009,andryukhin_phishing_2019,du_research_2013,wang_strider_2006}) exploits that internationalized domain names can contain Unicode characters. For domain registration, such domains are encoded in Punycode, but many email clients display the decoded Unicode form~\cite{Veit2024}. By substituting characters with homoglyphs, the attacker can craft a domain that is visually indistinguishable from a legitimate domain.

\begin{figure}[H]
     \centering
     \includegraphics[width=1\columnwidth]{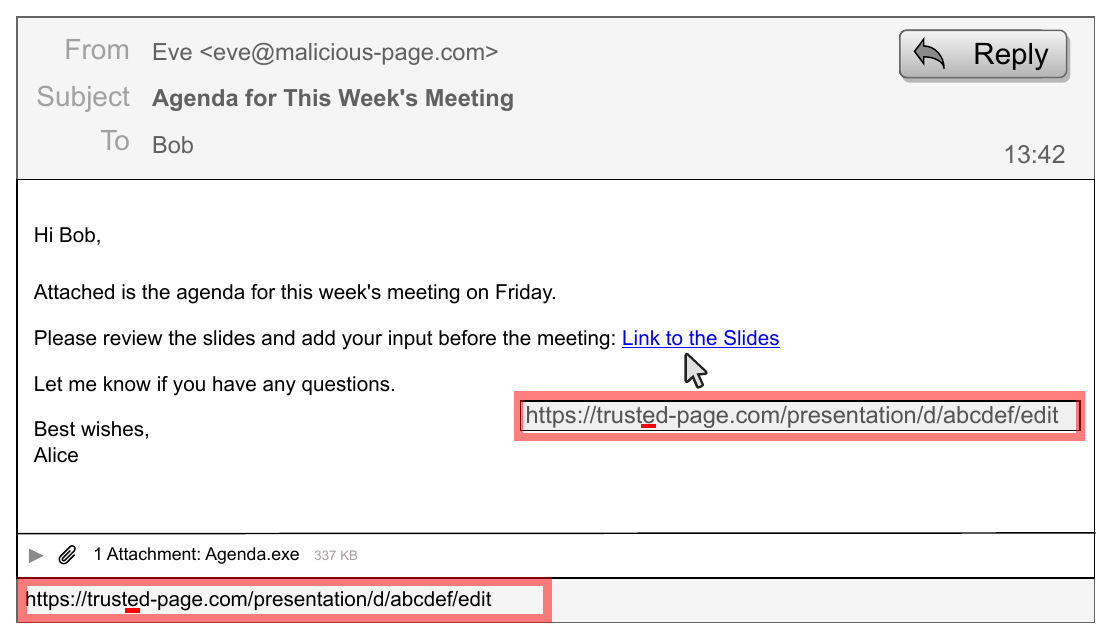}
     \caption{Example of Link URL Homographic Spoofing.}
     \label{fig:nlink-url-homographic-spoofing}
     \vspace{-1em}
 \end{figure}

 As shown in~\cref{fig:nlink-url-homographic-spoofing}, the attacker uses a link URL whose Registrable Domain is rendered in Unicode as \foreignlanguage{russian}{trustеd-page.com}. This domain is distinct from \texttt{trusted-page.com} because the first \texttt{e} is replaced with the Cyrillic character \foreignlanguage{russian}{е}. 

\subsection{Link URL Unrecognizable Domain}
\label{sec:link-url-unrecognizable-domain}

In this deception technique, the attacker aims to mislead recipients about the destination of a link by using an unrecognizable or unfamiliar-looking domain or IP in the link URL.

To achieve this, the attacker uses a link URL whose Registrable Domain does not provide meaningful cues to recipients (e.g., because it is a numeric address or resembles a random character sequence). Since such domains are difficult to interpret at a glance, recipients may rely on other email cues when deciding whether the link is trustworthy.

We describe the following implementations of this deception technique.

\subsubsection{Implementation: IP Address}
\label{sec:link-url-unrecognizable-domain-ip-address}

This implementation (mentioned in literature such as~\cite{sankhwar_novel_2018,soni_phishing_2011,qabajeh_experimental_2014,suriya_integrated_2009,vrbancic_datasets_2020,li_detection_2020,lee_d-fence_2021,jampen_dont_2020,zhu_dtof-ann_2020,awasthi_generating_2021,priya_gravitational_2020,azeez_identifying_2020-1,abedin_phishing_2020,salahdine_phishing_2021,salloum_phishing_2021-2,singh_phishing_2020,swarnalatha_real-time_2021,du_research_2013,volkamer_user_2017,bhardwaj_why_2020}) exploits that a numeric address (e.g., IPv4 or IPv6) can be used in place of the Registrable Domain, which can appear unfamiliar and hard to assess.

\begin{figure}[H]
     \centering
     \includegraphics[width=1\columnwidth]{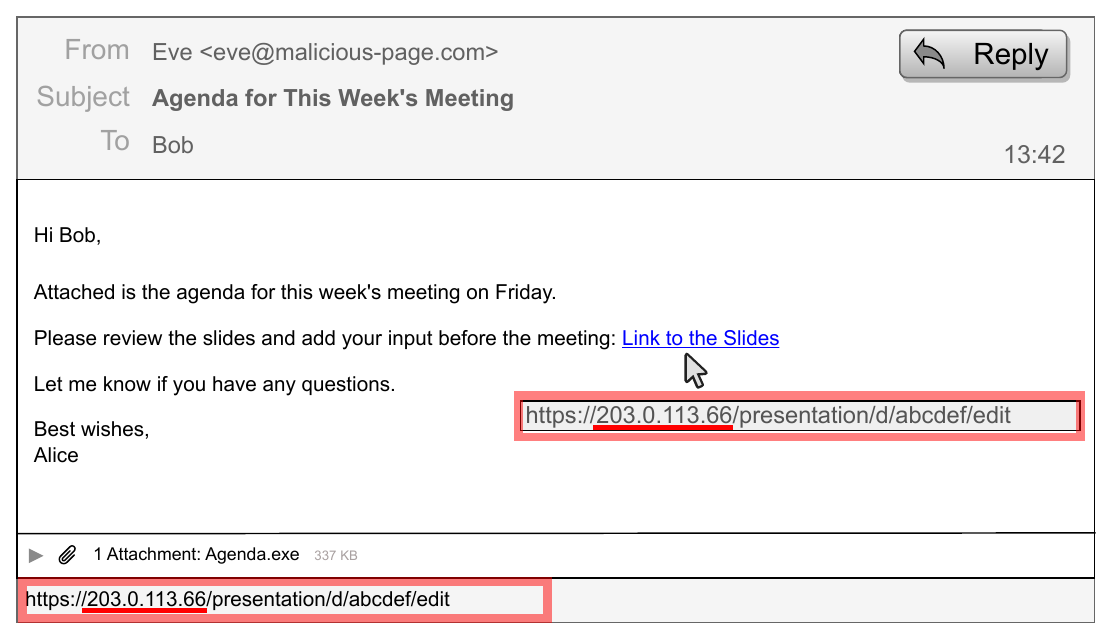}
     \caption{Example of Link URL Unrecognizable Domain (Implementation: IP Address).}
     \label{fig:nLink-Unrecognizable-Domain-IP-Address}
     \vspace{-1em}
 \end{figure}

 As shown in~\cref{fig:nLink-Unrecognizable-Domain-IP-Address}, the attacker uses a link URL with an IP address \texttt{203.0.113.66} instead of a Registrable Domain.

\subsubsection{Implementation: Random Characters}
\label{sec:link-url-unrecognizable-domain-random-characters}

This implementation (mentioned in literature such as~\cite{sankhwar_novel_2018,soni_phishing_2011,qabajeh_experimental_2014,suriya_integrated_2009,vrbancic_datasets_2020,li_detection_2020,lee_d-fence_2021,jampen_dont_2020,zhu_dtof-ann_2020,awasthi_generating_2021,priya_gravitational_2020,azeez_identifying_2020-1,abedin_phishing_2020,salahdine_phishing_2021,salloum_phishing_2021-2,singh_phishing_2020,swarnalatha_real-time_2021,du_research_2013,volkamer_user_2017,bhardwaj_why_2020}) exploits that a Registrable Domain can be chosen to resemble a random character sequence. Such domains provide little semantic context and may therefore be overlooked or under-scrutinized by recipients.

\begin{figure}[H]
     \centering
     \includegraphics[width=1\columnwidth]{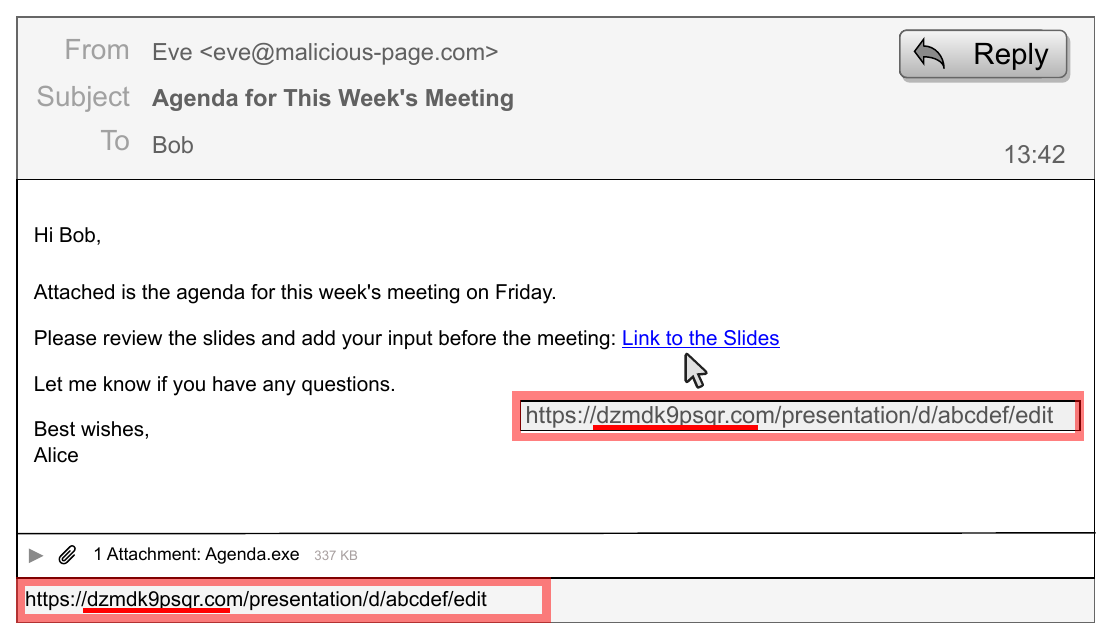}
    \caption{Example of Link URL Unrecognizable Domain (Implementation: Random Characters).}
     \label{fig:nLink-Unrecognizable-Domain-Random-Characters}
     \vspace{-1em}
 \end{figure}

 As shown in~\cref{fig:nLink-Unrecognizable-Domain-Random-Characters}, the attacker uses a link URL whose Registrable Domain is \texttt{dzmdk9psqr.com}.

\subsubsection{Implementation: Use of Fitting Keywords}
\label{sec:link-url-unrecognizable-domain-use-of-fitting-keywords}

This implementation (mentioned in literature such as~\cite{volkamer_user_2017}) exploits that an attacker-controlled domain can incorporate fitting keywords that make it appear plausible, despite being unrelated to the targeted service.

\begin{figure}[H]
     \centering
     \includegraphics[width=1\columnwidth]{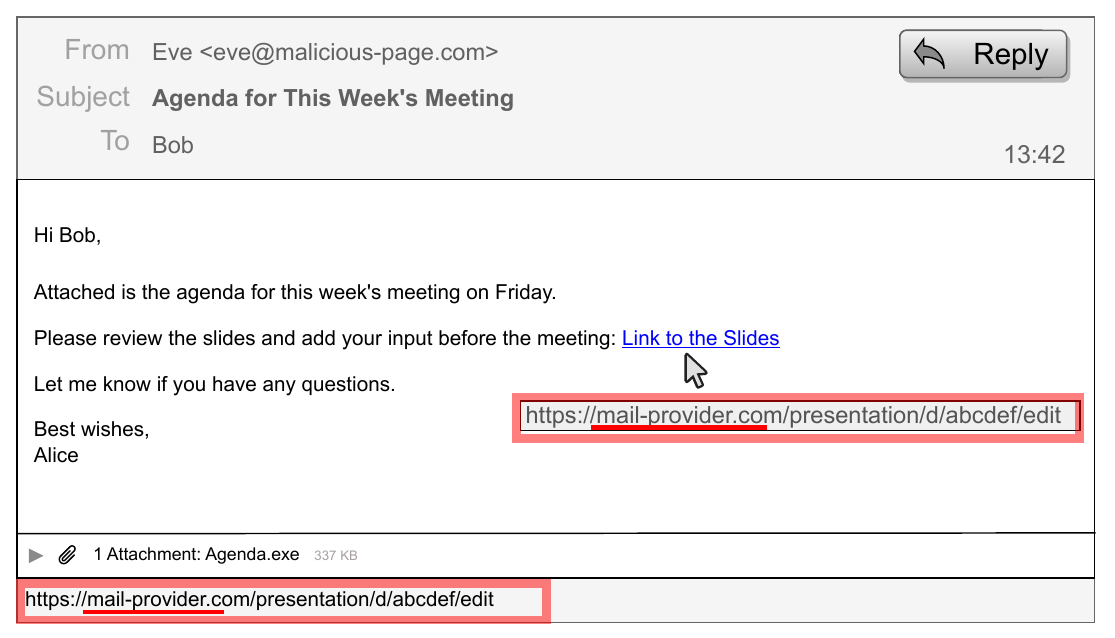}
    \caption{Example of Link URL Unrecognizable Domain (Implementation: Use of Fitting Keywords).}
     \label{fig:nLink-Unrecognizable-Domain-Use-of-Fitting-Keywords}
     \vspace{-1em}
 \end{figure}

 As shown in~\cref{fig:nLink-Unrecognizable-Domain-Use-of-Fitting-Keywords}, the attacker uses a domain such as \texttt{mail-provider.com} for \texttt{trusted-page.com}, which looks descriptive despite being unrelated to the real service.

\subsection{Link URL Registrable Domain Extension}
\label{sec:link-url-domain-extension}

In this deception technique, the attacker aims to mislead recipients about the destination of a link by using a URL whose Registrable Domain embeds the targeted brand name as part of an extended Registrable Domain.

To achieve this, the attacker registers a domain that contains the legitimate brand name plus additional terms (e.g., security-related words) so that the resulting link destination appears plausible at a glance. This exploits that recipients may have difficulty judging whether such an extended domain is actually owned by, or affiliated with, the legitimate brand when quickly inspecting the link security indicator.

A concrete implementation of this deception technique (mentioned in literature such as~\cite{sankhwar_novel_2018,suriya_integrated_2009,10.1186/s13673-020-00237-7,awasthi_generating_2021,azeez_identifying_2020-1,abedin_phishing_2020,bhardwaj_why_2020}) exploits that a domain can be registered that appends or prepends (i.e., add before) additional strings to a trusted brand name, and used as the Registrable Domain of the target URL.

\begin{figure}[H]
     \centering
     \includegraphics[width=1\columnwidth]{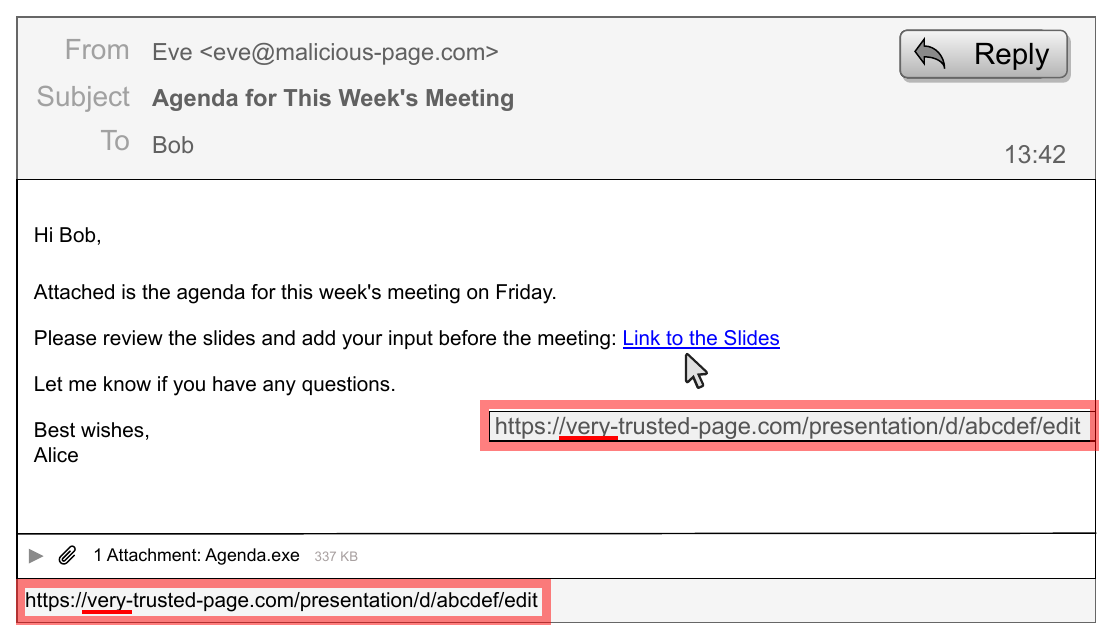}
    \caption{Example of Link URL Registrable Domain Extension.}
     \label{fig:nlink-url-domain-extension}
     \vspace{-1em}
\end{figure}

As shown in~\cref{fig:nlink-url-domain-extension}, the attacker replaces a revealing link destination with a URL whose Registrable Domain prepends the trusted brand domain \texttt{trusted-page.com} with the prefix \texttt{very-}, resulting in \texttt{very-trusted-page.com}. When the link is hovered, the link security indicator shows the plausible-looking domain \texttt{very-trusted-page.com}, although it is a different Registrable Domain controlled by the attacker.

\subsection{Link URL Subdomain}
\label{sec:link-url-subdomain}

In this deception technique, the attacker aims to mislead recipients about the destination of a link by placing a trusted-looking Registrable Domain string into the Subdomain of the target URL.

To achieve this, the attacker crafts the target URL such that a familiar brand or trusted domain appears in the Subdomain, while the actual Registrable Domain belongs to the attacker. This exploits that recipients may focus on the beginning of a domain-like string and may not fully inspect the Registrable Domain that determines the actual link destination.

A concrete implementation of this deception technique (mentioned in literature such as~\cite{t_n_business_2021,jampen_dont_2020,blum_lexical_2010,abedin_phishing_2020,swarnalatha_real-time_2021,bhardwaj_why_2020}) exploits that the attacker can place an arbitrary Subdomain under an attacker-controlled domain, allowing a targeted brand or trusted domain string to appear at the beginning of the link URL.

\begin{figure}[H]
    \centering
    \includegraphics[width=1\columnwidth]{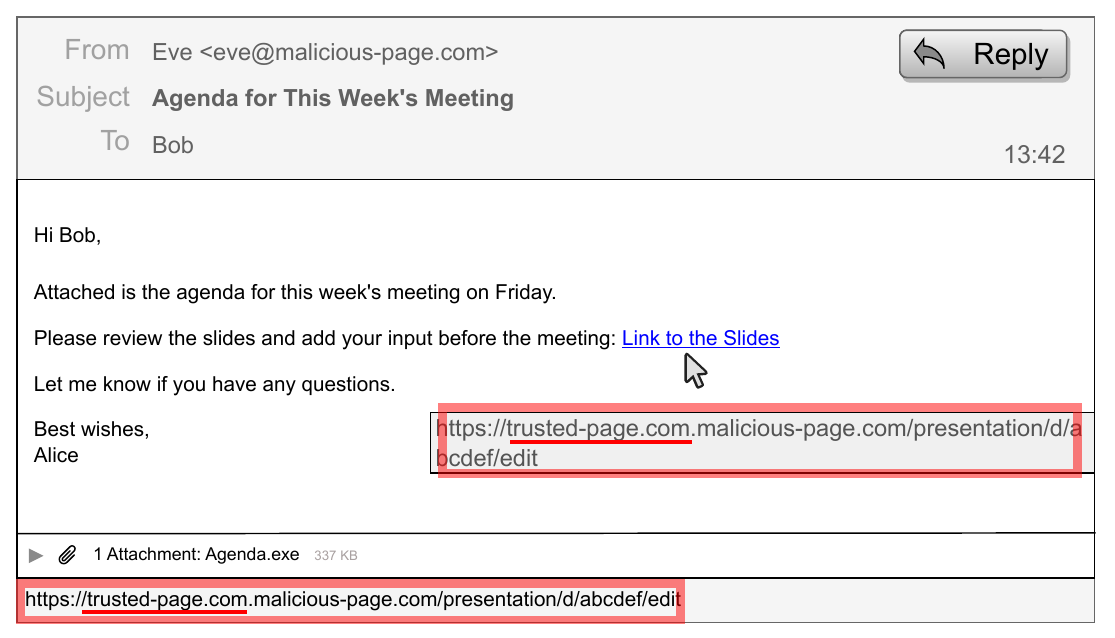}
    \caption{Example of Link URL Subdomain.}
    \label{fig:nlink-url-subdomain}
    \vspace{-1em}
\end{figure}

 As shown in~\cref{fig:nlink-url-subdomain}, the attacker can use a URL such as \texttt{https://trusted-page.com.malicious-page.com/[...]}, where the trusted domain string \texttt{trusted-page.com} appears as part of the Subdomain while the Registrable Domain \texttt{malicious-page.com} belongs to the attacker. This can make the URL appear legitimate at a glance, even though the actual link destination is different.

\subsection{Link URL Different eTLD}
\label{sec:link-url-different-etld}

In this deception technique, the attacker aims to mislead recipients about the destination of a link by using a Registrable Domain that contains the targeted brand name but uses a different eTLD.

To achieve this, the attacker registers a domain that matches the brand name while substituting the eTLD. This exploits that brands cannot practically register all available eTLD variants and that recipients may have difficulty judging which eTLDs are legitimately used by a given brand.

A concrete implementation of this deception technique (mentioned in literature such as~\cite{t_n_business_2021}) exploits that the targeted brand name can be registered under an alternative eTLD and used as the Registrable Domain of the target URL.

\begin{figure}[H]
    \centering
    \includegraphics[width=1\columnwidth]{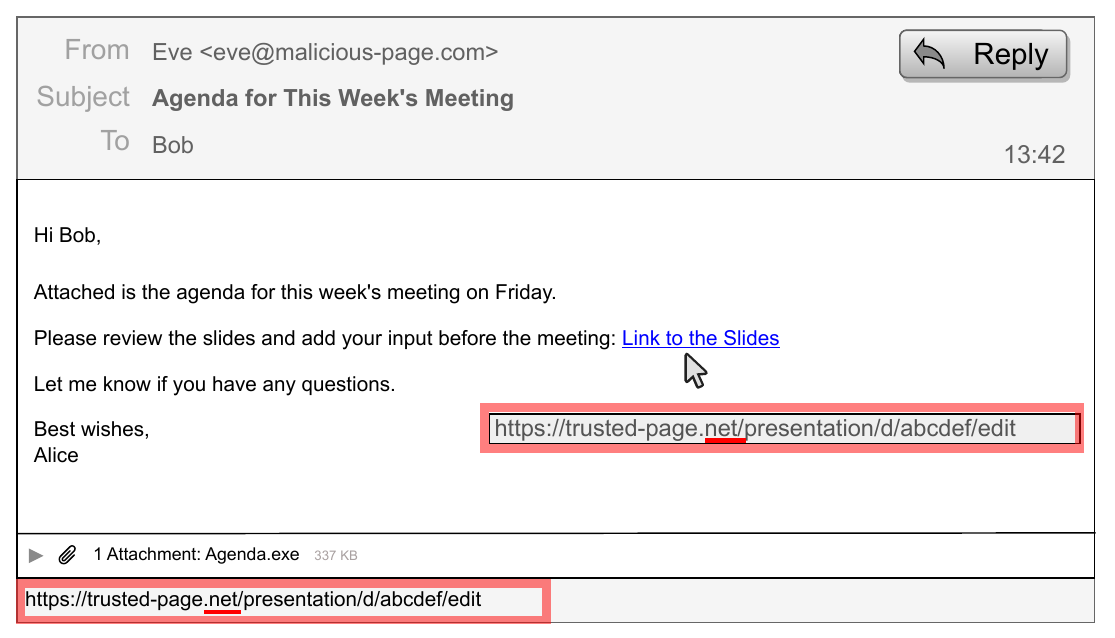}
    \caption{Example of Link URL Different eTLD.}
    \label{fig:nlink-url-different-etld}
    \vspace{-1em}
\end{figure}

 As shown in~\cref{fig:nlink-url-different-etld}, the attacker can register \texttt{trusted-page.net} instead of the legitimate domain \texttt{trusted-page.com} and use it as the Registrable Domain of the link URL.

\subsection{Link URL Mangle}
\label{sec:link-url-mangle}

In this deception technique, the attacker aims to mislead recipients about the destination of a link by using a slightly modified (mangled) variant of a legitimate domain in the target URL.

To achieve this, the attacker uses a domain whose spelling differs only subtly from the legitimate domain. This exploits that recipients may overlook small character differences when scanning the Registrable Domain of the target URL.

A concrete implementation of this deception technique (mentioned in literature such as~\cite{soni_phishing_2011,sankhwar_novel_2018,wang_strider_2006,agten_seven_2015,pearson_click_2017,spaulding_landscape_2016,suriya_integrated_2009,t_n_business_2021,pithawala_detecting_2021,jampen_dont_2020,pirocca_toolkit_2020,volkamer_user_2017,jakobsson_user_2016}) exploits that domains with minor variations of a trusted domain can be used as the Registrable Domain of the target URL. Common variants include intentional typos (e.g., \textit{mircosoft} instead of \textit{microsoft}), character substitutions with visually similar patterns (e.g., \textit{m} vs. \textit{rn} or \textit{p} vs. \textit{q}), or small edits such as doubling characters.

\begin{figure}[H]
    \centering
    \includegraphics[width=1\columnwidth]{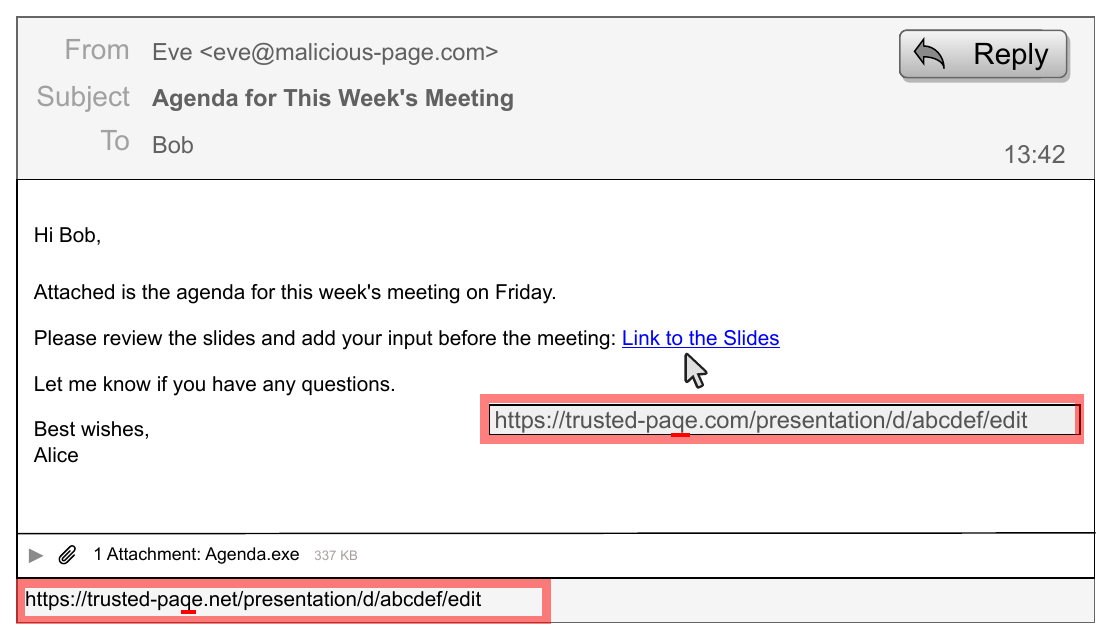}
    \caption{Example of Link URL Mangle.}
    \label{fig:nlink-url-mangle}
    \vspace{-1em}
\end{figure}

 As shown in~\cref{fig:nlink-url-mangle}, the attacker replaces a single character with a visually similar one (e.g., \texttt{g} with \texttt{q} in \texttt{trusted-page.com}, resulting in \texttt{trusted-paqe.com}) and uses the look-alike domain as the Registrable Domain of the target URL.

\subsection{Link URL Exceedingly Long}
\label{sec:link-url-exceedingly-long}

In this deception technique, the attacker aims to mislead recipients about the destination of a link by hiding the Registrable Domain of the target URL outside the visible area of the email client.

To achieve this, the attacker inflates the Subdomain so that the Registrable Domain of the target URL is pushed outside the visible area in the email client’s default view. At the same time, the visible prefix is crafted to resemble a trusted identity (e.g., by starting with a trusted-looking name or domain string). Since the non-visible Registrable Domain determines the actual website that is opened and is attacker-controlled, the attacker can freely choose the rest of the URL.

A concrete implementation of this deception technique (mentioned in literature such as~\cite{sankhwar_novel_2018,soni_phishing_2011,suriya_integrated_2009,zhu_dtof-ann_2020,priya_gravitational_2020,azeez_identifying_2020-1,abedin_phishing_2020,singh_phishing_2020,swarnalatha_real-time_2021,bhardwaj_why_2020}) exploits that the Subdomain can be padded with a long filler string and a trusted-looking prefix, so that the actual Registrable Domain is pushed outside the visible area.

\begin{figure}[H]
    \centering
    \includegraphics[width=1\columnwidth]{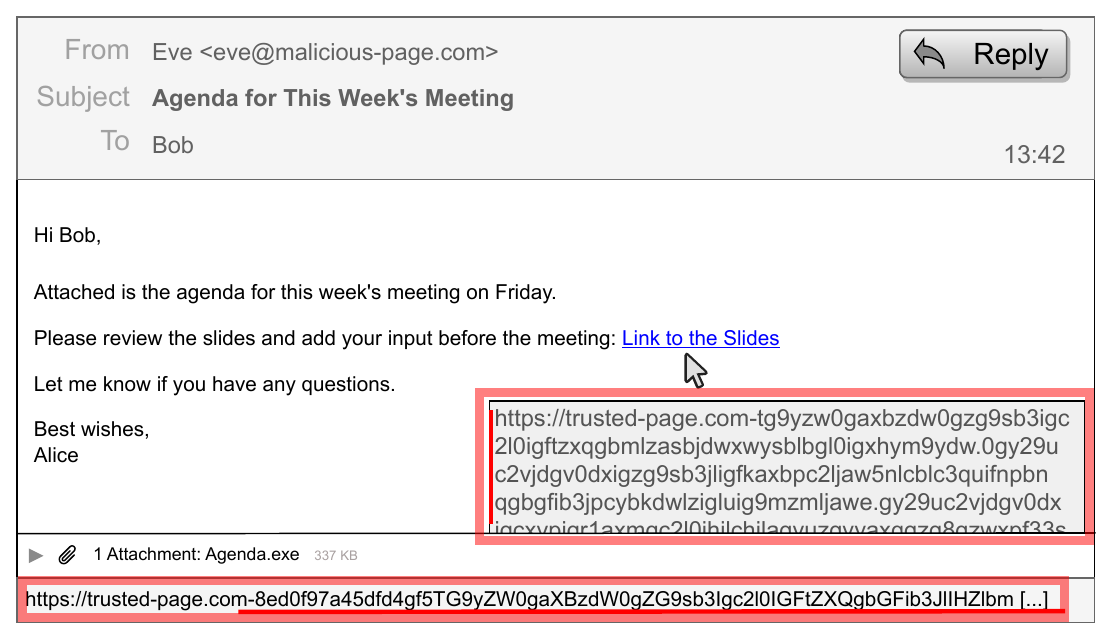}
    \caption{Example of Link URL Exceedingly Long.}
    \label{fig:nlink-url-exceedingly-long}
    \vspace{-1em}
\end{figure}

 As shown in~\cref{fig:nlink-url-exceedingly-long}, the attacker inserts a long filler string such as \texttt{-8ed0f97a45dfd4gf5[...]} and a trusted-looking prefix such as \texttt{trusted-page.com} into the Subdomain to push the actual Registrable Domain \texttt{malicious-page.com} outside the visible area of the email client.

\subsection{Link URL Tel Scheme}
\label{sec:link-tel-scheme}

In this deception technique, the attacker aims to trick recipients into initiating phone actions by using a link with the \texttt{tel:} Scheme instead of a web URL.

To achieve this, the attacker exploits two properties of \texttt{tel:} links. First, they do not contain a domain that recipients can inspect, which removes a common cue taught in security awareness training. Second, tapping a \texttt{tel:} link can trigger unexpected behavior (i.e., initiating a phone call or executing a phone code) rather than opening a browser, potentially without an explicit warning from the email client.

We describe the following implementations of this deception technique.

\subsubsection{Implementation: Phone Number}
\label{sec:link-tel-scheme-phone-number}

This implementation (identified during our examination) exploits that a \texttt{tel:} link can contain an attacker-chosen phone number, so that tapping the link opens the dialer with the number prefilled. The recipient may then unintentionally call the attacker (e.g., as part of a phone-based scam).

\begin{figure}[H]
    \centering
    \includegraphics[width=1\columnwidth]{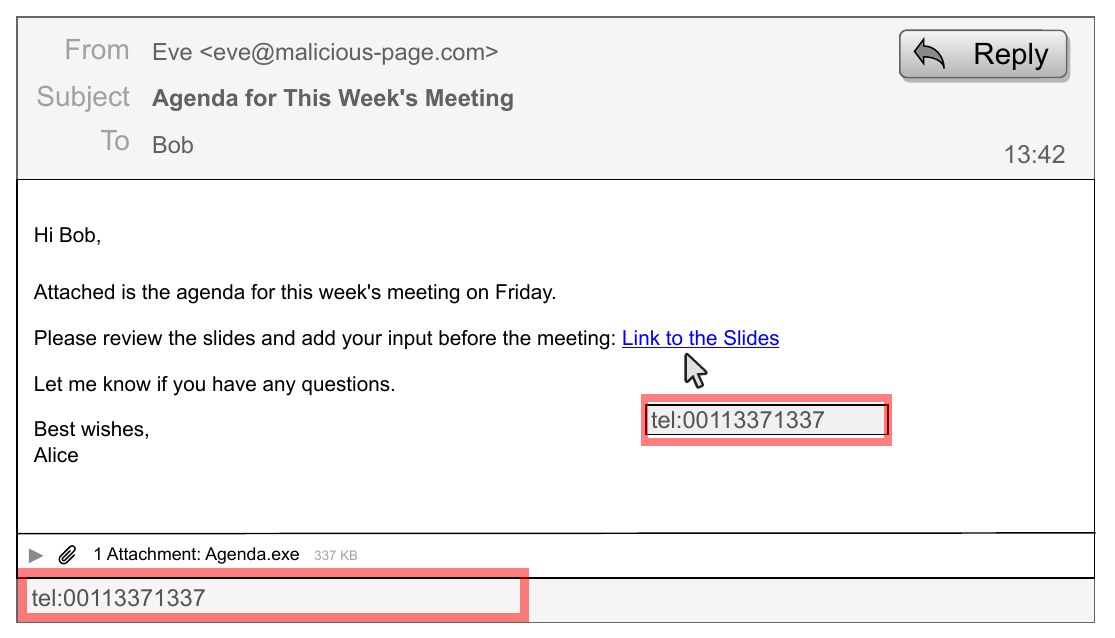}
    \caption{Example of Link URL Tel Scheme (Implementation: Phone Number).}
    \label{fig:nlink-tel-scheme-phone-number}
    \vspace{-1em}
\end{figure}

 As shown in~\cref{fig:nlink-tel-scheme-phone-number}, the attacker replaces a normal web link with a \texttt{tel:} link that contains an attacker-chosen phone number \texttt{00113371337}. When the recipient taps the link on a mobile device, the dialer is opened with the number prefilled.

\subsubsection{Implementation: USSD Code}
\label{sec:link-tel-scheme-ussd-code}

This implementation (identified during our examination) exploits that a USSD code can be placed in a \texttt{tel:} link. Depending on device and carrier behavior, tapping the link can send a supplementary-service request to the mobile network (e.g., to enable call forwarding). This can support follow-up attacks (e.g., intercepting calls used for account recovery or phone-based authentication).

\begin{figure}[H]
    \centering
    \includegraphics[width=1\columnwidth]{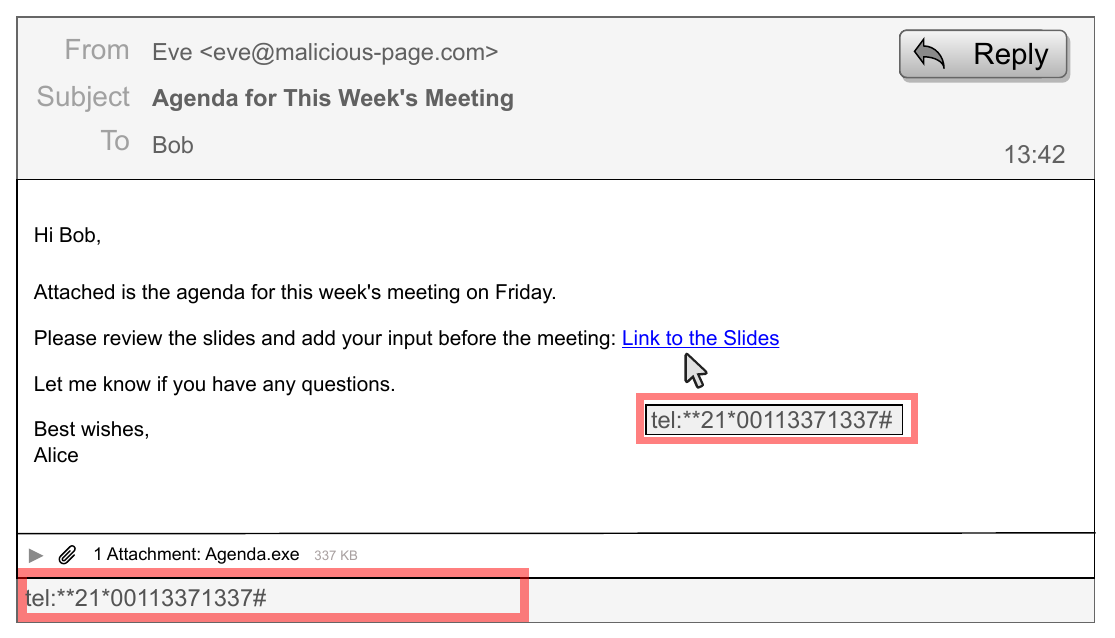}
    \caption{Example of Link URL Tel Scheme (Implementation: USSD Code).}
    \label{fig:nlink-tel-scheme-ussd-code}
    \vspace{-1em}
\end{figure}

 As shown in~\cref{fig:nlink-tel-scheme-ussd-code}, the attacker uses a \texttt{tel:} link that contains a USSD code such as \texttt{tel:**21*00113371337\#}. When the recipient taps the link on a mobile device and executes the code, call forwarding can be enabled so that incoming calls are redirected to the attacker-chosen number \texttt{00113371337}.

\subsubsection{Implementation: Premium-Rate Phone Number}
\label{sec:link-tel-scheme-premium-rate-number}

This implementation (identified during our examination) exploits that a \texttt{tel:} link can contain a premium-rate phone number, so that tapping the link opens the phone dialer with the number prefilled. This can lead to high charges if the call is placed, from which the attacker benefits.

\begin{figure}[H]
    \centering
    \includegraphics[width=1\columnwidth]{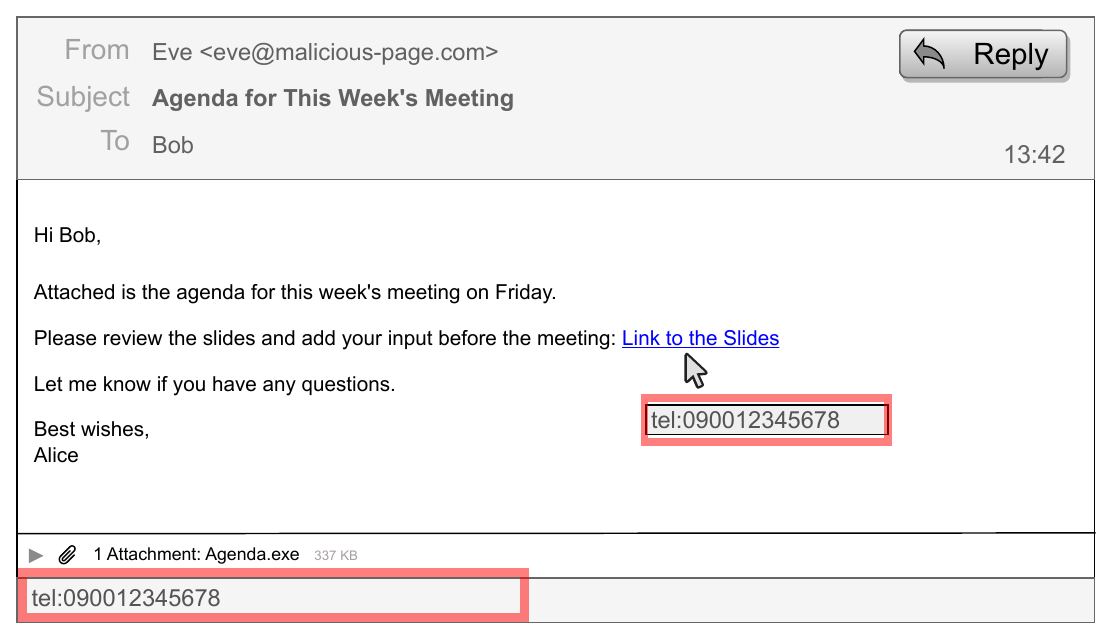}
    \caption{Example of Link URL Tel Scheme (Implementation: Premium-Rate Phone Number).}
    \label{fig:nlink-tel-scheme-premium-rate-number}
    \vspace{-1em}
\end{figure}

 As shown in~\cref{fig:nlink-tel-scheme-premium-rate-number}, the attacker uses a \texttt{tel:} link that contains a premium-rate number (e.g., \texttt{090012345678}). When the recipient taps the link on a mobile device and places the call, charges may be incurred.

\subsection{Link URL Data Scheme}
\label{sec:link-url-data-scheme}

In this deception technique, the attacker aims to mislead recipients by using a \texttt{data:} Scheme URL that embeds a phishing page directly in the link, rather than pointing to a website on a Registrable Domain.

To achieve this, the attacker crafts a \texttt{data:} Scheme URL. This exploits that \texttt{data:} Scheme URLs do not contain a Registrable Domain that recipients can verify, and that domain-based defenses (e.g., blocking or takedown of a hosting domain) is not effective because the content is carried within the link itself.

We describe the following implementations of this deception technique.

\subsubsection{Implementation: Embedded HTML Payload}
\label{sec:link-url-data-scheme-embedded-html-payload}

This implementation (identified during our examination) exploits that a \texttt{data:} Scheme URL can carry an HTML payload encoded as \texttt{base64}, which the browser renders as a website when the link is opened.

\begin{figure}[H]
    \centering
    \includegraphics[width=1\columnwidth]{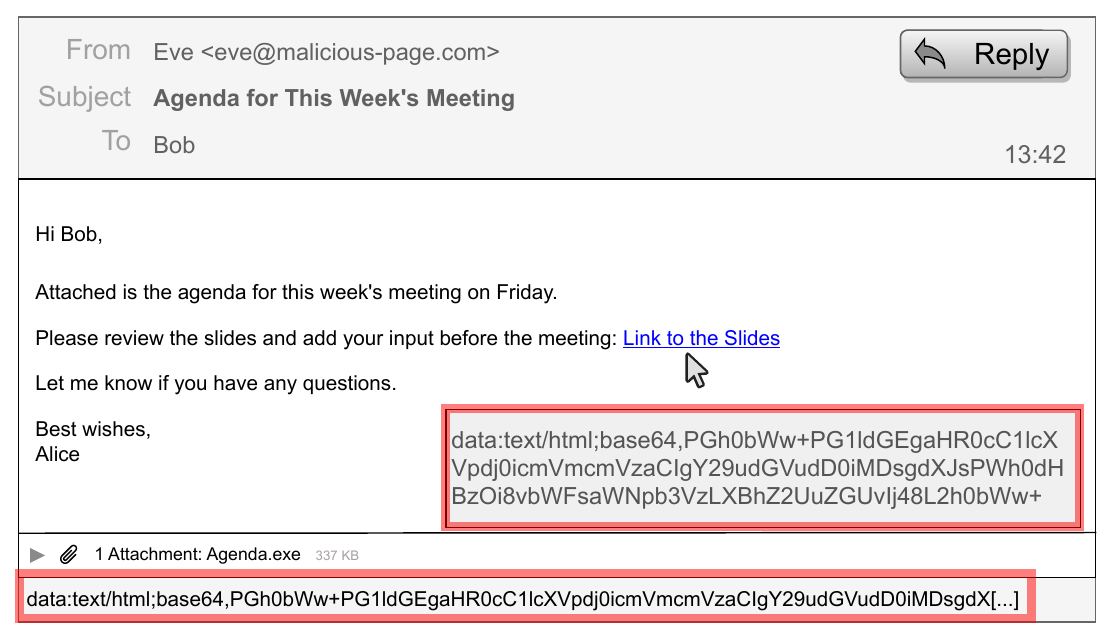}
    \caption{Example of Link URL Data Scheme (Implementation: Embedded HTML Payload).}
    \label{fig:nlink-url-data-scheme-embedded-html-payload}
    \vspace{-1em}
\end{figure}

 As shown in~\cref{fig:nlink-url-data-scheme-embedded-html-payload}, the attacker replaces a regular \texttt{https://} URL with \texttt{data:text/html;base64,\allowbreak{}PGh0bWw+P[...]}, where \texttt{PGh0bWw+P[...]} is a \texttt{base64}-encoded HTML document. When the link is clicked/tapped, the browser decodes the embedded HTML and renders it as a website.

\subsubsection{Implementation: Trusted-looking URL String}
\label{sec:link-url-data-scheme-trusted-looking-url-in-data-scheme}

This implementation (identified during our examination) exploits that a trusted-looking Registrable Domain or URL string can be placed early in the visible part of a \texttt{data:} URL (i.e., before the encoded payload), making the link appear more familiar at a glance, while the rendered page comes entirely from the embedded payload.

\begin{figure}[H]
    \centering
    \includegraphics[width=1\columnwidth]{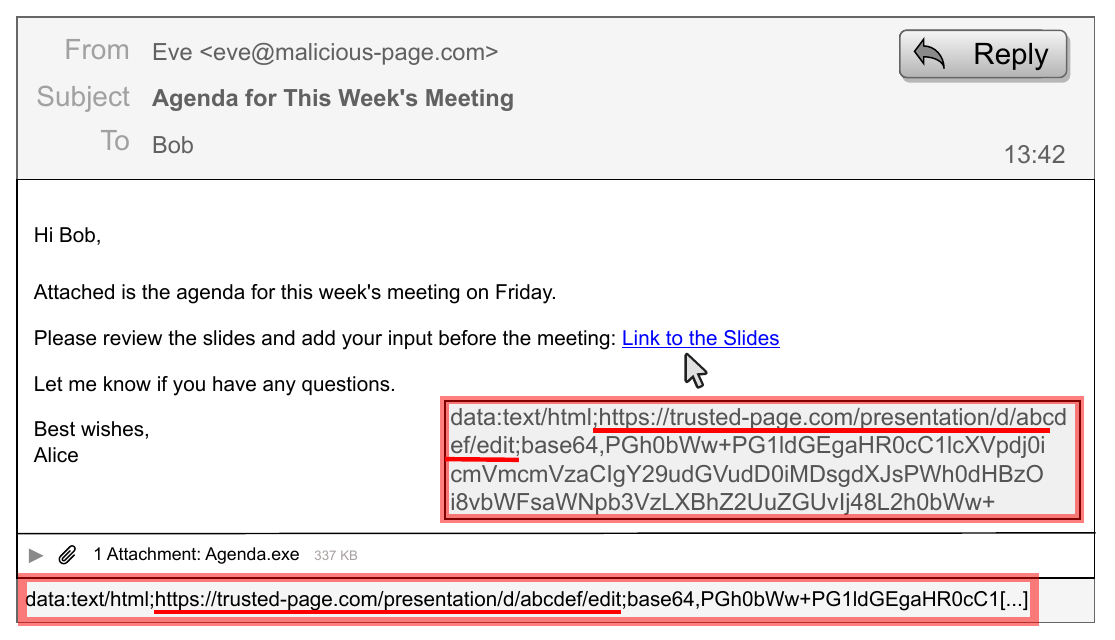}
    \caption{Example of Link URL Data Scheme (Implementation: Trusted-looking URL String in Data Scheme).}
    \label{fig:nlink-url-data-scheme-trusted-looking-url-in-data-scheme}
    \vspace{-1em}
\end{figure}

 As shown in~\cref{fig:nlink-url-data-scheme-trusted-looking-url-in-data-scheme}, the link begins with a trusted-looking Registrable Domain string (here: \texttt{trusted-page.com}) but does not point to a Registrable Domain destination. Instead, the link carries the website content in encoded form and renders it when opened.

\subsection{Link URL Mailto Scheme}
\label{sec:link-url-mailto-scheme}

In this deception technique, the attacker aims to obtain sensitive data by using a link with the \textit{mailto} Scheme that prepares an outgoing email to the attacker including a local file attachment.

To achieve this, the attacker embeds a \textit{mailto} URL that opens the recipient’s email client and prepares an outgoing email to the attacker including a local file as an attachment. This exploits that recipients may not notice that the link prepares an email with an attachment and may send it, thereby disclosing the attached file to the attacker.

A concrete implementation of this deception technique (mentioned in literature such as~\cite{muller_mailto_2020}) exploits that \textit{mailto} URLs can predefine email fields such as recipient, subject, and body, and may also preselect a local file attachment if supported. Once the link is clicked, the email client prepares an outgoing email to the attacker accordingly. The deception succeeds only if the recipient sends the prepared email.

\begin{figure}[H]
    \centering
    \includegraphics[width=1\columnwidth]{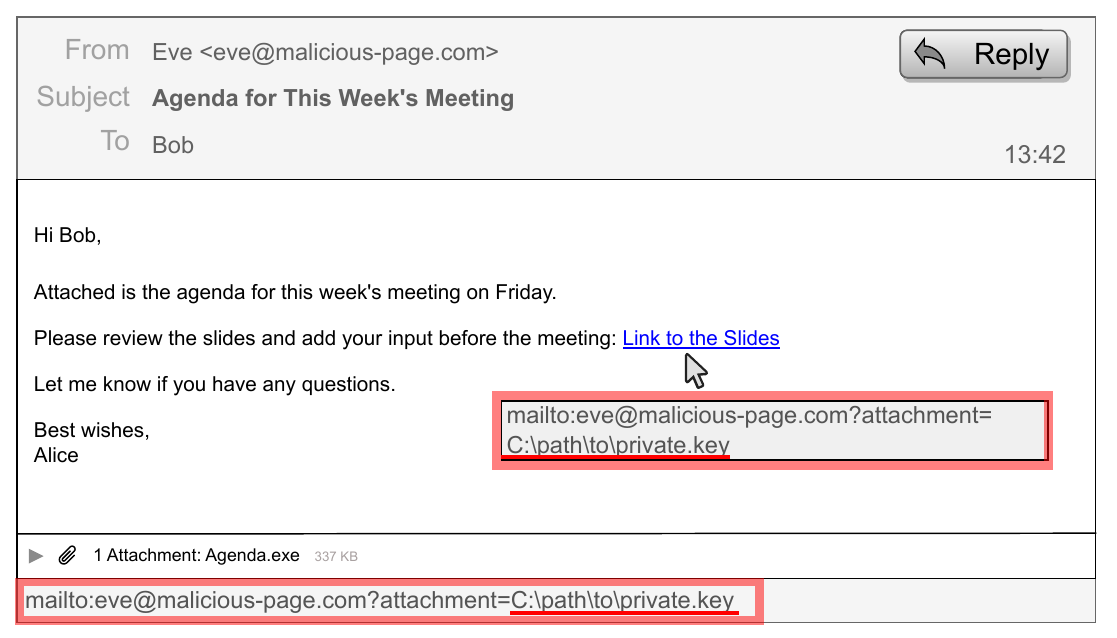}
    \caption{Example of Link URL Mailto Scheme.}
    \label{fig:nlink-url-mailto-scheme}
    \vspace{-1em}
\end{figure}

 As shown in~\cref{fig:nlink-url-mailto-scheme}, the attacker embeds a \textit{mailto} link that pre-fills an email addressed to the attacker, with the sensitive file \texttt{C:\textbackslash{}path\textbackslash{}to\textbackslash{}private.key}. When the recipient clicks it, the email client prepares an outgoing email to the attacker including the file.

\subsection{Link URL Custom Scheme and Deep Links}
\label{sec:link-url-custom-scheme-and-deep-links}

In this deception technique, the attacker aims to mislead recipients about the destination of a link by using a custom Scheme URL or deep link that opens an installed application instead of a web page.

To achieve this, the attacker embeds a link whose Scheme is registered by an application (e.g., \texttt{msteams://} or \texttt{zoommtg://}). This exploits that such links can look similar to legitimate meeting links, while the resulting behavior is handled inside the app and may be unclear to recipients.

A concrete implementation of this deception technique (identified during our examination) exploits that applications can register custom URL Schemes or deep links that are resolved by the operating system and opened directly by the corresponding app. By embedding such a link in an email, the attacker can cause the recipient to launch the app and follow an attacker-chosen deep link rather than a normal \texttt{https://} destination.

\begin{figure}[H]
    \centering
    \includegraphics[width=1\columnwidth]{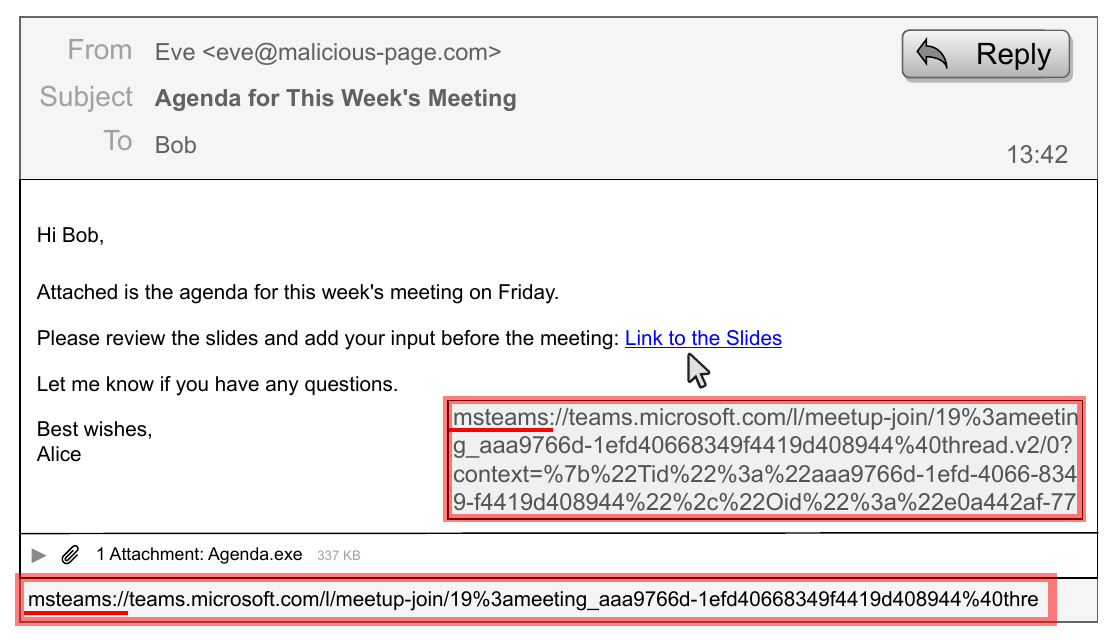}
    \caption{Example of Link URL Custom Scheme and Deep Links.}
    \label{fig:nlink-url-custom-scheme-and-deep-links}
    \vspace{-1em}
\end{figure}

 As shown in~\cref{fig:nlink-url-custom-scheme-and-deep-links}, the attacker uses a custom Scheme link (here: \texttt{msteams://}) that appears like a regular meeting link in the email. When the recipient clicks it, the installed application is opened and processes the deep link, which can lead to recipient actions outside the usual browser-based URL context.

\subsection{Link HTML QR Code}
\label{sec:link-qr-code}

In this deception technique, the attacker aims to mislead recipients about the destination of a link by embedding a QR code instead of a regular link.

To achieve this, the attacker encodes an attacker-chosen URL into a QR code image and places it in the email body. This has several properties that are relevant from a security perspective. First, since the URL is embedded in an image rather than present as a regular link, link-focused email security controls (e.g., spam or phishing filters) may not detect it. Second, since scanning a QR code typically involves a mobile device, the resulting connection may not be subject to network-level controls applied to corporate workstations. Third, it shifts the verification of the link destination from the email client to the QR code scanner. As a result, recipients may not inspect the destination at all, or may inspect it under less favorable conditions -- depending on whether the QR code scanner prominently displays the decoded URL before opening it and whether it provides additional warnings or checks.

A concrete implementation of this deception technique (identified during our examination) exploits that a QR code can be embedded as an inline image in the email body (e.g., via an HTML \texttt{<img>} element). When the recipient scans the QR code, the QR code scanner decodes the embedded URL and, after user confirmation, opens it in the phone's web browser.

\begin{figure}[H]
    \centering
    \includegraphics[width=1\columnwidth]{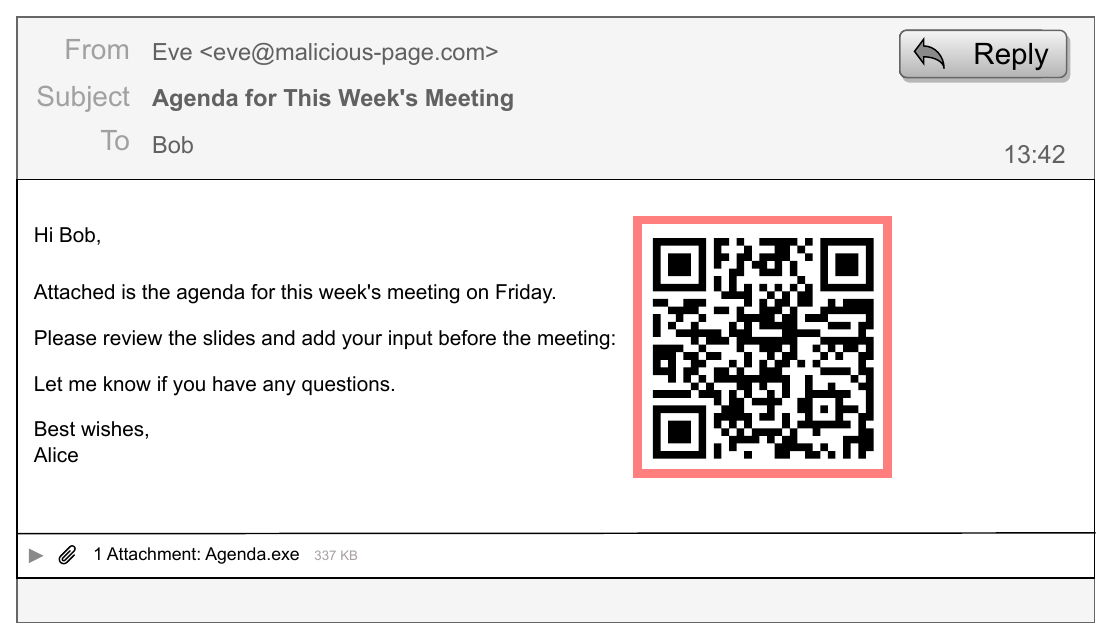}
    \caption{Example of Link HTML QR Code.}
    \label{fig:nlink-qr-code}
    \vspace{-1em}
\end{figure}

 As shown in~\cref{fig:nlink-qr-code}, the attacker replaces a normal link with a QR code that encodes \texttt{https://malicious-page.com/[...]}, and prompts the recipient to scan it to access the referenced document.

\section{Attachment-related Deception Techniques}
\label{sec:attachment-related-deception-techniques}

Security awareness training commonly instructs recipients to inspect the attachment security indicator -- in particular the file extension -- before opening an attachment~\cite{berensBetterTogheter}. Since the filename is set by the sender and can be freely chosen, an attacker can directly control how this indicator is presented. The attacker's goal, however, is to induce the recipient to open a file that executes attacker-controlled code, which constrains the choices: the actual file type must remain executable regardless of how the filename is manipulated. The following deception techniques demonstrate how attackers manipulate this indicator while satisfying this constraint.

\subsection{Attachment \DoubleFileExtension{}}
\label{sec:attachment-double-file-extension}

In this deception technique, the attacker aims to disguise a dangerous executable attachment by giving it a filename that suggests a benign file type.

To achieve this, the attacker uses multiple file extensions so that a low-risk extension (e.g., \texttt{.pdf}) appears before the actual executable extension (e.g., \texttt{.exe}). This exploits that recipients may judge the attachment based on the displayed filename or the first (visible) extension, and that some email clients do not prominently show the full extension of attachments.

A concrete implementation of this deception technique (mentioned in literature such as~\cite{alazab_spam_2016}) attaches an executable file whose filename includes an additional benign-looking extension before the executable one (e.g., \texttt{cv.pdf.exe}). While the attachment may appear like a document at a glance, the file type is still determined by the final executable extension. Note that the number of extensions is arbitrary, i.e., the attacker can use more than two (e.g., \texttt{cv.pdf.no.exe}).

\begin{figure}[H]
    \centering
    \includegraphics[width=1\columnwidth]{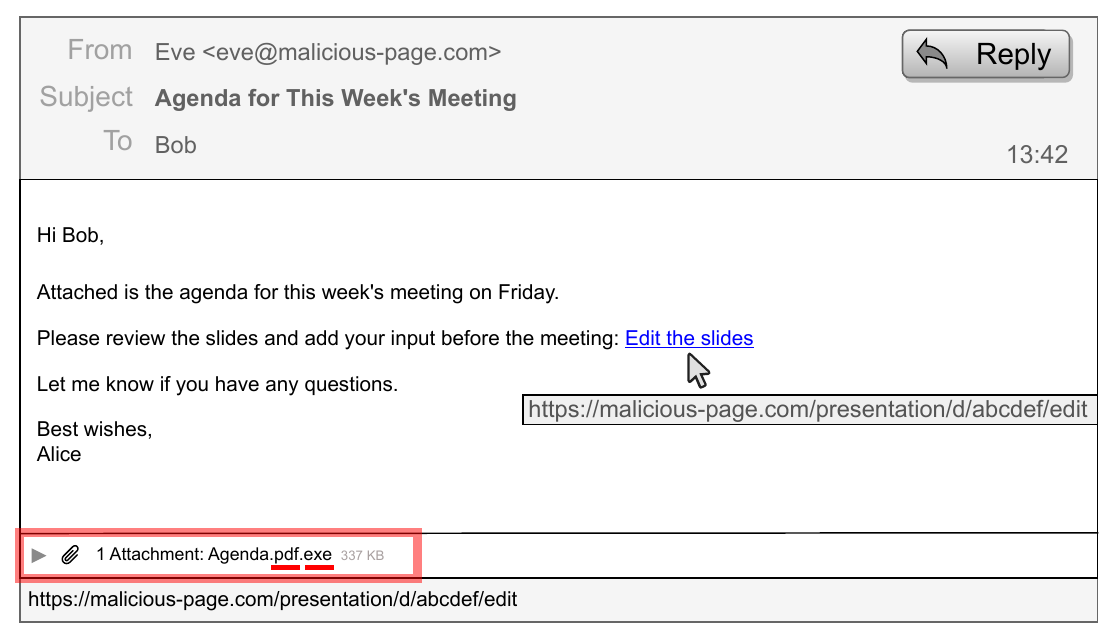}
    \caption{Example of Attachment Double File Extension.}
    \label{fig:nattachment-double-file-extension}
    \vspace{-1em}
\end{figure}

 As shown in~\cref{fig:nattachment-double-file-extension}, the attacker attaches an executable file named \texttt{Agenda.pdf.exe}. 

\subsection{Attachment Unknown File Extension}
\label{sec:attachment-unknown-file-extension}

In this deception technique, the attacker aims to disguise a dangerous executable attachment by using an uncommon or misleading file extension.

To achieve this, the attacker attaches an executable file whose file extension is unfamiliar or looks harmless in the attachment list. This exploits that many recipients recognize common dangerous extensions (e.g., \texttt{.exe}), but may not recognize other executable file types. If the recipient opens the attachment, the file type typically determines how it is handled by the operating system (e.g., execution), which can result in malware execution.

We describe the following implementations of this deception technique.

\subsubsection{Implementation: Uncommon Executable File Extension}
\label{sec:attachment-unknown-file-extension-uncommon-executable-file-extension}

This implementation (mentioned in literature such as~\cite{balan_detecting_2018,alazab_spam_2016}) exploits that a file extension that is executable on the recipient system can be uncommon or unknown to many recipients (e.g., \texttt{.jar}), making the attachment appear less suspicious than a typical \texttt{.exe} file.

\begin{figure}[H]
    \centering
    \includegraphics[width=1\columnwidth]{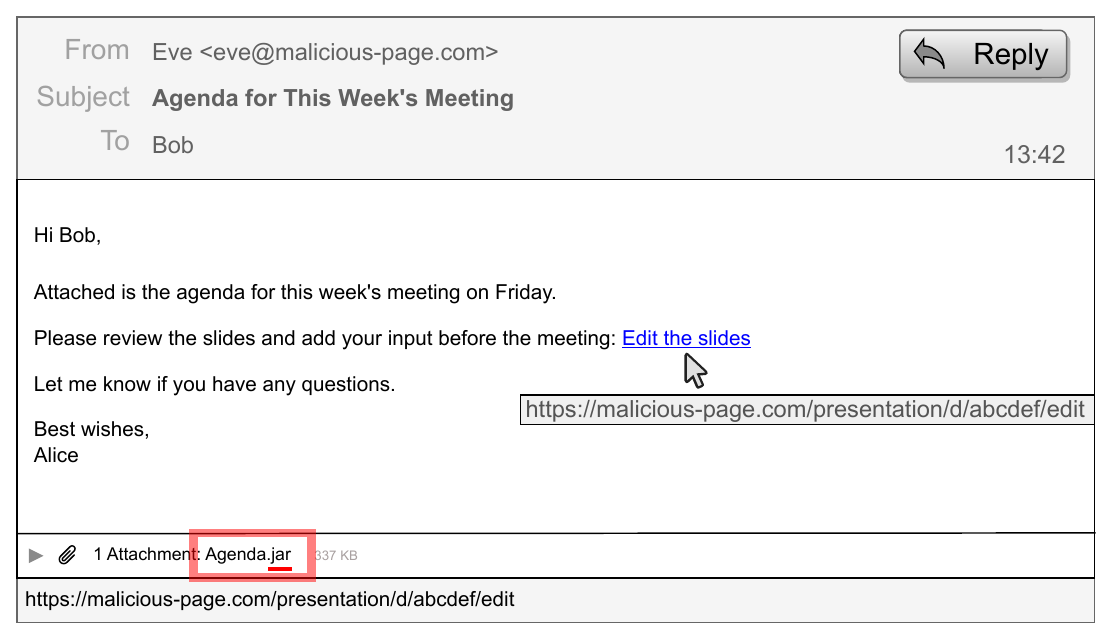}
    \caption{Example of Attachment \UnknownFileExtension{.} (Implementation: Uncommon Executable File Extension).}
    \label{fig:nattachment-unknown-file-extension-uncommon-executable-file-extension}
    \vspace{-1em}
\end{figure}

 As shown in~\cref{fig:nattachment-unknown-file-extension-uncommon-executable-file-extension}, the attacker attaches a file named \texttt{invoice.jar} so that the attachment appears less suspicious than a typical \texttt{.exe} file, although opening it can still execute attacker-controlled code (e.g., via an installed Java runtime).

\subsubsection{Implementation: Misleading File Extension}
\label{sec:attachment-unknown-file-extension-misleading-file-extension}

This implementation (mentioned in literature such as~\cite{balan_detecting_2018,alazab_spam_2016}) exploits that an executable file extension can be chosen to resemble an eTLD rather than an executable file type, so that the filename looks like a familiar domain-like string.

\begin{figure}[H]
    \centering
    \includegraphics[width=1\columnwidth]{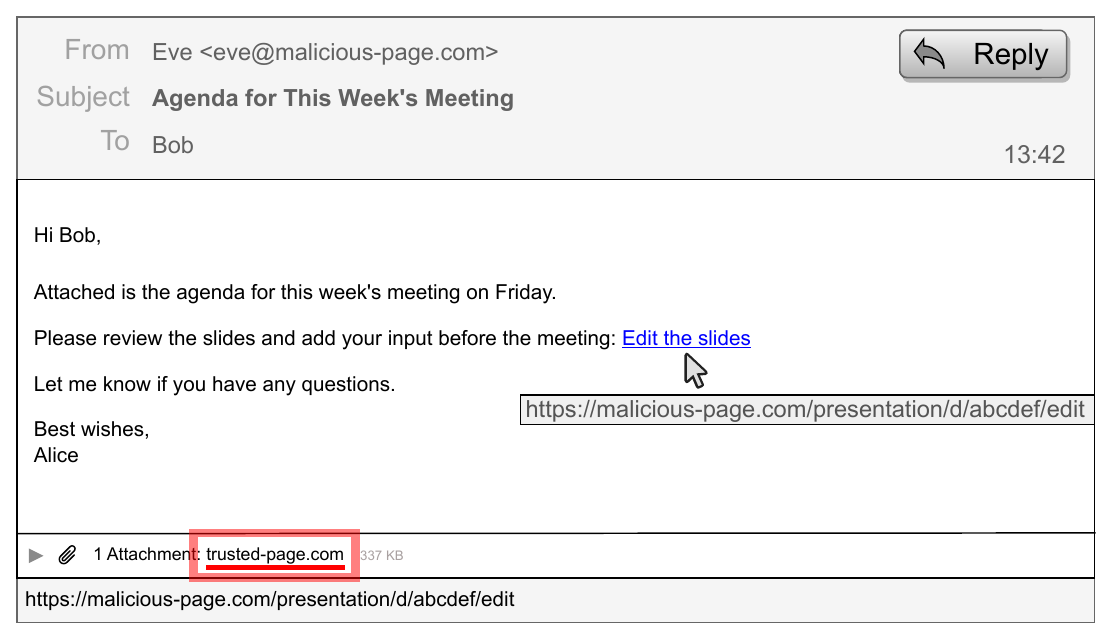}
    \caption{Example of Attachment \UnknownFileExtension{.} (Implementation: Misleading File Extension).}
    \label{fig:nattachment-unknown-file-extension-misleading-file-extension}
    \vspace{-1em}
\end{figure}

 As shown in~\cref{fig:nattachment-unknown-file-extension-misleading-file-extension}, the attacker attaches a file named \texttt{trusted-page.com}. The filename looks like the trusted website \texttt{trusted-page.com}, but the trailing \texttt{.com} is the file extension, which is treated as an executable file type on Windows.

\subsubsection{Implementation: Mangled File Extension}
\label{sec:attachment-unknown-file-extension-mangled-file-extension}

    This implementation (adapted from link-related deception techniques~\cref{sec:link-url-mangle}) exploits that a file extension can be chosen to look similar to a benign extension at a glance (e.g., \texttt{.pif} instead of \texttt{.pdf}), and that recipients may overlook small character differences when scanning attachment filenames.

\begin{figure}[H]
    \centering
    \includegraphics[width=1\columnwidth]{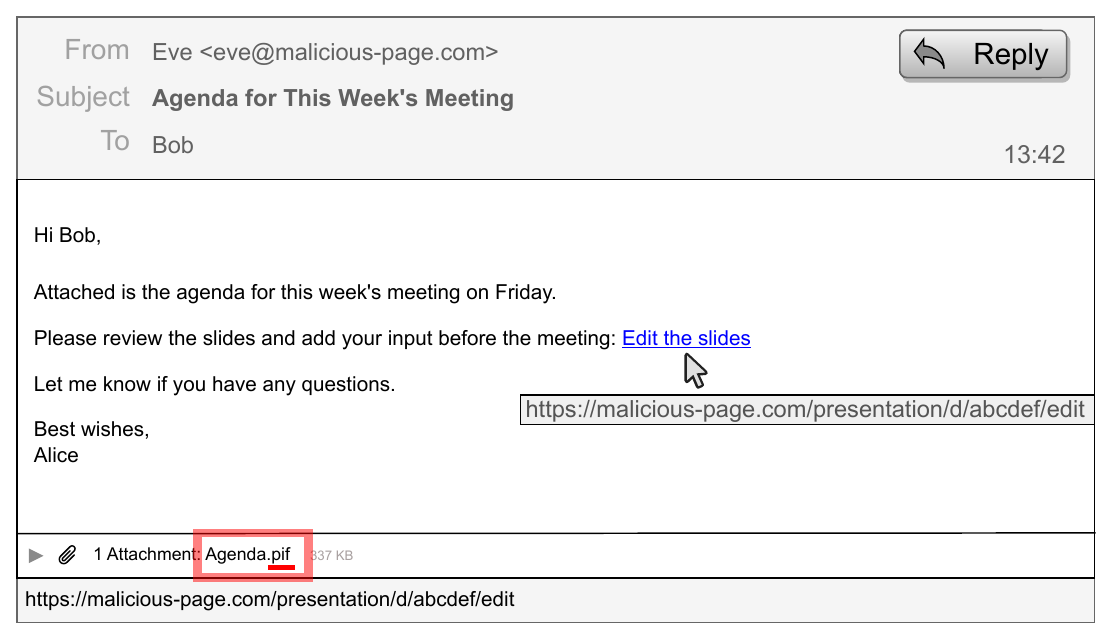}
    \caption{Example of Attachment \UnknownFileExtension{.} (Implementation: Mangled File Extension).}
    \label{fig:nattachment-unknown-file-extension-mangled-file-extension}
    \vspace{-1em}
\end{figure}

 As shown in~\cref{fig:nattachment-unknown-file-extension-mangled-file-extension}, the attacker attaches a file \texttt{Agenda.pif} that visually resembles the common pdf document file \texttt{Agenda.pdf}.

\subsection{Attachment Exceedingly Long}
\label{sec:attachment-exceedingly-long}

In this deception technique, the attacker aims to conceal the true file type of a malicious attachment by using an exceedingly long filename so that the actual file extension is pushed outside the visible area of the email client.

To achieve this, the attacker assigns a filename that exceeds the attachment display space of the email client. This exploits that recipients often rely on the visible portion of the filename when judging the attachment type. If the filename is truncated, the true (potentially dangerous) file extension may not be visible at all. 

A concrete implementation of this deception technique (mentioned in literature such as~\cite{alazab_spam_2016}) exploits that many email clients truncate long attachment filenames. The attacker therefore inserts a long filler segment between a benign-looking filename prefix and the real extension. As a result, the attachment can appear to be a harmless document in the visible UI, while the underlying file type is still determined by the (hidden) trailing extension and can be executed when opened.

\begin{figure}[H]
    \centering
    \includegraphics[width=1\columnwidth]{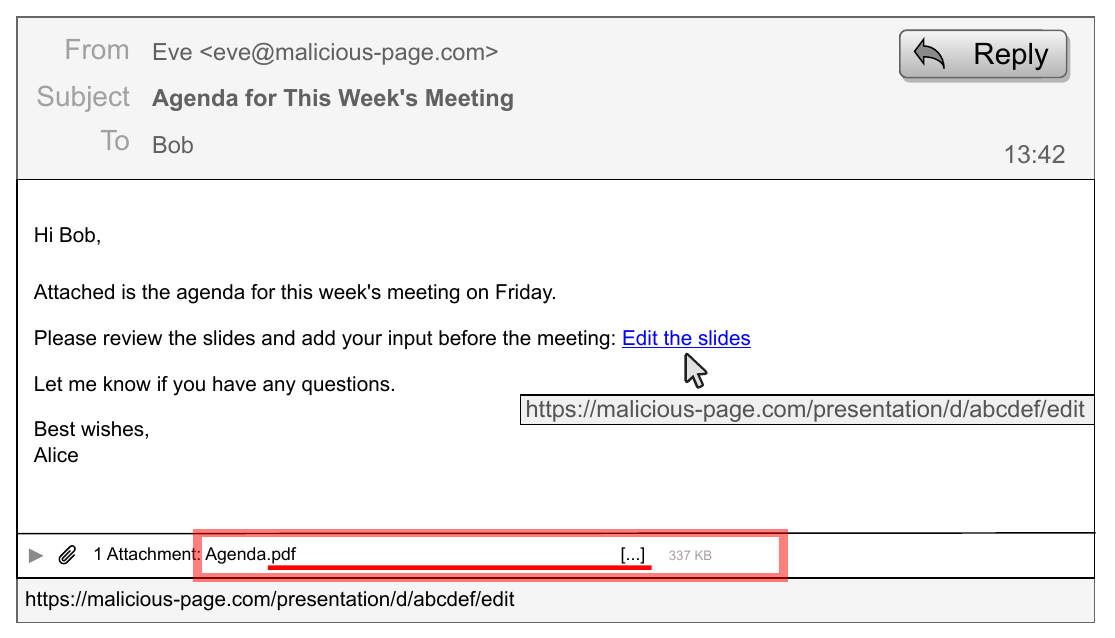}
    \caption{Example of Attachment Exceedingly Long.}
    \label{fig:nattachment-exceedingly-long}
    \vspace{-1em}
\end{figure}

 As shown in~\cref{fig:nattachment-exceedingly-long}, the attacker attaches an executable whose filename contains a visible benign-looking suffix (e.g., \texttt{Agenda.pdf}) and an exceedingly long padding segment (with whitespaces), while the real extension \texttt{.exe} is placed at the end of the filename. 

\subsection{Attachment Right-to-Left Override}
\label{sec:attachment-right-to-left-override}

In this deception technique, the attacker aims to disguise a dangerous attachment by manipulating the text direction in the attachment file name so that the displayed file extension appears benign.

To achieve this, the attacker inserts a Unicode right-to-left override control character into the attachment file name so that parts of the name are rendered in reverse order. This exploits that recipients typically judge attachment risk based on the displayed file extension and that many email clients do not make bidirectional control characters visible~\cite{Veit2024}.

A concrete implementation of this deception technique (mentioned in literature such as~\cite{alazab_spam_2016,marczak_when_2014}) exploits the Unicode \textit{Right-to-Left Override} (RLO, U+202E). By placing the RLO character before a chosen character sequence, the email client may render the following characters right-to-left, which can make an executable attachment appear as a harmless document type.

\begin{figure}[H]
    \centering
    \includegraphics[width=1\columnwidth]{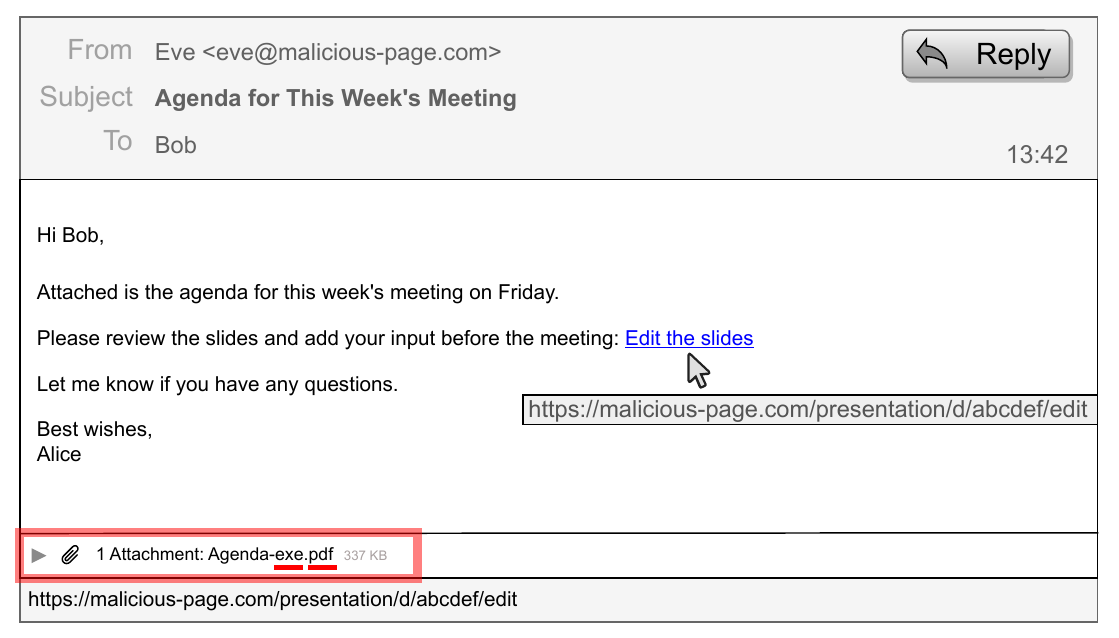}
    \caption{Example of Attachment Right-to-Left Override.}
    \label{fig:nattachment-right-to-left-override}
    \vspace{-1em}
\end{figure}

 As shown in~\cref{fig:nattachment-right-to-left-override}, the attacker attaches a file named \texttt{Agenda-[RLO]fdp.exe}. In the attachment list, the email client displays the file name as \texttt{Agenda-exe.pdf}, making the attachment appear like a PDF document even though the actual file extension remains \texttt{.exe}.

\section{Other Deception Techniques}
\label{sec:other-deception-techniques}

The following deception techniques do not target a specific security indicator but instead exploit properties of the email rendering environment or the recipient's device.

\subsection{Visual Integrity}
\label{sec:visual-integrity}

In this deception technique, the attacker aims to deceive recipients by making attacker-controlled content inside the email look like user interface elements of the email client.

To achieve this, the attacker embeds visual elements in the email body (e.g., images and HTML/CSS) that closely mimic the look of email-client UI components (e.g., dialogs or scrollbars) and aligns interactive elements (e.g., form fields or links) with these visuals. This exploits that recipients may not reliably distinguish between content that is part of the email body and elements that are part of the email client itself. 

We describe the following implementations of this deception technique.

\subsubsection{Implementation: Fake Password Prompt by the Email Client}
\label{sec:visual-integrity-fake-password-prompt}

This implementation (identified during our examination) exploits that the email body can render arbitrary HTML/CSS, including visual imitations of email-client UI elements. The attacker embeds a visual imitation of an email-client password prompt into the email body and places username/password input fields beneath or inside the prompt. The prompt is typically framed by contextual text that makes the dialog plausible (e.g., a claimed connection loss and a request to re-enter credentials). If the recipient enters credentials, the attacker can exfiltrate them via form submission, exploiting the general capability of HTML emails to render form elements.

\begin{figure}[H]
    \centering
    \includegraphics[width=1\columnwidth]{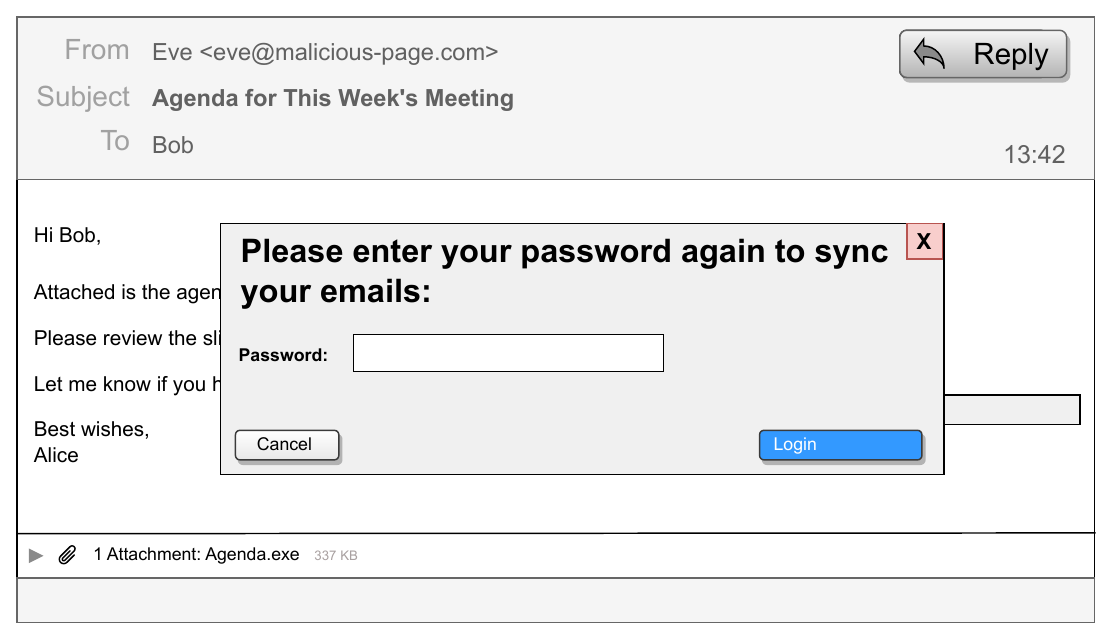}
    \caption{Example of Visual Integrity (Implementation: Fake Password Prompt by the Email Client).}
    \label{fig:nvisual-integrity-fake-password-prompt}
    \vspace{-1em}
\end{figure}

 As shown in~\cref{fig:nvisual-integrity-fake-password-prompt}, the email displays a box that resembles a genuine password prompt of the email client. The attacker places password input fields in the email body so that the prompt appears to request the recipient’s email credentials.

\subsubsection{Implementation: Clickjacking Using Fake User Interface Elements}
\label{sec:visual-integrity-clickjacking-fake-ui-elements}

This implementation (identified during our examination) exploits that interactive email-body elements (e.g., links) can be visually disguised as UI elements of the email client. The attacker embeds a visual imitation of a UI element (e.g., a scrollbar) into the email body and makes this imitation clickable (i.e., it is itself a link). The email text is crafted to suggest that interacting with the UI element is necessary (e.g., scroll down to read the full message). When the recipient attempts to interact with the fake UI element, the click is captured by the attacker-controlled link and triggers navigation to an attacker-chosen destination.

\begin{figure}[H]
    \centering
    \includegraphics[width=1\columnwidth]{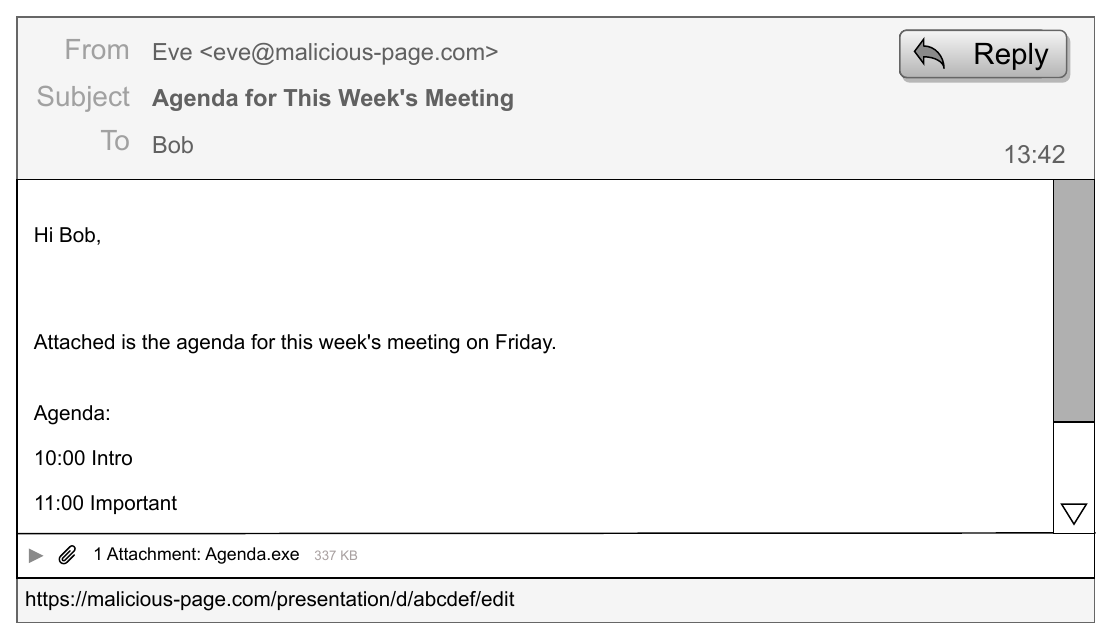}
    \caption{Example of Visual Integrity (Implementation: Clickjacking Using Fake User Interface Elements).}
    \label{fig:nvisual-integrity-clickjacking-fake-ui-elements}
    \vspace{-1em}
\end{figure}

 As shown in~\cref{fig:nvisual-integrity-clickjacking-fake-ui-elements}, the email contains a fake scrollbar rendered as an image and positioned so that clicking it feels like attempting to scroll. When the recipient clicks the fake scrollbar, the click activates an embedded link and opens the attacker-controlled destination.

\subsection{Local Resource Inclusion}
\label{sec:local-resource-inclusion}

In this deception technique, the attacker aims to create the impression of having access to the recipient’s local files (e.g., private photos) or to induce disclosure of such files by referencing local resources from within the email content.

To achieve this, the attacker embeds references to local files (e.g., images) using absolute paths or \texttt{file:} URLs inside the HTML email body. If the recipient’s email client resolves such local references during rendering, the referenced content is loaded from the recipient’s own device and displayed inside the email. 

We describe the following implementations of this deception technique.

\subsubsection{Implementation: Guessable Local Image Paths (On-screen Intimidation)}
\label{sec:local-resource-inclusion-guessable-local-image-paths}

This implementation (identified during our examination) exploits that some email clients (or configurations) resolve local file references in HTML emails when rendering images. The attacker guesses likely local paths and filenames (e.g., default picture folders and camera naming patterns) and embeds them as image sources. This can be scaled by including multiple \texttt{<img>} elements (i.e., brute-forcing a small set of common paths/filenames) until at least one resolves on the recipient’s device.

\begin{figure}[H]
    \centering
     \includegraphics[width=1\columnwidth]{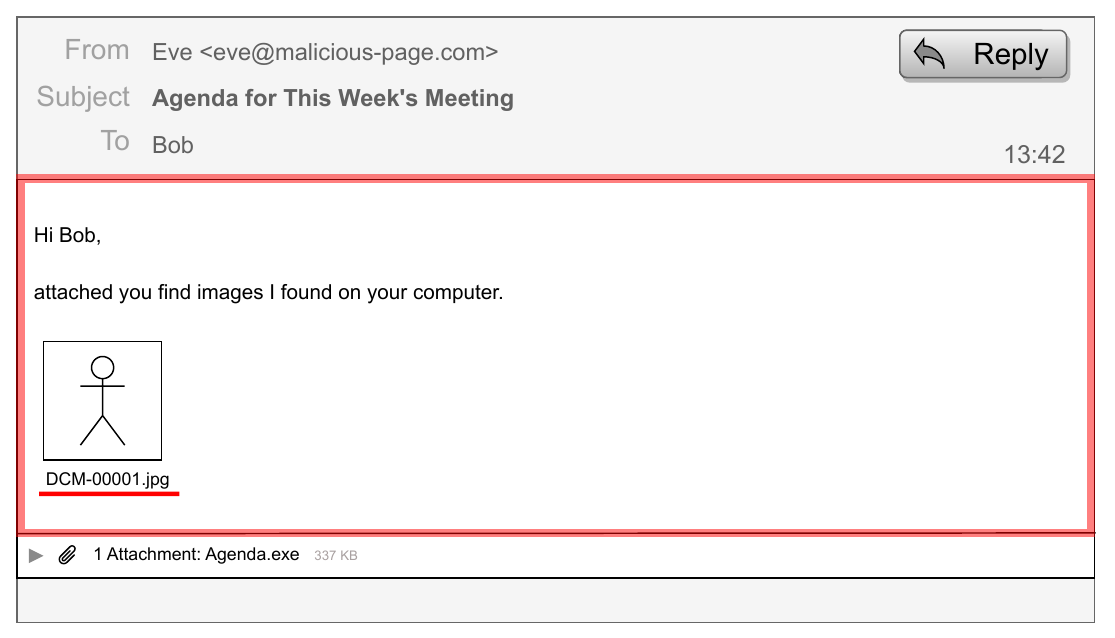}
     \caption{Example of Local Resource Inclusion (Implementation: Guessable Local Image Paths).}
     \label{fig:nlocal-resource-inclusion-guessable-local-image-paths}
     \vspace{-1em}
 \end{figure}

 As shown in~\cref{fig:nlocal-resource-inclusion-guessable-local-image-paths}, the attacker embeds an image reference such as
\texttt{file:///C:/Users/Bob\allowbreak{}/Pictures/DCM-00001.jpg}.
When Bob opens the email, the email client loads and displays the image from Bob’s local \texttt{Pictures} folder. The attacker uses the displayed local image to claim access to Bob’s private files (e.g., for extortion).

\subsubsection{Implementation: Reply-induced Embedding of Local Resources}
\label{sec:local-resource-inclusion-reply-induced-embedding}

This implementation (identified during our examination) exploits that some email clients, when replying to or forwarding an HTML email, may resolve referenced resources and embed them into the outgoing message (e.g., as inline MIME parts), rather than preserving the original local file path reference. The attacker therefore first sends an email that references local resources, and then socially engineers the recipient to reply (or forward) so that the email client includes the resolved local content in the response, which the attacker can receive.

\begin{figure}[H]
     \centering
     \includegraphics[width=1\columnwidth]{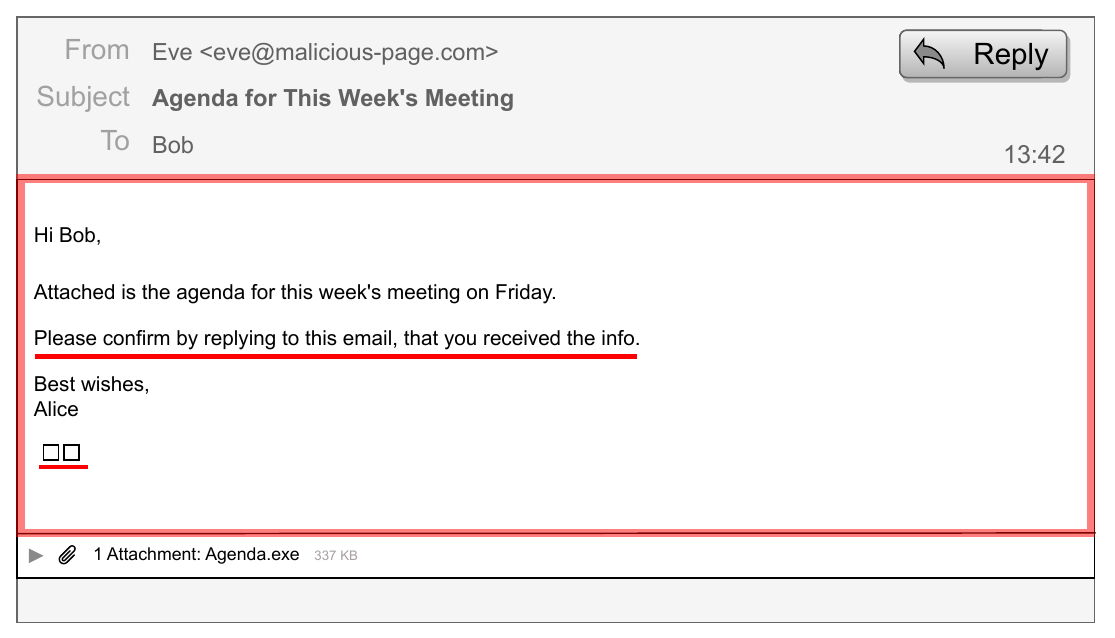}
     \caption{Example of Local Resource Inclusion (Implementation: Reply-induced Embedding of Local Resources).}
     \label{fig:nlocal-resource-inclusion-reply-induced-embedding}
     \vspace{-1em}
 \end{figure}

As shown in~\cref{fig:nlocal-resource-inclusion-reply-induced-embedding}, the attacker's email contains a local image reference:
\begin{center}
\texttt{\textless{}img src=\textquotedbl{}file:///C:/Users/Bob/\allowbreak{}Pictures/\allowbreak{}DCM-00001.jpg\textquotedbl{}\textgreater{}}
\end{center}
The attacker prompts Bob to reply (e.g., ``Please confirm by replying to this email.''). When Bob replies, the email client embeds the resolved image into the outgoing email, so the attacker receives the image content as part of Bob's response (instead of the original local path).

%% file: sections/040_conclusion.tex
\section{Conclusion}
\label{sec:conclusion}

This research note presents a comprehensive list of email-based deception techniques targeting the sender, link, and attachment security indicators, as well as techniques that exploit properties of the email rendering environment more broadly. For each technique, we describe the underlying mechanism in its pure form — demonstrating how the deception works in isolation using a consistent narrative and example implementations. While Veit et al.~\cite{Veit2024} provide a structured taxonomy and evaluate email client susceptibility, this note focuses on an extended list of deception techniques with an emphasis on how each works conceptually, rather than on evaluating effectiveness or real-world severity. Deliberately, we do not explore combinations of techniques or particularly optimized variants: each technique is documented as an individual building block, intended as a starting point for addressing them systematically.

Preventing email-based deception is unlikely to be solved by a single measure. For each technique, the most effective countermeasure may differ — some are more naturally addressed at the infrastructure level (e.g., authentication protocols), others through improvements to email client user interfaces, and others through security awareness training. In practice, these layers overlap and share responsibility, and addressing any one of them contributes to reducing the overall attack surface. Future work should propose, apply, and evaluate concrete solutions across these dimensions. One area warranting particular attention is the use of QR codes embedded in emails: by shifting the link interaction to a QR code scanner, the attack surface moves outside the email client entirely, and the scanner itself may face analogous deception problems that deserve dedicated investigation.

%% file: sections/050_acks.tex
\section*{Acknowledgments}
This work was supported by funding from the topic Engineering Secure Systems, subtopic 46.23.01 Methods for Engineering Secure Systems, of the Helmholtz Association (HGF) and by KASTEL Security Research Labs.

%% file: sections/060_appendix.tex
This work was supported by funding from the topic Engineering Secure Systems, subtopic 46.23.01 Methods for Engineering Secure Systems, of the Helmholtz Association (HGF) and by KASTEL Security Research Labs.